%% file: main.tex
\documentclass[11pt]{article}
\usepackage{fullpage}

\usepackage[natbib=true,
            style=numeric,
            citestyle=numeric,
            maxbibnames=10,
            doi=false,
            isbn=false,
            url=false]{biblatex}
\usepackage{amsmath,amssymb,amsthm}
\usepackage{booktabs} % For formal tables
\usepackage{xspace}
\usepackage[hidelinks]{hyperref}
\usepackage{accents} % for underbar
\usepackage{pgf} % for pgf figure; could be removed
\usepackage{xparse} % for DeclareDocumentCommand
\usepackage{bbm}
\usepackage{enumitem}
\usepackage{array}
\usepackage{multirow}

\usepackage{fixme}
\fxsetup{multiuser,layout={inline,index}}
\fxusetheme{colorsig}
\FXRegisterAuthor{bm}{abm}{JBM}
\FXRegisterAuthor{sc}{asc}{SC}
\FXRegisterAuthor{kg}{akg}{KG}
\FXRegisterAuthor{mp}{amp}{MP}

\newtheorem{lemma}{Lemma}
\newtheorem{theorem}{Theorem}

\theoremstyle{definition}
\newtheorem{definition}{Definition}
\newtheorem{example}{Example}

\newenvironment{proofof}[1]{\vspace{0.1in}\noindent{\sc Proof of #1.}}{\hfill\qed}

\newenvironment{numberedtheorem}[1]{%
\renewcommand{\thetheorem}{#1}%
\begin{theorem}}{\end{theorem}\addtocounter{theorem}{-1}}

\newenvironment{numberedlemma}[1]{%
\renewcommand{\thetheorem}{#1}%
\begin{lemma}}{\end{lemma}\addtocounter{theorem}{-1}}

\input{macros}

\begin{document}
\title{Revenue Maximization with an Uncertainty-Averse Buyer}
\author{Shuchi Chawla%
\thanks{%
    {University of
      Wisconsin-Madison}. {\tt shuchi@cs.wisc.edu}. Supported in part
    by NSF awards CCF-1320854 and CCF-1617505.}
\and Kira Goldner%
\thanks{%
    {University of Washington},
    {\tt kgoldner@cs.washington.edu}. Supported in part by NSF award
    CCF-1420381 and by a Microsoft Research PhD Fellowship.}
\and J. Benjamin Miller%
\thanks{%
    {University of Wisconsin-Madison},
    {\tt bmiller@cs.wisc.edu}. Supported by a Cisco graduate fellowship.}
\and Emmanouil Pountourakis%
\thanks{%
    {University of Texas at Austin},
    {\tt manolis@utexas.edu}. Supported in part by NSF awards CCF-1733832,
    CCF-1350823, CCF-1216103, and CCF-1331863.}}

\maketitle

\begin{abstract}
\input{abstract}
\end{abstract}
\newpage

\input{intro}
\input{model}
\input{one-shot}
\input{dynamic}

\section*{Acknowledgements}
We thank Evdokia Nikolova for comments on our model of risk.

% Bibliography
%\bibliographystyle{plainnat}
%\bibliography{risk}
\printbibliography

\newpage
\appendix
\input{app-proofs}

\end{document}

%% file: macros.tex
\newcommand{\R}{\mathbb{R}}
\newcommand{\half}{\frac12}
\newcommand{\ubar}[1]{\underaccent{\bar}{#1}}

\newcommand{\supdif}{{\bar{\partial}}}
\newcommand{\subdif}{{\ubar{\partial}}}
\newcommand{\argmax}{\operatorname{arg\,max}}

\newcommand{\prob}[2][]{\textrm{\bf Pr}\ifthenelse{\not\equal{}{#1}}{_{#1}}{}\!\left[#2\right]}
\newcommand{\expect}[2][]{\textrm{\bf E}\ifthenelse{\not\equal{}{#1}}{_{#1}}{}\!\left[#2\right]}
\newcommand{\given}{\,\middle|\,}

\newcommand{\rprof}{y}

\newcommand{\convU}{\ubar{u}}
\newcommand{\wt}{\rprof}
\newcommand{\wte}[1][\epsilon]{\rprof_{#1}}
\newcommand{\wtinv}{\rprof}

\newcommand{\rae}[1][\wt]{\expect[#1]}
\DeclareDocumentCommand\rau{O{\wt} m}{u_{#1}(#2)}
\DeclareDocumentCommand\drau{O{\wt} m}{u'_{#1}(#2)}
\DeclareDocumentCommand\ddrau{O{\wt} m}{u''_{#1}(#2)}
\DeclareDocumentCommand\lrau{O{\wt} m G{X, P}}{u_{#1}(#2, (#3))}
\DeclareDocumentCommand\dlrau{O{\wt} m G{X, P}}{u'_{#1}(#2, (#3))}
\newcommand{\lsupp}{L}

\newcommand{\dist}{F}

\newcommand{\mech}{\mathcal{M}}
\newcommand{\allocs}{\mathbf{X}}
\newcommand{\paymts}{\mathbf{P}}
\newcommand{\alloc}[1][]{X\ifthenelse{\not\equal{}{#1}}{_{#1}}{}}
\newcommand{\paymt}[1][]{P\ifthenelse{\not\equal{}{#1}}{_{#1}}{}}
\newcommand{\lottery}[1][]{\alloc[#1], \paymt[#1]}
\newcommand{\lotteryp}[1][]{\left(\lottery[#1]\right)}
\newcommand{\pdist}{\dist_{\paymt}}
\newcommand{\ppmf}{f_{\paymt}}

\newcommand{\rev}{{\normalfont\textsc{Rev}}}
\newcommand{\opt}{{\normalfont\textsc{OPT}}}
\newcommand{\myer}{{\normalfont\textsc{Mye}}}
\newcommand{\menu}{\mathcal{M}}
\newcommand{\profiles}{\mathcal{Y}}

\newcommand{\vareps}{\varepsilon}
\newcommand{\eps}{\varepsilon}

\newcommand{\mechl}{\texttt{PP}_{\ell}}
\newcommand{\lp}[1][p]{\ell({#1})}
\newcommand{\lpp}[1][p]{\ell'({#1})}
\newcommand{\lpb}[1][p]{\tilde\ell({#1})}
\newcommand{\lpbp}[1][p]{\tilde\ell'({#1})}
\newcommand{\g}{\texttt{U}}%_2

\newcommand{\peff}{\mathcal{P}_{\textrm{eff}}}

\newcommand{\pmin}{p_{\min}}

\newcommand{\wtfam}{\profiles}

%% file: abstract.tex
% Most work in mechanism design assumes that buyers are risk neutral; some
% considers risk aversion arising due to a non-linear utility for money. Yet
% behavioral studies have established that real agents exhibit risk attitudes
% which cannot be captured by any expected utility model. We initiate the study
% of revenue-optimal mechanisms under behavioral models beyond expected utility
% theory. We adopt a model from prospect theory which arose to explain these
% discrepancies and incorporates agents under-weighting uncertain outcomes. In
% our model, an event occurring with probability $x < 1$ is worth strictly less to
% the agent than $x$ times the value of the event when it occurs with certainty.
% 
% We present three main results. First, we characterize optimal mechanisms as
% menus of two-outcome lotteries. Second, we show that under a reasonable
% bounded-risk-aversion assumption, posted pricing obtains a constant
% approximation to the optimal revenue. Notably, this result is ``risk-robust''
% in that it does not depend on the details of the buyer's risk attitude. Third,
% we consider dynamic settings in which the buyer's uncertainty about his future
% value may allow the seller to extract more revenue. In contrast to the positive
% result above, here we show it is not possible to achieve any constant-factor
% approximation to revenue using deterministic mechanisms in a risk-robust
% manner.

Most work in mechanism design assumes that buyers are risk neutral; some
considers risk aversion arising due to a non-linear utility for money. Yet
behavioral studies have established that real agents exhibit risk attitudes
which cannot be captured by any expected utility model. We initiate the study
of revenue-optimal mechanisms under buyer behavioral models beyond expected
utility theory. We adopt a model from prospect theory which arose to explain
these discrepancies and incorporates agents under-weighting uncertain outcomes.
In our model, an event occurring with probability $x < 1$ is worth strictly
less to the agent than $x$ times the value of the event when it occurs with
certainty.

In contrast to the risk-neutral setting, the optimal mechanism may be
randomized and appears challenging to find, even for a single buyer and a
single item for sale. Nevertheless, we give a characterization of the optimal
mechanism which enables positive approximation results. In particular, we show
that under a reasonable bounded-risk-aversion assumption, posted pricing
obtains a constant approximation. Notably, this result is “risk-robust” in that
it does not depend on the details of the buyer's risk attitude. Finally, we
examine a dynamic setting in which the buyer is uncertain about his future
value. In contrast to positive results for a risk-neutral buyer, we show that
the buyer's risk aversion may prevent the seller from approximating the optimal
revenue in a risk-robust manner.

%% file: intro.tex
\section{Introduction}

Most work in mechanism design is based on the assumptions that agents
have quasilinear utilities and are expected utility maximizers. %  In
% particular, if the monetary value of the outcome of a process is
% (random variable) $v$, then the agent derives utility $v$ from the
% process, and behaves in a manner so as to maximize its expectation
% $\expect{v}$.
Owing to its linearity, this model of behavior is mathematically
simple and allows for a number of beautiful characterizations of
optimal and near-optimal mechanisms. The few exceptions to this line
of work employ \citet{neumann44a}'s %von Neumann-Morgenstern
expected utility theory (EUT): assuming that agents are expected
utility maximizers, although they may have a non-linear utility for
money.\footnote{In particular, if the monetary value of the outcome of
  a process is (random variable) $v$, then the agent derives a utility
  $u(v)$, where $u$ is some non-linear function, and aims to maximize
  $\expect{u(v)}$. In the risk-neutral case, $u$ is the
  identity function.} Reality, however, is more complex. Numerous
experiments, including the famous Allais paradox \citep{Allais53},\footnote{Subjects are
  asked to choose between two options: option A rewards \$1M with
  certainty; option B rewards \$1M with probability 89\%, increases
  the reward to \$5M with probability 10\%, but rewards nothing with
  the remaining 1\% probability. Most people prefer the certain reward
  to the small-probability risk of getting nothing, even though the
  latter is counterbalanced by the possiblity of a much larger
  reward. Subjects are then asked to choose between two other options
  that are modifications of the first two: option C has a reward of
  \$1M with 11\% probability and nothing otherwise; option D rewards
  \$5M with probability 10\%, or nothing with the remaining 90\%
  probability. Each of C and D are obtained from options A and B by
  reducing the probability of receiving the \$1M reward by a fixed
  amount and replacing it with no reward.  The paradox is that in the
  second case, most people choose option D over option C. No
  assignment of utilities to the two amounts can explain this
  ``switch'' in preferences across the two experiments. This paradox
  can be explained through prospect theory, including the special case
  that we study.} have uncovered behavior that cannot be captured by EUT.

Among the work in behavioral game theory that attempts to fit a
mathematical model onto observed attitudes, the most accepted non-EUT
theory is \citet{prospect}'s Prospect Theory (PT) and its
extensions. Prospect theory hypothesizes that risk attitudes arise not
only from a non-linear utility for money but also from agents'
perception of random outcomes. For example, the risk attitude of a
buyer who prefers \$50 with certainty over \$100 with probability 1/2
can be explained in two ways: as in EUT, the buyer may value \$100 at
less than twice \$50; alternately, the buyer may value obtaining \$100
with probability 1/2 at less than half his value for \$100. Thus, {\em
  aversion to uncertainty}\footnote{In this paper, the term
  ``uncertainty'' refers to chance or randomness rather than Knightian
  uncertainty. We sometimes use the term ``uncertainty aversion'' to
  distinguish prospect-theoretic risk attitudes from EUT-style risk
  attitudes. Elsewhere in the paper, however, risk will refer to PT-style
  risk.}
% Throughout the paper, we use the
%   terminology ``chance-aversion'' to denote this latter type of agent
%   behavior, to distinguish it from standard EUT risk-aversion. The
%   term ``uncertaintly-aversion'' usually refers to Knightian
%   uncertainty.}
provides an alternate explanation for this example,
with or without a concave utility for money.

A basic premise of PT is that agents
fundamentally misvalue random events. On the other hand, much of
optimal mechanism design relies on the use of randomness, both in
exploiting the agent's uncertainty about the environment and
other agents' types, as well as in offering randomized outcomes or
lotteries. As such, it is imperative to understand whether the main
insights and results of mechanism design continue or fail to hold
under risk attitudes described by prospect theory.

% {\em The goal of this paper is to revisit some basic mechanism design
% questions for risk-averse agents, and examine whether the main
% insights and results of mechanism design continue to hold in those
% settings.}
%
% I don't like this sentence:
%   1. "The goal of this paper"
%   2. As Kira points out, we're not the first to revisit MD for risk-averse
%   agents, just the first for prospect theory.
%   3. We don't touch on a lot of basic results, such as multiple buyers.
%

%{\em The goal of this paper is to revisit some basic mechanism design questions for risk-averse agents who underweight uncertain outcomes,  and examine whether the main insights and results of mechanism design  continue to hold in those settings.}

In this paper we revisit some basic mechanism design questions for
risk-averse agents who undervalue random outcomes. While PT has been
studied extensively within behavioral economics, to our knowledge
ours is the first work to consider its implications for mechanism
design.\footnote{See a complete discussion of related work at the end
  of this section.}  We consider the setting of selling one or two
items to a single buyer, with the goal of maximizing the seller's
revenue. 
% First, we ask whether uncertainty aversion affects the extent
% to which randomization is employed in optimal mechanisms. Second, in
% many mechanism design settings, the seller can exploit the buyer's
% lack of knowledge about certain parameters (e.g., other agents' values
% or the buyer's own future value) to extract more revenue. Does
% uncertainty aversion help or hinder such revenue extraction?

% The goal of this paper is to understand the implications of risk
% attitudes as described by prospect theory on mechanism design. We
% consider the setting of selling one or two items to a single buyer,
% with the goal of maximizing the seller's revenue. We consider two
% aspects of this question. First, randomization is an important tool in
% the design of optimal mechanisms. How does aversion to uncertainty
% affect the extent to which randomization helps? Second, in many
% mechanism design settings, the seller can exploit the buyer's lack of
% knowledge about certain parameters (e.g., other agents' values or the
% buyer's own future value) to extract more revenue. Does uncertainty
% aversion help or hinder such revenue extraction?

% The goal of this paper is to revisit basic questions in mechanism
% design when buyers are averse to uncertain outcomes, and to examine
% whether the main insights and results of mechanism design continue or
% fail to hold in these settings.

%under a novel model of risk aversion, which is derived from prospect theory. and examine whether the main insights and results of mechanism design continue to hold in those settings.}

\subsubsection*{A basic model for aversion to risk.}

In general, prospect theory incorporates both an agent's non-linear utility for 
value and a non-linear attitude toward probability. The buyer's net utility from 
a random value is then the weighted (non-linearly over outcomes) expectation 
of the non-linear utility
function applied to the value. We aim to isolate the effect of this non-linear
weighting of probability on mechanism design, and therefore consider a
simpler special case in which the utility function is linear but the buyer maximizes
weighted expected utility.

In our model, if the agent is offered a gain of $u$ with probability $x$,
his utility from this random outcome is given by $u\times
\wt(x)$. The {\em weighting function} $\wt$ maps probabilities to
weighted probabilities, $\wt:[0,1]\rightarrow [0,1]$, with $\wt(0)=0$
and $\wt(1)=1$. It encodes how strongly the agent dislikes
randomness. If $\wt$ is convex, the agent displays risk-averse
behavior---the less likely an event is, the more the agent
discounts his value. On the other hand, concavity captures
risk-seeking behavior. We focus on risk aversion and assume throughout
the paper that % $\wt(x)\le x$ for all $x\in[0,1]$. Further, for
% simplicity, we assume that
the weighting function is convex.

More generally, we define the notion of weighted expectation, a.k.a.
risk-averse expectation, of a random variable taking on many different
values.\footnote{In the context of a multi-stage mechanism or
  protocol, we define the agent's utility as the risk averse
  expectation of the random variable that represents the agent's net
  earnings at the end of the process. Importantly, risk averse
  expectation is not linear, and so does not separate neatly into
  contributions from each step in the process.}  Our definition has a
clean mathematical formulation in terms of the random variable's
distribution. For example, just as the expectation of a non-negative
random variable with c.d.f. $F$ can be written as an integral over
$1-F(x)$, the weighted expectation is an integral over
$\wt(1-F(x))$. This definition follows from requiring that certain
events do not affect the agent's evaluation of random events. In
particular, for a constant $c$ and random variable $X$, the weighted
expectation of $c+X$ is just $c$ plus the weighted expectation of
$X$. The useful implication for mechanism design is that the agent's
wealth does not affect his preferences within the mechanism.

% In other words, events that
% happen with certainty don't affect the agent's evaluation of uncertain
% events. On the other hand, if $X_1$ and $X_2$ are independent
% Bernoulli r.v.s, then the weighted expectation of $X_1+X_2$ can be
% much larger (but no smaller) than the sum of the two weighted
% expectations---the sum is more concentrated than the individual
% variables, and so its utility is discounted to a lesser extent.

\subsubsection*{Single-shot mechanism design.}
 
We begin our investigation with the simplest mechanism design setting,
namely a monopolist selling a single item to a single buyer. When the
buyer is risk-neutral, it is well known that the revenue-optimal
mechanism is a deterministic mechanism, namely a posted
price. Relative to a risk-neutral buyer, a risk-averse buyer reacts to
a deterministic posted price the same way, but undervalues mechanisms
that employ randomness. Therefore one might expect that the optimal
single-item mechanism for a risk-averse buyer continues to be a posted
price. This turns out to not be the case --- by offering lotteries
alongside certain outcomes, the mechanism can exploit the buyer's aversion to
risk to extract more revenue for the sure outcomes.  As risk-aversion
grows, the seller can extract nearly all of the buyer's
value.\footnote{Similar results are also known to arise from extreme
  aversion to risk within EUT, as in the work of \citet{Matthews83},
  for example.}

What do optimal mechanisms look like? We think of mechanisms as menus
where each option corresponds to a (correlated) random allocation and
random payment. We show that revenue-optimal menus are composed of
{\em binary or two-outcome lotteries}, where each lottery sells the item at a certain
price with some probability.\footnote{Note that in our model, in
  contrast to the risk-neutral setting, charging a price upfront for a
  randomized allocation and charging a price only when the item is
  allocated are not equivalent. The former is generally
  worse than the latter in terms of the revenue it generates.} In
particular, it doesn't help the mechanism to offer menu options with
multiple random outcomes. Given this format, we can express the
payment as a function of the allocation rule in the form of a payment
identity. Unfortunately, this payment identity is not linear in the
allocation rule, and so does not permit a closed form solution for the
optimal mechanism.

\subsubsection*{Risk robustness.} The theory of optimal mechanism
design is often criticized for being too detail-oriented, and
consequently impractical. Designing optimal mechanisms in the settings
we consider is even less practical and realistic in this regard---the
seller needs to know not only the buyer's value distribution but also
his weighting function exactly. Indeed it is difficult to imagine that
a buyer can describe his own weighting function in any manner other
than reacting to options presented to him. We therefore consider the
problem of designing mechanisms that achieve revenue guarantees robust
to the precise risk model.

Formally, given a family of weighting functions, we ask for a single
mechanism that for every weighting function in the family is
approximately optimal with respect to the revenue-optimal mechanism
specific to that weighting function.
%Here we measure ``closeness'' in
%terms of the multiplicative loss in revenue or the approximation
%factor.
While it appears challenging to obtain a constant-factor risk-robust
approximation\footnote{In Section~\ref{sec:risk-robust-oneshot} we
  give an $O(\log\log H)$ risk-robust approximation under certain
  assumptions, where $H$ is the ratio of the maximum to the minimum
  value in the buyer's value distribution.}  for arbitrary families of
weighting functions, we obtain a simple approximation under a boundedness
condition on the buyer's risk attitude. Specifically, we show that such a
result is possible under a certain boundedness condition on the buyer's risk
attitude. As long as there is some probability $x = 1-\Theta(1)$ at which the
buyer's weight is $y(x)=\Theta(1)$, then Myerson's optimal posted pricing
mechanism achieves an $O(1)$ approximation to the optimal revenue. In other
words, the only ``bad'' case for revenue maximization, where the optimal
revenue is far greater than that achievable from a risk neutral buyer, is when
the buyer values any event with probability bounded away from 1 at a weight
arbitrarily close to 0.

The implication for market design is that deterministic mechanisms continue to
perform well as long as the buyer's aversion to risk is bounded.

% Of course optimizing for revenue requires the seller to know the
% buyer's risk profile exactly. This is unrealistic. Buyers themselves
% may not know their uncertainty weighting function exactly. Motivate
% risk robustness. What revenue guarantee can we achieve in a
% risk-robust manner? Observe that Myerson's posted price is already a
% log H approximation. We show that under an assumption about the risk
% profiles, there exists a mechanism that achieves a log log H risk
% robust approximation, but we don't know how to compute it. Is there a
% simple risk robust approximation? Yes. Explain the guarantee achieved
% by Myerson. As long as there is some x=1-Theta(1) probability at which
% y(x) is Theta(1), we get a O(1) approximation. In other words, there
% is some probability bounded away from one, where the buyer gets
% utility bounded away from 0.  

% (Other results in one-shot setting?)

\subsubsection*{Dynamic mechanism design.}
We next consider a sequential sale of two items to the buyer.
Consider the following two-stage mechanism design problem. The seller has one
item to sell in each stage. During the first stage, the buyer knows his value
for the first item, but not for the second item. At this time, the seller may
charge the buyer a higher or lower price for the first item in exchange for a
second-stage mechanism that is more or less favorable, respectively, to the
buyer. In the second stage, the buyer's value for the second item is realized.
The seller follows through with his commitment and sells the second item
according to the mechanism promised in the first stage. Several recent works
have studied these kinds of dynamic mechanism design settings with risk-neutral
buyers.

The setting we consider is a special case of that introduced by~\citet{PPPR16}
and subsequently studied by \citet{ADH16} and \citet{MLTZ16}.
\citeauthor{ADH16} show that in the optimal mechanism the seller generally
charges higher prices in the first stage in exchange for a higher expected
utility promised to the buyer in the second stage. In fact in some cases it
becomes possible to extract the smaller of the buyer's expected values for the
two items\footnote{To be precise, if the buyer's values are denoted $v_1$
  and $v_2$ respectively, these mechanisms can extract
  $\expect{\min(v_1, \expect{v_2})}$ plus the single-shot revenues in
  the two stages.}. \citeauthor{ADH16}'s mechanisms have an unusual
format, however. The mechanisms offer a menu to the buyer in which the
buyer's net utility from every menu option is zero (even accounting
for future gains). Since the buyer is now indifferent between all of
the menu options, he by default picks the most expensive one he can
afford. Observe that the revenue guarantee of this mechanism is quite
fragile with respect to the buyer's risk attitude. Since different
menu options have different amounts of risk involved, if the
buyer's risk attitude changes, the options are no longer all
equivalent and the mechanism loses significant revenue. Is a risk-robust
approximation achievable in the dynamic setting?

%We ask: in the two-stage dynamic setting can the seller extract much
%more than the single shot revenue in each stage in a risk-robust manner? 

We first show that the kinds of mechanisms and revenue guarantees that
\citeauthor{ADH16} obtain in the risk-neutral setting continue to hold
in risk-averse settings, when the seller knows the buyer's weighting
function. We focus on the simple class of posted pricing mechanisms
where each menu option offers the buyer a fixed price in each stage,
but the second-stage price is a decreasing function of the first-stage
price. Beyond the single-shot revenues in both stages, posted price
mechanisms can obtain an additional revenue of
$O(\expect{\min(v_1, \rae{v_2})})$, where $\rae{v_2}$ is the
risk-averse or weighted expectation of $v_2$, but not much more.

We then explore whether it is possible to extract a constant fraction
of $\expect{\min(v_1, \rae{v_2})}$ via posted price mechanisms in a
risk-robust manner. We construct a family of value distributions and a
family of weighting functions such that the buyer's risk-averse
expectation of his second-stage value exhibits a large range under
different weighting functions, although all of the weighting functions
satisfy the boundedness condition discussed previously. We then show
that for any constant $\alpha>1$, there exists a value
distribution in the family such that no posted pricing mechanism can
obtain an $\alpha$ risk-robust approximation to revenue with respect
to all of the weighting functions we consider.

% The moral of these results is that in settings where the seller seeks to
% exploit the buyer's lack of information and the inherent risk in future
% outcomes for gains in revenue, the seller needs precise information about the
% buyer's attitude towards risk, without which such revenue extraction is not
% possible.

The moral of these results is that, in order to for a seller to increase her
revenue by exploiting the buyer's lack of information and the inherent risk in
future outcomes, she needs precise information about the buyer's attitude
towards risk, without which such revenue extraction is not possible.

\subsubsection*{A summary of our results.}

Our main results can be summarized as follows.
\begin{itemize}[leftmargin=*]
\item Optimal mechanisms in the single-shot setting are menus
  of binary lotteries. See Theorem~\ref{thm:single-shot-opt}.
\item In the single-shot setting, optimal mechanisms can obtain much
  more revenue than Myerson's mechanism; however, under a natural
  condition bounding the extent of risk-aversion, Myerson's mechanism
  extracts a constant fraction of the optimal revenue. See
  Theorems~\ref{thm:extract-sw} and \ref{thm:risk-robust-myer}.
\item For the dynamic two-stage setting, we present an upper bound as
    well as a $2$-approximation to the revenue achievable using
    posted-price mechanisms. See Theorems~\ref{lem:pp-ub} and
    \ref{lem:pp-lb}.
  \item We exhibit an example that shows that it is impossible to
    obtain any constant factor risk-robust approximation to revenue in
    the two-stage setting, even if the buyer's risk aversion is
    bounded. See Theorem~\ref{thm:2day-lb}. %Section~\ref{sec:risk-robust-lb}.
\end{itemize}

\subsubsection*{Related work.}

% 1. Prospect theory
Although empirically successful, prospect theory as defined by
\citet{prospect} suffers from a number of weaknesses, rectified in a
series of works subsumed by the cumulative prospect theory\footnote{In
  modern usage, ``prospect theory'' is understood to mean this
  improved theory and its extensions.} of \citet{Tversky1992}. Our
model is readily seen to be a special case of \citet{KSW11}'s
extension of cumulative prospect theory to continuous
values.\footnote{Other than assuming that the utility $u$ is the
  identity function, we also assume that the probability weighting
  function for losses is related to the probability weighting function
  for gains in a manner that satisfies the additivity axiom. See
  Section~\ref{sec:model}. In general, PT allows these to be different
  functions.}  See \citep{non-eut-survey} for a survey of non-EUT
theories.

% 2. Risk aversion and mechanism design

    % a. eut mechanism design
Despite the success of prospect theory, expected utility theory remains the
standard in mechanism design, where a large body of work studies
revenue-optimal mechanism design. For example, \citet{HMZ10} compare different
auction formats.  More relevant, \citet{Matthews83} and \citet{MR84}
characterize the optimal mechanism under certain expected-utility models; our
Theorems~\ref{thm:single-shot-opt} and \ref{thm:extract-sw} (characterization
of the optimal mechanism and full welfare extraction under extreme risk
aversion, respectively) have close analogues in their work.  Unsurprisingly,
these characterizations are complex and work in limited settings.

    % b. non-eut mechanism design (and game theory)
Non-EUT models have thus far attracted less attention in the mechanism design
community. \citet{FP10} study existence and computation of equilibria in a
variety of models, including special cases of prospect theory.  \citet{EG15}
explore the implications of a realistic, prospect-theoretic behavioral model in
contract design. \citet{AG12} look at designing gambles for buyers with
prospect-theoretic attitudes, but this is not relevant in problems where buyers
have a type that must be elicited.  To our knowledge, we are the first to
consider revenue maximization within any non-EUT model.

% 3. Robustness and risk aversion

Much recent work in algorithmic mechanism design has explored revenue
guarantees that are robust to finer details of the model. To our
knowledge, however, only three works consider robustness to the
buyer's risk attitude.  \citet{DP12} show that truthful-in-expectation
mechanisms can be implemented almost as-is in a manner robust to
risk attitudes. \citet{FHH13} and \citet{CDKS16} provide
risk-robust revenue guarantees in (different) stylized settings;
\citeauthor{FHH13}'s mechanism is additionally independent of the
buyers' value distributions.  Their techniques are unrelated to
ours.\footnote{\citet{CDKS16}'s guarantees work for any risk model
  where the utility drawn from a random outcome is no more than the
  expected value of the outcome. The other works consider EUT models
  of risk.}

%% file: model.tex
\section{A Model for Risk Aversion}
\label{sec:model}

%\subsection{The uncertainty weighting function}
A major premise of prospect theory, inherited from the rank-dependent expected
utility of \cite{Quiggin82}, is that the agent fundamentally misvalues random
events. % \footnote{Prospect theory also allows nonlinear utility for
  % money, and is thus a strict generalization of expected utility theory; we
  % focus on inherent aversion to randomness and therefore assume a linear
  % utility function.} 
While prospect theory allows for both risk-averse and
risk-seeking attitudes, we consider only risk-averse buyers. Thus, if the agent
gains value $v$ with probability $x$, his {\em risk-averse utility} from this
random event is $\wt(x) v$ where $\wt(x)\le x$. The function $\wt$ is called
a {\em probability weighting function}.  Prospect theory requires that the
weighting function satisfy the following properties: (1) $\wt : [0,1] \to
[0,1]$, and (2) $\wt(0) = 0$ and $\wt(1) = 1$. Because we are interested in
risk-averse behavior, we additionally assume (3) $\wt$ is weakly increasing and
convex.

Given such a weighting function $\wt$, we next describe how to compute the {\em
risk-averse expectation}, $\textrm{\bf E}_{\wt}$, of a random variable $V$ that
denotes the random value that an agent gains.  Suppose, for
example, that an agent gains a value of \$1 with probability $1/2$ and \$2 with
probability $1/2$. Then we observe that the agent gets a value of \$1 with
probability 1, and an increment of \$1 with probability $1/2$. So, the
risk-averse expectation of his value, a.k.a. his risk-averse utility, ought to
be $(\$1) + (\$1)\times \wt(1/2)$, because he faces no risk over the
first \$1. Our definition is therefore designed to satisfy the following {\em
additivity axiom}: 
\begin{align*}
  \text{For any constant $c$,} \quad \rae{c+V} & = c+\rae{V}. 
\end{align*}
The additivity axiom implies, in particular, that the utility of an agent from
participating in a mechanism does not depend on the wealth of the agent, but
rather only depends on how much the agent gains or loses in the mechanism.
Accordingly, we express the risk-averse expectation in the form of increments
from a base value.  Formally, the risk-averse expectation of a non-negative
random variable is defined as follows.
\begin{definition}
    Let $Z$ be a random variable supported over $[0,\infty)$ with c.d.f. $F$.
    Then the {\em risk-averse expectation} of the random variable with respect
    to weighting function $\wt$ is
    \label{def:rae}
    \[
       \rae{Z} = \int_0^\infty \wt(1-F(z)) dz.
   \]
\end{definition}
To understand this definition, observe that for any $z$, $dz$ is the
difference between $z-dz$ and $z$, and this difference is earned with
probability $1-F(z)$. Compare this against the standard definition of
expectation:
    \[
       \expect{Z} = \int_0^\infty z\, dF(z) = \int_0^\infty (1-F(z)) dz.
   \]

\subsubsection*{The weighting function for losses.}
\begin{figure}[htbp]
    \begin{center}
    \scalebox{1}{\input{figures/expectedVsWeighted_a.pgf}}~
    \scalebox{1}{\input{figures/expectedVsWeighted_b.pgf}}
    \caption{The transformation of a c.d.f. $F(z)$ by a weighting function
        $\wt$ as described in Definition~\ref{def:rae-both}. {\em Left:} The
        expected value of a random variable $Z$ drawn from $F$ is equal to the
        difference between the shaded areas: the area above the curve on the
        positive axis adds to the expectation, and the area below the curve on
        the negative axis subtracts.  {\em Right:} The risk-averse expectation
        is defined to be the expectation of the transformed curve. After
        transformation by $\wt$, the positive area has decreased while the
        negative area has increased. The risk-averse expectation is therefore
        less than the true expectation. The original c.d.f. is shown as a
        dotted line for comparison.}
    \label{fig:exp_vs_wted}
    \end{center}
\end{figure}
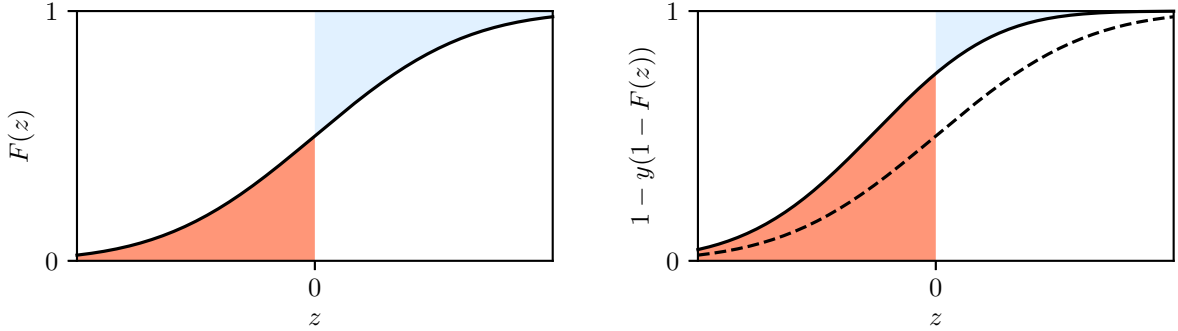

When the random variable $Z$ takes on negative values, we need to take extra
care in defining its risk-averse expectation. Once again, following the
additivity axiom stated above, for $\lsupp = \inf\{z : F(z) > 0\}$, we define 
\begin{align*}
    \rae{Z} &= \lsupp + \rae{Z-\lsupp} \\
        &= \lsupp + \int_0^\infty \wt(1-F(z+\lsupp))dz  \\
        &= \lsupp + \int_\lsupp^\infty \wt(1-F(z)) dz.
\end{align*}
It is convenient to express the contributions of the negative values that $Z$
takes to the risk-averse expectation in the form of decrements from the base
value of $0$. Accordingly, we obtain the following equivalent definition:
\begin{definition}
  Let $Z$ be a random variable supported over $(-\infty,\infty)$ with c.d.f.
  $F$. Then the {\em risk-averse expectation} of the random variable with
  respect to weighting function $\wt$ is
  \label{def:rae-both}
  \begin{align} \label{eq:rae}
      \rae{Z} \;=\; & -\int_{-\infty}^0(1-\wt(1-F(z)))dz
          \,+\,\int_0^\infty \wt(1-F(z))dz.
  \end{align}
%   \[
%   \rae{Z} = - \int_{-\infty}^0\nwt(F(z))dz + \int_0^\infty \wt(1-F(z)) dz,
% \]
%   where $\nwt(x) = 1-\wt(1-x)$.
\end{definition}
See Figure~\ref{fig:exp_vs_wted} for an illustration of this
definition. As an aside, it is well known that the quantile of a draw
from a distribution is itself distributed uniformly between 0 and 1. We note
that integrating \eqref{eq:rae} by parts shows our model can be interpreted as
sampling the distribution by drawing a quantile according to the distribution
with probability mass function $y'$. We skip the details.
%However, we do not make use of this
%observation.

%[width=.5\textwidth]

\subsubsection*{Some examples.}

\begin{example}
    If $\wt(x) = x$ for all $x\in [0,1]$, then the risk-averse expectation is
    exactly the expectation of the distribution. I.e., the agent is
    risk-neutral.
\end{example}

\begin{example}
    If the distribution of $Z$ is a point mass at $z$, its risk-averse
    expectation is exactly $z$.
\end{example}

% \begin{example}
%     \label{ex:binaryOutcome}
%     If $V=v$ with probability $q$ and $0$ with probability $1-q$, and
%     $v>0$, then its risk-averse expectation is $v \wt(q)$. If $v <
%     0$, the risk-averse expectation is $v(1 - \wt(1-q))$.
% \end{example}

% - \int_-infty^0 F(z)dz + \int_0^infty (1-F(z))dz
% = - \int_0^infty F(-z)dz + \int_0^infty (1-F(z))dz
% = 

\begin{example}
  Suppose that $Z$ takes on the value $v$ with probability $1/2$ and
  $-v$ with probability $1/2$, then its risk averse expectation is $-v
  + (2v)\wt(1/2) = -v(1-2\wt(1/2))$. Since we assume $\wt(1/2)\le
  1/2$, this quantity is non-positive. 

   More generally, if the distribution of $Z$ is symmetric around $0$, that is, for all $z>0$,
    $F(-z) = 1-F(z)$, then $\expect{Z}=0$.  The risk-averse expectation, on the
    other hand, is non-positive:
    \begin{align*}
        \int_0^\infty (\wt(1-F(z))+\wt(1-F(-z)) -1) dz
        &\;=\; \int_0^\infty (\wt(1-F(z))+\wt(F(z)) -1) dz \\
        &\;\le\; \int_0^\infty (1-F(z)+F(z) -1) dz \; = \; 0. 
    \end{align*}
\end{example}

\subsubsection*{Quantifying risk aversion.}

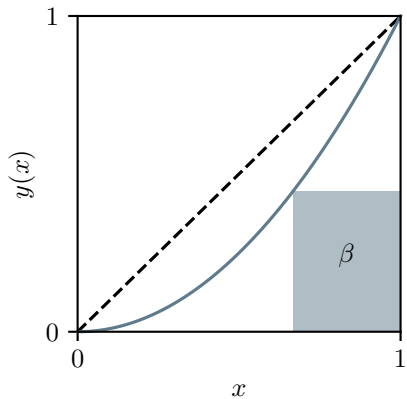
\begin{figure}[htbp]
  \begin{center}
    %\scalebox{0.5}{\input{figures/beta_bounded.pgf}}
    \scalebox{1}{\input{figures/beta_bounded.pgf}}
    \caption{A $\beta$-bounded weighting function: the area of the shaded
      rectangle is $\beta$, which is the maximal area of any rectangle
      contained under the curve.  This area gives a measure of the aversion to
      risk: a smaller $\beta$ corresponds to stronger aversion to risk.}
    \label{fig:beta-bounded}
  \end{center}
\end{figure}

The buyer's risk attitude in our model is described by a function
rather than by one (or a few) parameters. While this leads to a rich
set of behaviors, it makes it challenging to understand, for example,
whether one weighting function is more risk-averse than another. We
argue that a natural measure for the extent of risk aversion is the
gap between the function and the $x$ and $y$ axes. In other words, if
the function ``hugs'' the axes $x=0$ and $y=1$, then the buyer heavily
discounts all events that happen with probability bounded away from 1,
and is highly risk-averse. On the other hand, if the function is
bounded away from the axes, then the buyer is less risk-averse. See
Figure~\ref{fig:beta-bounded}.

\begin{definition}
  \label{def:boundedness}
    A weighting function is $\beta$-bounded if there exists an $x\in (0,1)$
    such that $\wt(x)(1-x)\ge \beta$. In other words, $\wt$ is $\beta$-bounded
    if we can fit a rectangle with area $\beta$ under the curve. 
\end{definition}
We will see in Section~\ref{sec:risk-robust-oneshot} that $\beta$-boundedness
affects the revenue approximation achievable for the given weighting function.

%% file: figures/expectedVsWeighted_a.pgf
%% Creator: Matplotlib, PGF backend
%%
%% To include the figure in your LaTeX document, write
%%   \input{<filename>.pgf}
%%
%% Make sure the required packages are loaded in your preamble
%%   \usepackage{pgf}
%%
%% Figures using additional raster images can only be included by \input if
%% they are in the same directory as the main LaTeX file. For loading figures
%% from other directories you can use the `import` package
%%   \usepackage{import}
%% and then include the figures with
%%   \import{<path to file>}{<filename>.pgf}
%%
%% Matplotlib used the following preamble
%%   \usepackage{fontspec}
%%
\begingroup%
\makeatletter%
\begin{pgfpicture}%
\pgfpathrectangle{\pgfpointorigin}{\pgfqpoint{3.150000in}{1.946807in}}%
\pgfusepath{use as bounding box, clip}%
\begin{pgfscope}%
\pgfsetbuttcap%
\pgfsetmiterjoin%
\pgfsetlinewidth{0.000000pt}%
\definecolor{currentstroke}{rgb}{1.000000,1.000000,1.000000}%
\pgfsetstrokecolor{currentstroke}%
\pgfsetstrokeopacity{0.000000}%
\pgfsetdash{}{0pt}%
\pgfpathmoveto{\pgfqpoint{0.000000in}{0.000000in}}%
\pgfpathlineto{\pgfqpoint{3.150000in}{0.000000in}}%
\pgfpathlineto{\pgfqpoint{3.150000in}{1.946807in}}%
\pgfpathlineto{\pgfqpoint{0.000000in}{1.946807in}}%
\pgfpathclose%
\pgfusepath{}%
\end{pgfscope}%
\begin{pgfscope}%
\pgfsetbuttcap%
\pgfsetmiterjoin%
\definecolor{currentfill}{rgb}{1.000000,1.000000,1.000000}%
\pgfsetfillcolor{currentfill}%
\pgfsetlinewidth{0.000000pt}%
\definecolor{currentstroke}{rgb}{0.000000,0.000000,0.000000}%
\pgfsetstrokecolor{currentstroke}%
\pgfsetstrokeopacity{0.000000}%
\pgfsetdash{}{0pt}%
\pgfpathmoveto{\pgfqpoint{0.306648in}{0.403361in}}%
\pgfpathlineto{\pgfqpoint{3.045556in}{0.403361in}}%
\pgfpathlineto{\pgfqpoint{3.045556in}{1.842363in}}%
\pgfpathlineto{\pgfqpoint{0.306648in}{1.842363in}}%
\pgfpathclose%
\pgfusepath{fill}%
\end{pgfscope}%
\begin{pgfscope}%
\pgfpathrectangle{\pgfqpoint{0.306648in}{0.403361in}}{\pgfqpoint{2.738907in}{1.439002in}} %
\pgfusepath{clip}%
\pgfsetbuttcap%
\pgfsetroundjoin%
\definecolor{currentfill}{rgb}{0.882353,0.945098,1.000000}%
\pgfsetfillcolor{currentfill}%
\pgfsetlinewidth{0.000000pt}%
\definecolor{currentstroke}{rgb}{0.000000,0.000000,0.000000}%
\pgfsetstrokecolor{currentstroke}%
\pgfsetdash{}{0pt}%
\pgfpathmoveto{\pgfqpoint{1.676102in}{1.122862in}}%
\pgfpathlineto{\pgfqpoint{1.676102in}{1.122862in}}%
\pgfpathlineto{\pgfqpoint{1.684402in}{1.129820in}}%
\pgfpathlineto{\pgfqpoint{1.692701in}{1.136777in}}%
\pgfpathlineto{\pgfqpoint{1.701001in}{1.143732in}}%
\pgfpathlineto{\pgfqpoint{1.709301in}{1.150683in}}%
\pgfpathlineto{\pgfqpoint{1.717601in}{1.157629in}}%
\pgfpathlineto{\pgfqpoint{1.725900in}{1.164568in}}%
\pgfpathlineto{\pgfqpoint{1.734200in}{1.171501in}}%
\pgfpathlineto{\pgfqpoint{1.742500in}{1.178425in}}%
\pgfpathlineto{\pgfqpoint{1.750799in}{1.185338in}}%
\pgfpathlineto{\pgfqpoint{1.759099in}{1.192241in}}%
\pgfpathlineto{\pgfqpoint{1.767399in}{1.199132in}}%
\pgfpathlineto{\pgfqpoint{1.775699in}{1.206009in}}%
\pgfpathlineto{\pgfqpoint{1.783998in}{1.212872in}}%
\pgfpathlineto{\pgfqpoint{1.792298in}{1.219718in}}%
\pgfpathlineto{\pgfqpoint{1.800598in}{1.226548in}}%
\pgfpathlineto{\pgfqpoint{1.808897in}{1.233359in}}%
\pgfpathlineto{\pgfqpoint{1.817197in}{1.240152in}}%
\pgfpathlineto{\pgfqpoint{1.825497in}{1.246923in}}%
\pgfpathlineto{\pgfqpoint{1.833797in}{1.253673in}}%
\pgfpathlineto{\pgfqpoint{1.842096in}{1.260400in}}%
\pgfpathlineto{\pgfqpoint{1.850396in}{1.267104in}}%
\pgfpathlineto{\pgfqpoint{1.858696in}{1.273782in}}%
\pgfpathlineto{\pgfqpoint{1.866995in}{1.280434in}}%
\pgfpathlineto{\pgfqpoint{1.875295in}{1.287059in}}%
\pgfpathlineto{\pgfqpoint{1.883595in}{1.293656in}}%
\pgfpathlineto{\pgfqpoint{1.891895in}{1.300224in}}%
\pgfpathlineto{\pgfqpoint{1.900194in}{1.306761in}}%
\pgfpathlineto{\pgfqpoint{1.908494in}{1.313268in}}%
\pgfpathlineto{\pgfqpoint{1.916794in}{1.319741in}}%
\pgfpathlineto{\pgfqpoint{1.925093in}{1.326182in}}%
\pgfpathlineto{\pgfqpoint{1.933393in}{1.332588in}}%
\pgfpathlineto{\pgfqpoint{1.941693in}{1.338959in}}%
\pgfpathlineto{\pgfqpoint{1.949993in}{1.345294in}}%
\pgfpathlineto{\pgfqpoint{1.958292in}{1.351592in}}%
\pgfpathlineto{\pgfqpoint{1.966592in}{1.357852in}}%
\pgfpathlineto{\pgfqpoint{1.974892in}{1.364073in}}%
\pgfpathlineto{\pgfqpoint{1.983192in}{1.370254in}}%
\pgfpathlineto{\pgfqpoint{1.991491in}{1.376395in}}%
\pgfpathlineto{\pgfqpoint{1.999791in}{1.382495in}}%
\pgfpathlineto{\pgfqpoint{2.008091in}{1.388552in}}%
\pgfpathlineto{\pgfqpoint{2.016390in}{1.394566in}}%
\pgfpathlineto{\pgfqpoint{2.024690in}{1.400537in}}%
\pgfpathlineto{\pgfqpoint{2.032990in}{1.406463in}}%
\pgfpathlineto{\pgfqpoint{2.041290in}{1.412345in}}%
\pgfpathlineto{\pgfqpoint{2.049589in}{1.418180in}}%
\pgfpathlineto{\pgfqpoint{2.057889in}{1.423969in}}%
\pgfpathlineto{\pgfqpoint{2.066189in}{1.429711in}}%
\pgfpathlineto{\pgfqpoint{2.074488in}{1.435405in}}%
\pgfpathlineto{\pgfqpoint{2.082788in}{1.441050in}}%
\pgfpathlineto{\pgfqpoint{2.091088in}{1.446647in}}%
\pgfpathlineto{\pgfqpoint{2.099388in}{1.452194in}}%
\pgfpathlineto{\pgfqpoint{2.107687in}{1.457691in}}%
\pgfpathlineto{\pgfqpoint{2.115987in}{1.463137in}}%
\pgfpathlineto{\pgfqpoint{2.124287in}{1.468533in}}%
\pgfpathlineto{\pgfqpoint{2.132586in}{1.473877in}}%
\pgfpathlineto{\pgfqpoint{2.140886in}{1.479169in}}%
\pgfpathlineto{\pgfqpoint{2.149186in}{1.484408in}}%
\pgfpathlineto{\pgfqpoint{2.157486in}{1.489595in}}%
\pgfpathlineto{\pgfqpoint{2.165785in}{1.494728in}}%
\pgfpathlineto{\pgfqpoint{2.174085in}{1.499808in}}%
\pgfpathlineto{\pgfqpoint{2.182385in}{1.504834in}}%
\pgfpathlineto{\pgfqpoint{2.190685in}{1.509805in}}%
\pgfpathlineto{\pgfqpoint{2.198984in}{1.514723in}}%
\pgfpathlineto{\pgfqpoint{2.207284in}{1.519585in}}%
\pgfpathlineto{\pgfqpoint{2.215584in}{1.524392in}}%
\pgfpathlineto{\pgfqpoint{2.223883in}{1.529145in}}%
\pgfpathlineto{\pgfqpoint{2.232183in}{1.533841in}}%
\pgfpathlineto{\pgfqpoint{2.240483in}{1.538483in}}%
\pgfpathlineto{\pgfqpoint{2.248783in}{1.543068in}}%
\pgfpathlineto{\pgfqpoint{2.257082in}{1.547598in}}%
\pgfpathlineto{\pgfqpoint{2.265382in}{1.552071in}}%
\pgfpathlineto{\pgfqpoint{2.273682in}{1.556488in}}%
\pgfpathlineto{\pgfqpoint{2.281981in}{1.560850in}}%
\pgfpathlineto{\pgfqpoint{2.290281in}{1.565155in}}%
\pgfpathlineto{\pgfqpoint{2.298581in}{1.569403in}}%
\pgfpathlineto{\pgfqpoint{2.306881in}{1.573596in}}%
\pgfpathlineto{\pgfqpoint{2.315180in}{1.577732in}}%
\pgfpathlineto{\pgfqpoint{2.323480in}{1.581812in}}%
\pgfpathlineto{\pgfqpoint{2.331780in}{1.585836in}}%
\pgfpathlineto{\pgfqpoint{2.340079in}{1.589803in}}%
\pgfpathlineto{\pgfqpoint{2.348379in}{1.593715in}}%
\pgfpathlineto{\pgfqpoint{2.356679in}{1.597570in}}%
\pgfpathlineto{\pgfqpoint{2.364979in}{1.601370in}}%
\pgfpathlineto{\pgfqpoint{2.373278in}{1.605115in}}%
\pgfpathlineto{\pgfqpoint{2.381578in}{1.608803in}}%
\pgfpathlineto{\pgfqpoint{2.389878in}{1.612437in}}%
\pgfpathlineto{\pgfqpoint{2.398177in}{1.616015in}}%
\pgfpathlineto{\pgfqpoint{2.406477in}{1.619538in}}%
\pgfpathlineto{\pgfqpoint{2.414777in}{1.623007in}}%
\pgfpathlineto{\pgfqpoint{2.423077in}{1.626421in}}%
\pgfpathlineto{\pgfqpoint{2.431376in}{1.629781in}}%
\pgfpathlineto{\pgfqpoint{2.439676in}{1.633087in}}%
\pgfpathlineto{\pgfqpoint{2.447976in}{1.636340in}}%
\pgfpathlineto{\pgfqpoint{2.456276in}{1.639539in}}%
\pgfpathlineto{\pgfqpoint{2.464575in}{1.642685in}}%
\pgfpathlineto{\pgfqpoint{2.472875in}{1.645779in}}%
\pgfpathlineto{\pgfqpoint{2.481175in}{1.648820in}}%
\pgfpathlineto{\pgfqpoint{2.489474in}{1.651809in}}%
\pgfpathlineto{\pgfqpoint{2.497774in}{1.654746in}}%
\pgfpathlineto{\pgfqpoint{2.506074in}{1.657632in}}%
\pgfpathlineto{\pgfqpoint{2.514374in}{1.660468in}}%
\pgfpathlineto{\pgfqpoint{2.522673in}{1.663253in}}%
\pgfpathlineto{\pgfqpoint{2.530973in}{1.665988in}}%
\pgfpathlineto{\pgfqpoint{2.539273in}{1.668673in}}%
\pgfpathlineto{\pgfqpoint{2.547572in}{1.671309in}}%
\pgfpathlineto{\pgfqpoint{2.555872in}{1.673896in}}%
\pgfpathlineto{\pgfqpoint{2.564172in}{1.676435in}}%
\pgfpathlineto{\pgfqpoint{2.572472in}{1.678926in}}%
\pgfpathlineto{\pgfqpoint{2.580771in}{1.681370in}}%
\pgfpathlineto{\pgfqpoint{2.589071in}{1.683767in}}%
\pgfpathlineto{\pgfqpoint{2.597371in}{1.686118in}}%
\pgfpathlineto{\pgfqpoint{2.605670in}{1.688422in}}%
\pgfpathlineto{\pgfqpoint{2.613970in}{1.690681in}}%
\pgfpathlineto{\pgfqpoint{2.622270in}{1.692895in}}%
\pgfpathlineto{\pgfqpoint{2.630570in}{1.695065in}}%
\pgfpathlineto{\pgfqpoint{2.638869in}{1.697191in}}%
\pgfpathlineto{\pgfqpoint{2.647169in}{1.699274in}}%
\pgfpathlineto{\pgfqpoint{2.655469in}{1.701313in}}%
\pgfpathlineto{\pgfqpoint{2.663768in}{1.703310in}}%
\pgfpathlineto{\pgfqpoint{2.672068in}{1.705266in}}%
\pgfpathlineto{\pgfqpoint{2.680368in}{1.707180in}}%
\pgfpathlineto{\pgfqpoint{2.688668in}{1.709054in}}%
\pgfpathlineto{\pgfqpoint{2.696967in}{1.710887in}}%
\pgfpathlineto{\pgfqpoint{2.705267in}{1.712680in}}%
\pgfpathlineto{\pgfqpoint{2.713567in}{1.714435in}}%
\pgfpathlineto{\pgfqpoint{2.721867in}{1.716151in}}%
\pgfpathlineto{\pgfqpoint{2.730166in}{1.717829in}}%
\pgfpathlineto{\pgfqpoint{2.738466in}{1.719469in}}%
\pgfpathlineto{\pgfqpoint{2.746766in}{1.721073in}}%
\pgfpathlineto{\pgfqpoint{2.755065in}{1.722640in}}%
\pgfpathlineto{\pgfqpoint{2.763365in}{1.724171in}}%
\pgfpathlineto{\pgfqpoint{2.771665in}{1.725668in}}%
\pgfpathlineto{\pgfqpoint{2.779965in}{1.727129in}}%
\pgfpathlineto{\pgfqpoint{2.788264in}{1.728556in}}%
\pgfpathlineto{\pgfqpoint{2.796564in}{1.729950in}}%
\pgfpathlineto{\pgfqpoint{2.804864in}{1.731311in}}%
\pgfpathlineto{\pgfqpoint{2.813163in}{1.732639in}}%
\pgfpathlineto{\pgfqpoint{2.821463in}{1.733935in}}%
\pgfpathlineto{\pgfqpoint{2.829763in}{1.735200in}}%
\pgfpathlineto{\pgfqpoint{2.838063in}{1.736434in}}%
\pgfpathlineto{\pgfqpoint{2.846362in}{1.737637in}}%
\pgfpathlineto{\pgfqpoint{2.854662in}{1.738811in}}%
\pgfpathlineto{\pgfqpoint{2.862962in}{1.739956in}}%
\pgfpathlineto{\pgfqpoint{2.871261in}{1.741071in}}%
\pgfpathlineto{\pgfqpoint{2.879561in}{1.742159in}}%
\pgfpathlineto{\pgfqpoint{2.887861in}{1.743219in}}%
\pgfpathlineto{\pgfqpoint{2.896161in}{1.744252in}}%
\pgfpathlineto{\pgfqpoint{2.904460in}{1.745258in}}%
\pgfpathlineto{\pgfqpoint{2.912760in}{1.746238in}}%
\pgfpathlineto{\pgfqpoint{2.921060in}{1.747192in}}%
\pgfpathlineto{\pgfqpoint{2.921060in}{1.747192in}}%
\pgfpathlineto{\pgfqpoint{2.921060in}{1.776953in}}%
\pgfpathlineto{\pgfqpoint{2.912760in}{1.776953in}}%
\pgfpathlineto{\pgfqpoint{2.904460in}{1.776953in}}%
\pgfpathlineto{\pgfqpoint{2.896161in}{1.776953in}}%
\pgfpathlineto{\pgfqpoint{2.887861in}{1.776953in}}%
\pgfpathlineto{\pgfqpoint{2.879561in}{1.776953in}}%
\pgfpathlineto{\pgfqpoint{2.871261in}{1.776953in}}%
\pgfpathlineto{\pgfqpoint{2.862962in}{1.776953in}}%
\pgfpathlineto{\pgfqpoint{2.854662in}{1.776953in}}%
\pgfpathlineto{\pgfqpoint{2.846362in}{1.776953in}}%
\pgfpathlineto{\pgfqpoint{2.838063in}{1.776953in}}%
\pgfpathlineto{\pgfqpoint{2.829763in}{1.776953in}}%
\pgfpathlineto{\pgfqpoint{2.821463in}{1.776953in}}%
\pgfpathlineto{\pgfqpoint{2.813163in}{1.776953in}}%
\pgfpathlineto{\pgfqpoint{2.804864in}{1.776953in}}%
\pgfpathlineto{\pgfqpoint{2.796564in}{1.776953in}}%
\pgfpathlineto{\pgfqpoint{2.788264in}{1.776953in}}%
\pgfpathlineto{\pgfqpoint{2.779965in}{1.776953in}}%
\pgfpathlineto{\pgfqpoint{2.771665in}{1.776953in}}%
\pgfpathlineto{\pgfqpoint{2.763365in}{1.776953in}}%
\pgfpathlineto{\pgfqpoint{2.755065in}{1.776953in}}%
\pgfpathlineto{\pgfqpoint{2.746766in}{1.776953in}}%
\pgfpathlineto{\pgfqpoint{2.738466in}{1.776953in}}%
\pgfpathlineto{\pgfqpoint{2.730166in}{1.776953in}}%
\pgfpathlineto{\pgfqpoint{2.721867in}{1.776953in}}%
\pgfpathlineto{\pgfqpoint{2.713567in}{1.776953in}}%
\pgfpathlineto{\pgfqpoint{2.705267in}{1.776953in}}%
\pgfpathlineto{\pgfqpoint{2.696967in}{1.776953in}}%
\pgfpathlineto{\pgfqpoint{2.688668in}{1.776953in}}%
\pgfpathlineto{\pgfqpoint{2.680368in}{1.776953in}}%
\pgfpathlineto{\pgfqpoint{2.672068in}{1.776953in}}%
\pgfpathlineto{\pgfqpoint{2.663768in}{1.776953in}}%
\pgfpathlineto{\pgfqpoint{2.655469in}{1.776953in}}%
\pgfpathlineto{\pgfqpoint{2.647169in}{1.776953in}}%
\pgfpathlineto{\pgfqpoint{2.638869in}{1.776953in}}%
\pgfpathlineto{\pgfqpoint{2.630570in}{1.776953in}}%
\pgfpathlineto{\pgfqpoint{2.622270in}{1.776953in}}%
\pgfpathlineto{\pgfqpoint{2.613970in}{1.776953in}}%
\pgfpathlineto{\pgfqpoint{2.605670in}{1.776953in}}%
\pgfpathlineto{\pgfqpoint{2.597371in}{1.776953in}}%
\pgfpathlineto{\pgfqpoint{2.589071in}{1.776953in}}%
\pgfpathlineto{\pgfqpoint{2.580771in}{1.776953in}}%
\pgfpathlineto{\pgfqpoint{2.572472in}{1.776953in}}%
\pgfpathlineto{\pgfqpoint{2.564172in}{1.776953in}}%
\pgfpathlineto{\pgfqpoint{2.555872in}{1.776953in}}%
\pgfpathlineto{\pgfqpoint{2.547572in}{1.776953in}}%
\pgfpathlineto{\pgfqpoint{2.539273in}{1.776953in}}%
\pgfpathlineto{\pgfqpoint{2.530973in}{1.776953in}}%
\pgfpathlineto{\pgfqpoint{2.522673in}{1.776953in}}%
\pgfpathlineto{\pgfqpoint{2.514374in}{1.776953in}}%
\pgfpathlineto{\pgfqpoint{2.506074in}{1.776953in}}%
\pgfpathlineto{\pgfqpoint{2.497774in}{1.776953in}}%
\pgfpathlineto{\pgfqpoint{2.489474in}{1.776953in}}%
\pgfpathlineto{\pgfqpoint{2.481175in}{1.776953in}}%
\pgfpathlineto{\pgfqpoint{2.472875in}{1.776953in}}%
\pgfpathlineto{\pgfqpoint{2.464575in}{1.776953in}}%
\pgfpathlineto{\pgfqpoint{2.456276in}{1.776953in}}%
\pgfpathlineto{\pgfqpoint{2.447976in}{1.776953in}}%
\pgfpathlineto{\pgfqpoint{2.439676in}{1.776953in}}%
\pgfpathlineto{\pgfqpoint{2.431376in}{1.776953in}}%
\pgfpathlineto{\pgfqpoint{2.423077in}{1.776953in}}%
\pgfpathlineto{\pgfqpoint{2.414777in}{1.776953in}}%
\pgfpathlineto{\pgfqpoint{2.406477in}{1.776953in}}%
\pgfpathlineto{\pgfqpoint{2.398177in}{1.776953in}}%
\pgfpathlineto{\pgfqpoint{2.389878in}{1.776953in}}%
\pgfpathlineto{\pgfqpoint{2.381578in}{1.776953in}}%
\pgfpathlineto{\pgfqpoint{2.373278in}{1.776953in}}%
\pgfpathlineto{\pgfqpoint{2.364979in}{1.776953in}}%
\pgfpathlineto{\pgfqpoint{2.356679in}{1.776953in}}%
\pgfpathlineto{\pgfqpoint{2.348379in}{1.776953in}}%
\pgfpathlineto{\pgfqpoint{2.340079in}{1.776953in}}%
\pgfpathlineto{\pgfqpoint{2.331780in}{1.776953in}}%
\pgfpathlineto{\pgfqpoint{2.323480in}{1.776953in}}%
\pgfpathlineto{\pgfqpoint{2.315180in}{1.776953in}}%
\pgfpathlineto{\pgfqpoint{2.306881in}{1.776953in}}%
\pgfpathlineto{\pgfqpoint{2.298581in}{1.776953in}}%
\pgfpathlineto{\pgfqpoint{2.290281in}{1.776953in}}%
\pgfpathlineto{\pgfqpoint{2.281981in}{1.776953in}}%
\pgfpathlineto{\pgfqpoint{2.273682in}{1.776953in}}%
\pgfpathlineto{\pgfqpoint{2.265382in}{1.776953in}}%
\pgfpathlineto{\pgfqpoint{2.257082in}{1.776953in}}%
\pgfpathlineto{\pgfqpoint{2.248783in}{1.776953in}}%
\pgfpathlineto{\pgfqpoint{2.240483in}{1.776953in}}%
\pgfpathlineto{\pgfqpoint{2.232183in}{1.776953in}}%
\pgfpathlineto{\pgfqpoint{2.223883in}{1.776953in}}%
\pgfpathlineto{\pgfqpoint{2.215584in}{1.776953in}}%
\pgfpathlineto{\pgfqpoint{2.207284in}{1.776953in}}%
\pgfpathlineto{\pgfqpoint{2.198984in}{1.776953in}}%
\pgfpathlineto{\pgfqpoint{2.190685in}{1.776953in}}%
\pgfpathlineto{\pgfqpoint{2.182385in}{1.776953in}}%
\pgfpathlineto{\pgfqpoint{2.174085in}{1.776953in}}%
\pgfpathlineto{\pgfqpoint{2.165785in}{1.776953in}}%
\pgfpathlineto{\pgfqpoint{2.157486in}{1.776953in}}%
\pgfpathlineto{\pgfqpoint{2.149186in}{1.776953in}}%
\pgfpathlineto{\pgfqpoint{2.140886in}{1.776953in}}%
\pgfpathlineto{\pgfqpoint{2.132586in}{1.776953in}}%
\pgfpathlineto{\pgfqpoint{2.124287in}{1.776953in}}%
\pgfpathlineto{\pgfqpoint{2.115987in}{1.776953in}}%
\pgfpathlineto{\pgfqpoint{2.107687in}{1.776953in}}%
\pgfpathlineto{\pgfqpoint{2.099388in}{1.776953in}}%
\pgfpathlineto{\pgfqpoint{2.091088in}{1.776953in}}%
\pgfpathlineto{\pgfqpoint{2.082788in}{1.776953in}}%
\pgfpathlineto{\pgfqpoint{2.074488in}{1.776953in}}%
\pgfpathlineto{\pgfqpoint{2.066189in}{1.776953in}}%
\pgfpathlineto{\pgfqpoint{2.057889in}{1.776953in}}%
\pgfpathlineto{\pgfqpoint{2.049589in}{1.776953in}}%
\pgfpathlineto{\pgfqpoint{2.041290in}{1.776953in}}%
\pgfpathlineto{\pgfqpoint{2.032990in}{1.776953in}}%
\pgfpathlineto{\pgfqpoint{2.024690in}{1.776953in}}%
\pgfpathlineto{\pgfqpoint{2.016390in}{1.776953in}}%
\pgfpathlineto{\pgfqpoint{2.008091in}{1.776953in}}%
\pgfpathlineto{\pgfqpoint{1.999791in}{1.776953in}}%
\pgfpathlineto{\pgfqpoint{1.991491in}{1.776953in}}%
\pgfpathlineto{\pgfqpoint{1.983192in}{1.776953in}}%
\pgfpathlineto{\pgfqpoint{1.974892in}{1.776953in}}%
\pgfpathlineto{\pgfqpoint{1.966592in}{1.776953in}}%
\pgfpathlineto{\pgfqpoint{1.958292in}{1.776953in}}%
\pgfpathlineto{\pgfqpoint{1.949993in}{1.776953in}}%
\pgfpathlineto{\pgfqpoint{1.941693in}{1.776953in}}%
\pgfpathlineto{\pgfqpoint{1.933393in}{1.776953in}}%
\pgfpathlineto{\pgfqpoint{1.925093in}{1.776953in}}%
\pgfpathlineto{\pgfqpoint{1.916794in}{1.776953in}}%
\pgfpathlineto{\pgfqpoint{1.908494in}{1.776953in}}%
\pgfpathlineto{\pgfqpoint{1.900194in}{1.776953in}}%
\pgfpathlineto{\pgfqpoint{1.891895in}{1.776953in}}%
\pgfpathlineto{\pgfqpoint{1.883595in}{1.776953in}}%
\pgfpathlineto{\pgfqpoint{1.875295in}{1.776953in}}%
\pgfpathlineto{\pgfqpoint{1.866995in}{1.776953in}}%
\pgfpathlineto{\pgfqpoint{1.858696in}{1.776953in}}%
\pgfpathlineto{\pgfqpoint{1.850396in}{1.776953in}}%
\pgfpathlineto{\pgfqpoint{1.842096in}{1.776953in}}%
\pgfpathlineto{\pgfqpoint{1.833797in}{1.776953in}}%
\pgfpathlineto{\pgfqpoint{1.825497in}{1.776953in}}%
\pgfpathlineto{\pgfqpoint{1.817197in}{1.776953in}}%
\pgfpathlineto{\pgfqpoint{1.808897in}{1.776953in}}%
\pgfpathlineto{\pgfqpoint{1.800598in}{1.776953in}}%
\pgfpathlineto{\pgfqpoint{1.792298in}{1.776953in}}%
\pgfpathlineto{\pgfqpoint{1.783998in}{1.776953in}}%
\pgfpathlineto{\pgfqpoint{1.775699in}{1.776953in}}%
\pgfpathlineto{\pgfqpoint{1.767399in}{1.776953in}}%
\pgfpathlineto{\pgfqpoint{1.759099in}{1.776953in}}%
\pgfpathlineto{\pgfqpoint{1.750799in}{1.776953in}}%
\pgfpathlineto{\pgfqpoint{1.742500in}{1.776953in}}%
\pgfpathlineto{\pgfqpoint{1.734200in}{1.776953in}}%
\pgfpathlineto{\pgfqpoint{1.725900in}{1.776953in}}%
\pgfpathlineto{\pgfqpoint{1.717601in}{1.776953in}}%
\pgfpathlineto{\pgfqpoint{1.709301in}{1.776953in}}%
\pgfpathlineto{\pgfqpoint{1.701001in}{1.776953in}}%
\pgfpathlineto{\pgfqpoint{1.692701in}{1.776953in}}%
\pgfpathlineto{\pgfqpoint{1.684402in}{1.776953in}}%
\pgfpathlineto{\pgfqpoint{1.676102in}{1.776953in}}%
\pgfpathclose%
\pgfusepath{fill}%
\end{pgfscope}%
\begin{pgfscope}%
\pgfpathrectangle{\pgfqpoint{0.306648in}{0.403361in}}{\pgfqpoint{2.738907in}{1.439002in}} %
\pgfusepath{clip}%
\pgfsetbuttcap%
\pgfsetroundjoin%
\definecolor{currentfill}{rgb}{1.000000,0.588235,0.470588}%
\pgfsetfillcolor{currentfill}%
\pgfsetlinewidth{0.000000pt}%
\definecolor{currentstroke}{rgb}{0.000000,0.000000,0.000000}%
\pgfsetstrokecolor{currentstroke}%
\pgfsetdash{}{0pt}%
\pgfpathmoveto{\pgfqpoint{0.431144in}{0.498531in}}%
\pgfpathlineto{\pgfqpoint{0.431144in}{0.498531in}}%
\pgfpathlineto{\pgfqpoint{0.439444in}{0.499486in}}%
\pgfpathlineto{\pgfqpoint{0.447744in}{0.500466in}}%
\pgfpathlineto{\pgfqpoint{0.456043in}{0.501472in}}%
\pgfpathlineto{\pgfqpoint{0.464343in}{0.502505in}}%
\pgfpathlineto{\pgfqpoint{0.472643in}{0.503564in}}%
\pgfpathlineto{\pgfqpoint{0.480942in}{0.504652in}}%
\pgfpathlineto{\pgfqpoint{0.489242in}{0.505768in}}%
\pgfpathlineto{\pgfqpoint{0.497542in}{0.506912in}}%
\pgfpathlineto{\pgfqpoint{0.505842in}{0.508086in}}%
\pgfpathlineto{\pgfqpoint{0.514141in}{0.509290in}}%
\pgfpathlineto{\pgfqpoint{0.522441in}{0.510524in}}%
\pgfpathlineto{\pgfqpoint{0.530741in}{0.511788in}}%
\pgfpathlineto{\pgfqpoint{0.539040in}{0.513085in}}%
\pgfpathlineto{\pgfqpoint{0.547340in}{0.514413in}}%
\pgfpathlineto{\pgfqpoint{0.555640in}{0.515773in}}%
\pgfpathlineto{\pgfqpoint{0.563940in}{0.517167in}}%
\pgfpathlineto{\pgfqpoint{0.572239in}{0.518594in}}%
\pgfpathlineto{\pgfqpoint{0.580539in}{0.520056in}}%
\pgfpathlineto{\pgfqpoint{0.588839in}{0.521552in}}%
\pgfpathlineto{\pgfqpoint{0.597138in}{0.523083in}}%
\pgfpathlineto{\pgfqpoint{0.605438in}{0.524651in}}%
\pgfpathlineto{\pgfqpoint{0.613738in}{0.526254in}}%
\pgfpathlineto{\pgfqpoint{0.622038in}{0.527895in}}%
\pgfpathlineto{\pgfqpoint{0.630337in}{0.529573in}}%
\pgfpathlineto{\pgfqpoint{0.638637in}{0.531289in}}%
\pgfpathlineto{\pgfqpoint{0.646937in}{0.533043in}}%
\pgfpathlineto{\pgfqpoint{0.655237in}{0.534837in}}%
\pgfpathlineto{\pgfqpoint{0.663536in}{0.536670in}}%
\pgfpathlineto{\pgfqpoint{0.671836in}{0.538543in}}%
\pgfpathlineto{\pgfqpoint{0.680136in}{0.540457in}}%
\pgfpathlineto{\pgfqpoint{0.688435in}{0.542413in}}%
\pgfpathlineto{\pgfqpoint{0.696735in}{0.544410in}}%
\pgfpathlineto{\pgfqpoint{0.705035in}{0.546450in}}%
\pgfpathlineto{\pgfqpoint{0.713335in}{0.548532in}}%
\pgfpathlineto{\pgfqpoint{0.721634in}{0.550658in}}%
\pgfpathlineto{\pgfqpoint{0.729934in}{0.552828in}}%
\pgfpathlineto{\pgfqpoint{0.738234in}{0.555042in}}%
\pgfpathlineto{\pgfqpoint{0.746533in}{0.557301in}}%
\pgfpathlineto{\pgfqpoint{0.754833in}{0.559606in}}%
\pgfpathlineto{\pgfqpoint{0.763133in}{0.561956in}}%
\pgfpathlineto{\pgfqpoint{0.771433in}{0.564353in}}%
\pgfpathlineto{\pgfqpoint{0.779732in}{0.566797in}}%
\pgfpathlineto{\pgfqpoint{0.788032in}{0.569288in}}%
\pgfpathlineto{\pgfqpoint{0.796332in}{0.571827in}}%
\pgfpathlineto{\pgfqpoint{0.804631in}{0.574414in}}%
\pgfpathlineto{\pgfqpoint{0.812931in}{0.577051in}}%
\pgfpathlineto{\pgfqpoint{0.821231in}{0.579736in}}%
\pgfpathlineto{\pgfqpoint{0.829531in}{0.582471in}}%
\pgfpathlineto{\pgfqpoint{0.837830in}{0.585256in}}%
\pgfpathlineto{\pgfqpoint{0.846130in}{0.588091in}}%
\pgfpathlineto{\pgfqpoint{0.854430in}{0.590977in}}%
\pgfpathlineto{\pgfqpoint{0.862729in}{0.593915in}}%
\pgfpathlineto{\pgfqpoint{0.871029in}{0.596904in}}%
\pgfpathlineto{\pgfqpoint{0.879329in}{0.599945in}}%
\pgfpathlineto{\pgfqpoint{0.887629in}{0.603038in}}%
\pgfpathlineto{\pgfqpoint{0.895928in}{0.606184in}}%
\pgfpathlineto{\pgfqpoint{0.904228in}{0.609384in}}%
\pgfpathlineto{\pgfqpoint{0.912528in}{0.612636in}}%
\pgfpathlineto{\pgfqpoint{0.920828in}{0.615942in}}%
\pgfpathlineto{\pgfqpoint{0.929127in}{0.619302in}}%
\pgfpathlineto{\pgfqpoint{0.937427in}{0.622716in}}%
\pgfpathlineto{\pgfqpoint{0.945727in}{0.626185in}}%
\pgfpathlineto{\pgfqpoint{0.954026in}{0.629708in}}%
\pgfpathlineto{\pgfqpoint{0.962326in}{0.633287in}}%
\pgfpathlineto{\pgfqpoint{0.970626in}{0.636920in}}%
\pgfpathlineto{\pgfqpoint{0.978926in}{0.640609in}}%
\pgfpathlineto{\pgfqpoint{0.987225in}{0.644353in}}%
\pgfpathlineto{\pgfqpoint{0.995525in}{0.648153in}}%
\pgfpathlineto{\pgfqpoint{1.003825in}{0.652009in}}%
\pgfpathlineto{\pgfqpoint{1.012124in}{0.655920in}}%
\pgfpathlineto{\pgfqpoint{1.020424in}{0.659888in}}%
\pgfpathlineto{\pgfqpoint{1.028724in}{0.663912in}}%
\pgfpathlineto{\pgfqpoint{1.037024in}{0.667992in}}%
\pgfpathlineto{\pgfqpoint{1.045323in}{0.672128in}}%
\pgfpathlineto{\pgfqpoint{1.053623in}{0.676320in}}%
\pgfpathlineto{\pgfqpoint{1.061923in}{0.680569in}}%
\pgfpathlineto{\pgfqpoint{1.070222in}{0.684874in}}%
\pgfpathlineto{\pgfqpoint{1.078522in}{0.689235in}}%
\pgfpathlineto{\pgfqpoint{1.086822in}{0.693652in}}%
\pgfpathlineto{\pgfqpoint{1.095122in}{0.698126in}}%
\pgfpathlineto{\pgfqpoint{1.103421in}{0.702655in}}%
\pgfpathlineto{\pgfqpoint{1.111721in}{0.707241in}}%
\pgfpathlineto{\pgfqpoint{1.120021in}{0.711882in}}%
\pgfpathlineto{\pgfqpoint{1.128320in}{0.716579in}}%
\pgfpathlineto{\pgfqpoint{1.136620in}{0.721331in}}%
\pgfpathlineto{\pgfqpoint{1.144920in}{0.726138in}}%
\pgfpathlineto{\pgfqpoint{1.153220in}{0.731001in}}%
\pgfpathlineto{\pgfqpoint{1.161519in}{0.735918in}}%
\pgfpathlineto{\pgfqpoint{1.169819in}{0.740890in}}%
\pgfpathlineto{\pgfqpoint{1.178119in}{0.745916in}}%
\pgfpathlineto{\pgfqpoint{1.186419in}{0.750996in}}%
\pgfpathlineto{\pgfqpoint{1.194718in}{0.756129in}}%
\pgfpathlineto{\pgfqpoint{1.203018in}{0.761315in}}%
\pgfpathlineto{\pgfqpoint{1.211318in}{0.766555in}}%
\pgfpathlineto{\pgfqpoint{1.219617in}{0.771847in}}%
\pgfpathlineto{\pgfqpoint{1.227917in}{0.777191in}}%
\pgfpathlineto{\pgfqpoint{1.236217in}{0.782586in}}%
\pgfpathlineto{\pgfqpoint{1.244517in}{0.788032in}}%
\pgfpathlineto{\pgfqpoint{1.252816in}{0.793529in}}%
\pgfpathlineto{\pgfqpoint{1.261116in}{0.799077in}}%
\pgfpathlineto{\pgfqpoint{1.269416in}{0.804673in}}%
\pgfpathlineto{\pgfqpoint{1.277715in}{0.810319in}}%
\pgfpathlineto{\pgfqpoint{1.286015in}{0.816013in}}%
\pgfpathlineto{\pgfqpoint{1.294315in}{0.821754in}}%
\pgfpathlineto{\pgfqpoint{1.302615in}{0.827543in}}%
\pgfpathlineto{\pgfqpoint{1.310914in}{0.833379in}}%
\pgfpathlineto{\pgfqpoint{1.319214in}{0.839260in}}%
\pgfpathlineto{\pgfqpoint{1.327514in}{0.845186in}}%
\pgfpathlineto{\pgfqpoint{1.335813in}{0.851157in}}%
\pgfpathlineto{\pgfqpoint{1.344113in}{0.857172in}}%
\pgfpathlineto{\pgfqpoint{1.352413in}{0.863229in}}%
\pgfpathlineto{\pgfqpoint{1.360713in}{0.869328in}}%
\pgfpathlineto{\pgfqpoint{1.369012in}{0.875469in}}%
\pgfpathlineto{\pgfqpoint{1.377312in}{0.881651in}}%
\pgfpathlineto{\pgfqpoint{1.385612in}{0.887872in}}%
\pgfpathlineto{\pgfqpoint{1.393911in}{0.894132in}}%
\pgfpathlineto{\pgfqpoint{1.402211in}{0.900429in}}%
\pgfpathlineto{\pgfqpoint{1.410511in}{0.906764in}}%
\pgfpathlineto{\pgfqpoint{1.418811in}{0.913135in}}%
\pgfpathlineto{\pgfqpoint{1.427110in}{0.919542in}}%
\pgfpathlineto{\pgfqpoint{1.435410in}{0.925982in}}%
\pgfpathlineto{\pgfqpoint{1.443710in}{0.932456in}}%
\pgfpathlineto{\pgfqpoint{1.452010in}{0.938962in}}%
\pgfpathlineto{\pgfqpoint{1.460309in}{0.945499in}}%
\pgfpathlineto{\pgfqpoint{1.468609in}{0.952067in}}%
\pgfpathlineto{\pgfqpoint{1.476909in}{0.958664in}}%
\pgfpathlineto{\pgfqpoint{1.485208in}{0.965289in}}%
\pgfpathlineto{\pgfqpoint{1.493508in}{0.971941in}}%
\pgfpathlineto{\pgfqpoint{1.501808in}{0.978620in}}%
\pgfpathlineto{\pgfqpoint{1.510108in}{0.985323in}}%
\pgfpathlineto{\pgfqpoint{1.518407in}{0.992050in}}%
\pgfpathlineto{\pgfqpoint{1.526707in}{0.998800in}}%
\pgfpathlineto{\pgfqpoint{1.535007in}{1.005572in}}%
\pgfpathlineto{\pgfqpoint{1.543306in}{1.012364in}}%
\pgfpathlineto{\pgfqpoint{1.551606in}{1.019175in}}%
\pgfpathlineto{\pgfqpoint{1.559906in}{1.026005in}}%
\pgfpathlineto{\pgfqpoint{1.568206in}{1.032852in}}%
\pgfpathlineto{\pgfqpoint{1.576505in}{1.039714in}}%
\pgfpathlineto{\pgfqpoint{1.584805in}{1.046591in}}%
\pgfpathlineto{\pgfqpoint{1.593105in}{1.053482in}}%
\pgfpathlineto{\pgfqpoint{1.601404in}{1.060385in}}%
\pgfpathlineto{\pgfqpoint{1.609704in}{1.067299in}}%
\pgfpathlineto{\pgfqpoint{1.618004in}{1.074223in}}%
\pgfpathlineto{\pgfqpoint{1.626304in}{1.081155in}}%
\pgfpathlineto{\pgfqpoint{1.634603in}{1.088095in}}%
\pgfpathlineto{\pgfqpoint{1.642903in}{1.095041in}}%
\pgfpathlineto{\pgfqpoint{1.651203in}{1.101992in}}%
\pgfpathlineto{\pgfqpoint{1.659502in}{1.108946in}}%
\pgfpathlineto{\pgfqpoint{1.667802in}{1.115903in}}%
\pgfpathlineto{\pgfqpoint{1.676102in}{1.122862in}}%
\pgfpathlineto{\pgfqpoint{1.676102in}{1.122862in}}%
\pgfpathlineto{\pgfqpoint{1.676102in}{0.468770in}}%
\pgfpathlineto{\pgfqpoint{1.667802in}{0.468770in}}%
\pgfpathlineto{\pgfqpoint{1.659502in}{0.468770in}}%
\pgfpathlineto{\pgfqpoint{1.651203in}{0.468770in}}%
\pgfpathlineto{\pgfqpoint{1.642903in}{0.468770in}}%
\pgfpathlineto{\pgfqpoint{1.634603in}{0.468770in}}%
\pgfpathlineto{\pgfqpoint{1.626304in}{0.468770in}}%
\pgfpathlineto{\pgfqpoint{1.618004in}{0.468770in}}%
\pgfpathlineto{\pgfqpoint{1.609704in}{0.468770in}}%
\pgfpathlineto{\pgfqpoint{1.601404in}{0.468770in}}%
\pgfpathlineto{\pgfqpoint{1.593105in}{0.468770in}}%
\pgfpathlineto{\pgfqpoint{1.584805in}{0.468770in}}%
\pgfpathlineto{\pgfqpoint{1.576505in}{0.468770in}}%
\pgfpathlineto{\pgfqpoint{1.568206in}{0.468770in}}%
\pgfpathlineto{\pgfqpoint{1.559906in}{0.468770in}}%
\pgfpathlineto{\pgfqpoint{1.551606in}{0.468770in}}%
\pgfpathlineto{\pgfqpoint{1.543306in}{0.468770in}}%
\pgfpathlineto{\pgfqpoint{1.535007in}{0.468770in}}%
\pgfpathlineto{\pgfqpoint{1.526707in}{0.468770in}}%
\pgfpathlineto{\pgfqpoint{1.518407in}{0.468770in}}%
\pgfpathlineto{\pgfqpoint{1.510108in}{0.468770in}}%
\pgfpathlineto{\pgfqpoint{1.501808in}{0.468770in}}%
\pgfpathlineto{\pgfqpoint{1.493508in}{0.468770in}}%
\pgfpathlineto{\pgfqpoint{1.485208in}{0.468770in}}%
\pgfpathlineto{\pgfqpoint{1.476909in}{0.468770in}}%
\pgfpathlineto{\pgfqpoint{1.468609in}{0.468770in}}%
\pgfpathlineto{\pgfqpoint{1.460309in}{0.468770in}}%
\pgfpathlineto{\pgfqpoint{1.452010in}{0.468770in}}%
\pgfpathlineto{\pgfqpoint{1.443710in}{0.468770in}}%
\pgfpathlineto{\pgfqpoint{1.435410in}{0.468770in}}%
\pgfpathlineto{\pgfqpoint{1.427110in}{0.468770in}}%
\pgfpathlineto{\pgfqpoint{1.418811in}{0.468770in}}%
\pgfpathlineto{\pgfqpoint{1.410511in}{0.468770in}}%
\pgfpathlineto{\pgfqpoint{1.402211in}{0.468770in}}%
\pgfpathlineto{\pgfqpoint{1.393911in}{0.468770in}}%
\pgfpathlineto{\pgfqpoint{1.385612in}{0.468770in}}%
\pgfpathlineto{\pgfqpoint{1.377312in}{0.468770in}}%
\pgfpathlineto{\pgfqpoint{1.369012in}{0.468770in}}%
\pgfpathlineto{\pgfqpoint{1.360713in}{0.468770in}}%
\pgfpathlineto{\pgfqpoint{1.352413in}{0.468770in}}%
\pgfpathlineto{\pgfqpoint{1.344113in}{0.468770in}}%
\pgfpathlineto{\pgfqpoint{1.335813in}{0.468770in}}%
\pgfpathlineto{\pgfqpoint{1.327514in}{0.468770in}}%
\pgfpathlineto{\pgfqpoint{1.319214in}{0.468770in}}%
\pgfpathlineto{\pgfqpoint{1.310914in}{0.468770in}}%
\pgfpathlineto{\pgfqpoint{1.302615in}{0.468770in}}%
\pgfpathlineto{\pgfqpoint{1.294315in}{0.468770in}}%
\pgfpathlineto{\pgfqpoint{1.286015in}{0.468770in}}%
\pgfpathlineto{\pgfqpoint{1.277715in}{0.468770in}}%
\pgfpathlineto{\pgfqpoint{1.269416in}{0.468770in}}%
\pgfpathlineto{\pgfqpoint{1.261116in}{0.468770in}}%
\pgfpathlineto{\pgfqpoint{1.252816in}{0.468770in}}%
\pgfpathlineto{\pgfqpoint{1.244517in}{0.468770in}}%
\pgfpathlineto{\pgfqpoint{1.236217in}{0.468770in}}%
\pgfpathlineto{\pgfqpoint{1.227917in}{0.468770in}}%
\pgfpathlineto{\pgfqpoint{1.219617in}{0.468770in}}%
\pgfpathlineto{\pgfqpoint{1.211318in}{0.468770in}}%
\pgfpathlineto{\pgfqpoint{1.203018in}{0.468770in}}%
\pgfpathlineto{\pgfqpoint{1.194718in}{0.468770in}}%
\pgfpathlineto{\pgfqpoint{1.186419in}{0.468770in}}%
\pgfpathlineto{\pgfqpoint{1.178119in}{0.468770in}}%
\pgfpathlineto{\pgfqpoint{1.169819in}{0.468770in}}%
\pgfpathlineto{\pgfqpoint{1.161519in}{0.468770in}}%
\pgfpathlineto{\pgfqpoint{1.153220in}{0.468770in}}%
\pgfpathlineto{\pgfqpoint{1.144920in}{0.468770in}}%
\pgfpathlineto{\pgfqpoint{1.136620in}{0.468770in}}%
\pgfpathlineto{\pgfqpoint{1.128320in}{0.468770in}}%
\pgfpathlineto{\pgfqpoint{1.120021in}{0.468770in}}%
\pgfpathlineto{\pgfqpoint{1.111721in}{0.468770in}}%
\pgfpathlineto{\pgfqpoint{1.103421in}{0.468770in}}%
\pgfpathlineto{\pgfqpoint{1.095122in}{0.468770in}}%
\pgfpathlineto{\pgfqpoint{1.086822in}{0.468770in}}%
\pgfpathlineto{\pgfqpoint{1.078522in}{0.468770in}}%
\pgfpathlineto{\pgfqpoint{1.070222in}{0.468770in}}%
\pgfpathlineto{\pgfqpoint{1.061923in}{0.468770in}}%
\pgfpathlineto{\pgfqpoint{1.053623in}{0.468770in}}%
\pgfpathlineto{\pgfqpoint{1.045323in}{0.468770in}}%
\pgfpathlineto{\pgfqpoint{1.037024in}{0.468770in}}%
\pgfpathlineto{\pgfqpoint{1.028724in}{0.468770in}}%
\pgfpathlineto{\pgfqpoint{1.020424in}{0.468770in}}%
\pgfpathlineto{\pgfqpoint{1.012124in}{0.468770in}}%
\pgfpathlineto{\pgfqpoint{1.003825in}{0.468770in}}%
\pgfpathlineto{\pgfqpoint{0.995525in}{0.468770in}}%
\pgfpathlineto{\pgfqpoint{0.987225in}{0.468770in}}%
\pgfpathlineto{\pgfqpoint{0.978926in}{0.468770in}}%
\pgfpathlineto{\pgfqpoint{0.970626in}{0.468770in}}%
\pgfpathlineto{\pgfqpoint{0.962326in}{0.468770in}}%
\pgfpathlineto{\pgfqpoint{0.954026in}{0.468770in}}%
\pgfpathlineto{\pgfqpoint{0.945727in}{0.468770in}}%
\pgfpathlineto{\pgfqpoint{0.937427in}{0.468770in}}%
\pgfpathlineto{\pgfqpoint{0.929127in}{0.468770in}}%
\pgfpathlineto{\pgfqpoint{0.920828in}{0.468770in}}%
\pgfpathlineto{\pgfqpoint{0.912528in}{0.468770in}}%
\pgfpathlineto{\pgfqpoint{0.904228in}{0.468770in}}%
\pgfpathlineto{\pgfqpoint{0.895928in}{0.468770in}}%
\pgfpathlineto{\pgfqpoint{0.887629in}{0.468770in}}%
\pgfpathlineto{\pgfqpoint{0.879329in}{0.468770in}}%
\pgfpathlineto{\pgfqpoint{0.871029in}{0.468770in}}%
\pgfpathlineto{\pgfqpoint{0.862729in}{0.468770in}}%
\pgfpathlineto{\pgfqpoint{0.854430in}{0.468770in}}%
\pgfpathlineto{\pgfqpoint{0.846130in}{0.468770in}}%
\pgfpathlineto{\pgfqpoint{0.837830in}{0.468770in}}%
\pgfpathlineto{\pgfqpoint{0.829531in}{0.468770in}}%
\pgfpathlineto{\pgfqpoint{0.821231in}{0.468770in}}%
\pgfpathlineto{\pgfqpoint{0.812931in}{0.468770in}}%
\pgfpathlineto{\pgfqpoint{0.804631in}{0.468770in}}%
\pgfpathlineto{\pgfqpoint{0.796332in}{0.468770in}}%
\pgfpathlineto{\pgfqpoint{0.788032in}{0.468770in}}%
\pgfpathlineto{\pgfqpoint{0.779732in}{0.468770in}}%
\pgfpathlineto{\pgfqpoint{0.771433in}{0.468770in}}%
\pgfpathlineto{\pgfqpoint{0.763133in}{0.468770in}}%
\pgfpathlineto{\pgfqpoint{0.754833in}{0.468770in}}%
\pgfpathlineto{\pgfqpoint{0.746533in}{0.468770in}}%
\pgfpathlineto{\pgfqpoint{0.738234in}{0.468770in}}%
\pgfpathlineto{\pgfqpoint{0.729934in}{0.468770in}}%
\pgfpathlineto{\pgfqpoint{0.721634in}{0.468770in}}%
\pgfpathlineto{\pgfqpoint{0.713335in}{0.468770in}}%
\pgfpathlineto{\pgfqpoint{0.705035in}{0.468770in}}%
\pgfpathlineto{\pgfqpoint{0.696735in}{0.468770in}}%
\pgfpathlineto{\pgfqpoint{0.688435in}{0.468770in}}%
\pgfpathlineto{\pgfqpoint{0.680136in}{0.468770in}}%
\pgfpathlineto{\pgfqpoint{0.671836in}{0.468770in}}%
\pgfpathlineto{\pgfqpoint{0.663536in}{0.468770in}}%
\pgfpathlineto{\pgfqpoint{0.655237in}{0.468770in}}%
\pgfpathlineto{\pgfqpoint{0.646937in}{0.468770in}}%
\pgfpathlineto{\pgfqpoint{0.638637in}{0.468770in}}%
\pgfpathlineto{\pgfqpoint{0.630337in}{0.468770in}}%
\pgfpathlineto{\pgfqpoint{0.622038in}{0.468770in}}%
\pgfpathlineto{\pgfqpoint{0.613738in}{0.468770in}}%
\pgfpathlineto{\pgfqpoint{0.605438in}{0.468770in}}%
\pgfpathlineto{\pgfqpoint{0.597138in}{0.468770in}}%
\pgfpathlineto{\pgfqpoint{0.588839in}{0.468770in}}%
\pgfpathlineto{\pgfqpoint{0.580539in}{0.468770in}}%
\pgfpathlineto{\pgfqpoint{0.572239in}{0.468770in}}%
\pgfpathlineto{\pgfqpoint{0.563940in}{0.468770in}}%
\pgfpathlineto{\pgfqpoint{0.555640in}{0.468770in}}%
\pgfpathlineto{\pgfqpoint{0.547340in}{0.468770in}}%
\pgfpathlineto{\pgfqpoint{0.539040in}{0.468770in}}%
\pgfpathlineto{\pgfqpoint{0.530741in}{0.468770in}}%
\pgfpathlineto{\pgfqpoint{0.522441in}{0.468770in}}%
\pgfpathlineto{\pgfqpoint{0.514141in}{0.468770in}}%
\pgfpathlineto{\pgfqpoint{0.505842in}{0.468770in}}%
\pgfpathlineto{\pgfqpoint{0.497542in}{0.468770in}}%
\pgfpathlineto{\pgfqpoint{0.489242in}{0.468770in}}%
\pgfpathlineto{\pgfqpoint{0.480942in}{0.468770in}}%
\pgfpathlineto{\pgfqpoint{0.472643in}{0.468770in}}%
\pgfpathlineto{\pgfqpoint{0.464343in}{0.468770in}}%
\pgfpathlineto{\pgfqpoint{0.456043in}{0.468770in}}%
\pgfpathlineto{\pgfqpoint{0.447744in}{0.468770in}}%
\pgfpathlineto{\pgfqpoint{0.439444in}{0.468770in}}%
\pgfpathlineto{\pgfqpoint{0.431144in}{0.468770in}}%
\pgfpathclose%
\pgfusepath{fill}%
\end{pgfscope}%
\begin{pgfscope}%
\pgfsetbuttcap%
\pgfsetroundjoin%
\definecolor{currentfill}{rgb}{0.000000,0.000000,0.000000}%
\pgfsetfillcolor{currentfill}%
\pgfsetlinewidth{0.803000pt}%
\definecolor{currentstroke}{rgb}{0.000000,0.000000,0.000000}%
\pgfsetstrokecolor{currentstroke}%
\pgfsetdash{}{0pt}%
\pgfsys@defobject{currentmarker}{\pgfqpoint{0.000000in}{-0.048611in}}{\pgfqpoint{0.000000in}{0.000000in}}{%
\pgfpathmoveto{\pgfqpoint{0.000000in}{0.000000in}}%
\pgfpathlineto{\pgfqpoint{0.000000in}{-0.048611in}}%
\pgfusepath{stroke,fill}%
}%
\begin{pgfscope}%
\pgfsys@transformshift{1.676102in}{0.468770in}%
\pgfsys@useobject{currentmarker}{}%
\end{pgfscope}%
\end{pgfscope}%
\begin{pgfscope}%
\pgftext[x=1.676102in,y=0.371548in,,top]{\sffamily\fontsize{10.000000}{12.000000}\selectfont \(\displaystyle 0\)}%
\end{pgfscope}%
\begin{pgfscope}%
\pgftext[x=1.676102in,y=0.192659in,,top]{\sffamily\fontsize{10.000000}{12.000000}\selectfont \(\displaystyle z\)}%
\end{pgfscope}%
\begin{pgfscope}%
\pgfsetbuttcap%
\pgfsetroundjoin%
\definecolor{currentfill}{rgb}{0.000000,0.000000,0.000000}%
\pgfsetfillcolor{currentfill}%
\pgfsetlinewidth{0.803000pt}%
\definecolor{currentstroke}{rgb}{0.000000,0.000000,0.000000}%
\pgfsetstrokecolor{currentstroke}%
\pgfsetdash{}{0pt}%
\pgfsys@defobject{currentmarker}{\pgfqpoint{-0.048611in}{0.000000in}}{\pgfqpoint{0.000000in}{0.000000in}}{%
\pgfpathmoveto{\pgfqpoint{0.000000in}{0.000000in}}%
\pgfpathlineto{\pgfqpoint{-0.048611in}{0.000000in}}%
\pgfusepath{stroke,fill}%
}%
\begin{pgfscope}%
\pgfsys@transformshift{0.431144in}{0.468770in}%
\pgfsys@useobject{currentmarker}{}%
\end{pgfscope}%
\end{pgfscope}%
\begin{pgfscope}%
\pgftext[x=0.264477in,y=0.420576in,left,base]{\sffamily\fontsize{10.000000}{12.000000}\selectfont \(\displaystyle 0\)}%
\end{pgfscope}%
\begin{pgfscope}%
\pgfsetbuttcap%
\pgfsetroundjoin%
\definecolor{currentfill}{rgb}{0.000000,0.000000,0.000000}%
\pgfsetfillcolor{currentfill}%
\pgfsetlinewidth{0.803000pt}%
\definecolor{currentstroke}{rgb}{0.000000,0.000000,0.000000}%
\pgfsetstrokecolor{currentstroke}%
\pgfsetdash{}{0pt}%
\pgfsys@defobject{currentmarker}{\pgfqpoint{-0.048611in}{0.000000in}}{\pgfqpoint{0.000000in}{0.000000in}}{%
\pgfpathmoveto{\pgfqpoint{0.000000in}{0.000000in}}%
\pgfpathlineto{\pgfqpoint{-0.048611in}{0.000000in}}%
\pgfusepath{stroke,fill}%
}%
\begin{pgfscope}%
\pgfsys@transformshift{0.431144in}{1.776953in}%
\pgfsys@useobject{currentmarker}{}%
\end{pgfscope}%
\end{pgfscope}%
\begin{pgfscope}%
\pgftext[x=0.264477in,y=1.728759in,left,base]{\sffamily\fontsize{10.000000}{12.000000}\selectfont \(\displaystyle 1\)}%
\end{pgfscope}%
\begin{pgfscope}%
\pgftext[x=0.208922in,y=1.122862in,,bottom,rotate=90.000000]{\sffamily\fontsize{10.000000}{12.000000}\selectfont \(\displaystyle F(z)\)}%
\end{pgfscope}%
\begin{pgfscope}%
\pgfpathrectangle{\pgfqpoint{0.306648in}{0.403361in}}{\pgfqpoint{2.738907in}{1.439002in}} %
\pgfusepath{clip}%
\pgfsetrectcap%
\pgfsetroundjoin%
\pgfsetlinewidth{1.204500pt}%
\definecolor{currentstroke}{rgb}{0.000000,0.000000,0.000000}%
\pgfsetstrokecolor{currentstroke}%
\pgfsetdash{}{0pt}%
\pgfpathmoveto{\pgfqpoint{0.431144in}{0.498531in}}%
\pgfpathlineto{\pgfqpoint{0.505842in}{0.508086in}}%
\pgfpathlineto{\pgfqpoint{0.572239in}{0.518594in}}%
\pgfpathlineto{\pgfqpoint{0.638637in}{0.531289in}}%
\pgfpathlineto{\pgfqpoint{0.705035in}{0.546450in}}%
\pgfpathlineto{\pgfqpoint{0.763133in}{0.561956in}}%
\pgfpathlineto{\pgfqpoint{0.821231in}{0.579736in}}%
\pgfpathlineto{\pgfqpoint{0.879329in}{0.599945in}}%
\pgfpathlineto{\pgfqpoint{0.937427in}{0.622716in}}%
\pgfpathlineto{\pgfqpoint{0.995525in}{0.648153in}}%
\pgfpathlineto{\pgfqpoint{1.053623in}{0.676320in}}%
\pgfpathlineto{\pgfqpoint{1.111721in}{0.707241in}}%
\pgfpathlineto{\pgfqpoint{1.169819in}{0.740890in}}%
\pgfpathlineto{\pgfqpoint{1.227917in}{0.777191in}}%
\pgfpathlineto{\pgfqpoint{1.286015in}{0.816013in}}%
\pgfpathlineto{\pgfqpoint{1.352413in}{0.863229in}}%
\pgfpathlineto{\pgfqpoint{1.418811in}{0.913135in}}%
\pgfpathlineto{\pgfqpoint{1.493508in}{0.971941in}}%
\pgfpathlineto{\pgfqpoint{1.593105in}{1.053482in}}%
\pgfpathlineto{\pgfqpoint{1.883595in}{1.293656in}}%
\pgfpathlineto{\pgfqpoint{1.958292in}{1.351592in}}%
\pgfpathlineto{\pgfqpoint{2.024690in}{1.400537in}}%
\pgfpathlineto{\pgfqpoint{2.091088in}{1.446647in}}%
\pgfpathlineto{\pgfqpoint{2.149186in}{1.484408in}}%
\pgfpathlineto{\pgfqpoint{2.207284in}{1.519585in}}%
\pgfpathlineto{\pgfqpoint{2.265382in}{1.552071in}}%
\pgfpathlineto{\pgfqpoint{2.323480in}{1.581812in}}%
\pgfpathlineto{\pgfqpoint{2.381578in}{1.608803in}}%
\pgfpathlineto{\pgfqpoint{2.439676in}{1.633087in}}%
\pgfpathlineto{\pgfqpoint{2.497774in}{1.654746in}}%
\pgfpathlineto{\pgfqpoint{2.555872in}{1.673896in}}%
\pgfpathlineto{\pgfqpoint{2.613970in}{1.690681in}}%
\pgfpathlineto{\pgfqpoint{2.672068in}{1.705266in}}%
\pgfpathlineto{\pgfqpoint{2.738466in}{1.719469in}}%
\pgfpathlineto{\pgfqpoint{2.804864in}{1.731311in}}%
\pgfpathlineto{\pgfqpoint{2.879561in}{1.742159in}}%
\pgfpathlineto{\pgfqpoint{2.921060in}{1.747192in}}%
\pgfpathlineto{\pgfqpoint{2.921060in}{1.747192in}}%
\pgfusepath{stroke}%
\end{pgfscope}%
\begin{pgfscope}%
\pgfsetrectcap%
\pgfsetmiterjoin%
\pgfsetlinewidth{0.803000pt}%
\definecolor{currentstroke}{rgb}{0.000000,0.000000,0.000000}%
\pgfsetstrokecolor{currentstroke}%
\pgfsetdash{}{0pt}%
\pgfpathmoveto{\pgfqpoint{0.431144in}{0.468770in}}%
\pgfpathlineto{\pgfqpoint{0.431144in}{1.776953in}}%
\pgfusepath{stroke}%
\end{pgfscope}%
\begin{pgfscope}%
\pgfsetrectcap%
\pgfsetmiterjoin%
\pgfsetlinewidth{0.803000pt}%
\definecolor{currentstroke}{rgb}{0.000000,0.000000,0.000000}%
\pgfsetstrokecolor{currentstroke}%
\pgfsetdash{}{0pt}%
\pgfpathmoveto{\pgfqpoint{2.921060in}{0.468770in}}%
\pgfpathlineto{\pgfqpoint{2.921060in}{1.776953in}}%
\pgfusepath{stroke}%
\end{pgfscope}%
\begin{pgfscope}%
\pgfsetrectcap%
\pgfsetmiterjoin%
\pgfsetlinewidth{0.803000pt}%
\definecolor{currentstroke}{rgb}{0.000000,0.000000,0.000000}%
\pgfsetstrokecolor{currentstroke}%
\pgfsetdash{}{0pt}%
\pgfpathmoveto{\pgfqpoint{0.431144in}{0.468770in}}%
\pgfpathlineto{\pgfqpoint{2.921060in}{0.468770in}}%
\pgfusepath{stroke}%
\end{pgfscope}%
\begin{pgfscope}%
\pgfsetrectcap%
\pgfsetmiterjoin%
\pgfsetlinewidth{0.803000pt}%
\definecolor{currentstroke}{rgb}{0.000000,0.000000,0.000000}%
\pgfsetstrokecolor{currentstroke}%
\pgfsetdash{}{0pt}%
\pgfpathmoveto{\pgfqpoint{0.431144in}{1.776953in}}%
\pgfpathlineto{\pgfqpoint{2.921060in}{1.776953in}}%
\pgfusepath{stroke}%
\end{pgfscope}%
\end{pgfpicture}%
\makeatother%
\endgroup%

%% file: figures/expectedVsWeighted_b.pgf
%% Creator: Matplotlib, PGF backend
%%
%% To include the figure in your LaTeX document, write
%%   \input{<filename>.pgf}
%%
%% Make sure the required packages are loaded in your preamble
%%   \usepackage{pgf}
%%
%% Figures using additional raster images can only be included by \input if
%% they are in the same directory as the main LaTeX file. For loading figures
%% from other directories you can use the `import` package
%%   \usepackage{import}
%% and then include the figures with
%%   \import{<path to file>}{<filename>.pgf}
%%
%% Matplotlib used the following preamble
%%   \usepackage{fontspec}
%%
\begingroup%
\makeatletter%
\begin{pgfpicture}%
\pgfpathrectangle{\pgfpointorigin}{\pgfqpoint{3.150000in}{1.946807in}}%
\pgfusepath{use as bounding box, clip}%
\begin{pgfscope}%
\pgfsetbuttcap%
\pgfsetmiterjoin%
\pgfsetlinewidth{0.000000pt}%
\definecolor{currentstroke}{rgb}{1.000000,1.000000,1.000000}%
\pgfsetstrokecolor{currentstroke}%
\pgfsetstrokeopacity{0.000000}%
\pgfsetdash{}{0pt}%
\pgfpathmoveto{\pgfqpoint{0.000000in}{0.000000in}}%
\pgfpathlineto{\pgfqpoint{3.150000in}{0.000000in}}%
\pgfpathlineto{\pgfqpoint{3.150000in}{1.946807in}}%
\pgfpathlineto{\pgfqpoint{0.000000in}{1.946807in}}%
\pgfpathclose%
\pgfusepath{}%
\end{pgfscope}%
\begin{pgfscope}%
\pgfsetbuttcap%
\pgfsetmiterjoin%
\definecolor{currentfill}{rgb}{1.000000,1.000000,1.000000}%
\pgfsetfillcolor{currentfill}%
\pgfsetlinewidth{0.000000pt}%
\definecolor{currentstroke}{rgb}{0.000000,0.000000,0.000000}%
\pgfsetstrokecolor{currentstroke}%
\pgfsetstrokeopacity{0.000000}%
\pgfsetdash{}{0pt}%
\pgfpathmoveto{\pgfqpoint{0.306648in}{0.403361in}}%
\pgfpathlineto{\pgfqpoint{3.045556in}{0.403361in}}%
\pgfpathlineto{\pgfqpoint{3.045556in}{1.842363in}}%
\pgfpathlineto{\pgfqpoint{0.306648in}{1.842363in}}%
\pgfpathclose%
\pgfusepath{fill}%
\end{pgfscope}%
\begin{pgfscope}%
\pgfpathrectangle{\pgfqpoint{0.306648in}{0.403361in}}{\pgfqpoint{2.738907in}{1.439002in}} %
\pgfusepath{clip}%
\pgfsetbuttcap%
\pgfsetroundjoin%
\definecolor{currentfill}{rgb}{0.882353,0.945098,1.000000}%
\pgfsetfillcolor{currentfill}%
\pgfsetlinewidth{0.000000pt}%
\definecolor{currentstroke}{rgb}{0.000000,0.000000,0.000000}%
\pgfsetstrokecolor{currentstroke}%
\pgfsetdash{}{0pt}%
\pgfpathmoveto{\pgfqpoint{1.676102in}{1.449908in}}%
\pgfpathlineto{\pgfqpoint{1.676102in}{1.449908in}}%
\pgfpathlineto{\pgfqpoint{1.684402in}{1.456829in}}%
\pgfpathlineto{\pgfqpoint{1.692701in}{1.463675in}}%
\pgfpathlineto{\pgfqpoint{1.701001in}{1.470445in}}%
\pgfpathlineto{\pgfqpoint{1.709301in}{1.477137in}}%
\pgfpathlineto{\pgfqpoint{1.717601in}{1.483750in}}%
\pgfpathlineto{\pgfqpoint{1.725900in}{1.490285in}}%
\pgfpathlineto{\pgfqpoint{1.734200in}{1.496738in}}%
\pgfpathlineto{\pgfqpoint{1.742500in}{1.503110in}}%
\pgfpathlineto{\pgfqpoint{1.750799in}{1.509401in}}%
\pgfpathlineto{\pgfqpoint{1.759099in}{1.515608in}}%
\pgfpathlineto{\pgfqpoint{1.767399in}{1.521731in}}%
\pgfpathlineto{\pgfqpoint{1.775699in}{1.527770in}}%
\pgfpathlineto{\pgfqpoint{1.783998in}{1.533724in}}%
\pgfpathlineto{\pgfqpoint{1.792298in}{1.539593in}}%
\pgfpathlineto{\pgfqpoint{1.800598in}{1.545376in}}%
\pgfpathlineto{\pgfqpoint{1.808897in}{1.551072in}}%
\pgfpathlineto{\pgfqpoint{1.817197in}{1.556681in}}%
\pgfpathlineto{\pgfqpoint{1.825497in}{1.562204in}}%
\pgfpathlineto{\pgfqpoint{1.833797in}{1.567639in}}%
\pgfpathlineto{\pgfqpoint{1.842096in}{1.572986in}}%
\pgfpathlineto{\pgfqpoint{1.850396in}{1.578245in}}%
\pgfpathlineto{\pgfqpoint{1.858696in}{1.583417in}}%
\pgfpathlineto{\pgfqpoint{1.866995in}{1.588500in}}%
\pgfpathlineto{\pgfqpoint{1.875295in}{1.593496in}}%
\pgfpathlineto{\pgfqpoint{1.883595in}{1.598404in}}%
\pgfpathlineto{\pgfqpoint{1.891895in}{1.603223in}}%
\pgfpathlineto{\pgfqpoint{1.900194in}{1.607955in}}%
\pgfpathlineto{\pgfqpoint{1.908494in}{1.612600in}}%
\pgfpathlineto{\pgfqpoint{1.916794in}{1.617157in}}%
\pgfpathlineto{\pgfqpoint{1.925093in}{1.621627in}}%
\pgfpathlineto{\pgfqpoint{1.933393in}{1.626011in}}%
\pgfpathlineto{\pgfqpoint{1.941693in}{1.630308in}}%
\pgfpathlineto{\pgfqpoint{1.949993in}{1.634519in}}%
\pgfpathlineto{\pgfqpoint{1.958292in}{1.638645in}}%
\pgfpathlineto{\pgfqpoint{1.966592in}{1.642686in}}%
\pgfpathlineto{\pgfqpoint{1.974892in}{1.646643in}}%
\pgfpathlineto{\pgfqpoint{1.983192in}{1.650515in}}%
\pgfpathlineto{\pgfqpoint{1.991491in}{1.654305in}}%
\pgfpathlineto{\pgfqpoint{1.999791in}{1.658012in}}%
\pgfpathlineto{\pgfqpoint{2.008091in}{1.661636in}}%
\pgfpathlineto{\pgfqpoint{2.016390in}{1.665180in}}%
\pgfpathlineto{\pgfqpoint{2.024690in}{1.668644in}}%
\pgfpathlineto{\pgfqpoint{2.032990in}{1.672027in}}%
\pgfpathlineto{\pgfqpoint{2.041290in}{1.675332in}}%
\pgfpathlineto{\pgfqpoint{2.049589in}{1.678559in}}%
\pgfpathlineto{\pgfqpoint{2.057889in}{1.681708in}}%
\pgfpathlineto{\pgfqpoint{2.066189in}{1.684782in}}%
\pgfpathlineto{\pgfqpoint{2.074488in}{1.687780in}}%
\pgfpathlineto{\pgfqpoint{2.082788in}{1.690703in}}%
\pgfpathlineto{\pgfqpoint{2.091088in}{1.693554in}}%
\pgfpathlineto{\pgfqpoint{2.099388in}{1.696331in}}%
\pgfpathlineto{\pgfqpoint{2.107687in}{1.699037in}}%
\pgfpathlineto{\pgfqpoint{2.115987in}{1.701673in}}%
\pgfpathlineto{\pgfqpoint{2.124287in}{1.704239in}}%
\pgfpathlineto{\pgfqpoint{2.132586in}{1.706737in}}%
\pgfpathlineto{\pgfqpoint{2.140886in}{1.709168in}}%
\pgfpathlineto{\pgfqpoint{2.149186in}{1.711532in}}%
\pgfpathlineto{\pgfqpoint{2.157486in}{1.713831in}}%
\pgfpathlineto{\pgfqpoint{2.165785in}{1.716067in}}%
\pgfpathlineto{\pgfqpoint{2.174085in}{1.718239in}}%
\pgfpathlineto{\pgfqpoint{2.182385in}{1.720349in}}%
\pgfpathlineto{\pgfqpoint{2.190685in}{1.722398in}}%
\pgfpathlineto{\pgfqpoint{2.198984in}{1.724388in}}%
\pgfpathlineto{\pgfqpoint{2.207284in}{1.726319in}}%
\pgfpathlineto{\pgfqpoint{2.215584in}{1.728193in}}%
\pgfpathlineto{\pgfqpoint{2.223883in}{1.730011in}}%
\pgfpathlineto{\pgfqpoint{2.232183in}{1.731774in}}%
\pgfpathlineto{\pgfqpoint{2.240483in}{1.733482in}}%
\pgfpathlineto{\pgfqpoint{2.248783in}{1.735138in}}%
\pgfpathlineto{\pgfqpoint{2.257082in}{1.736742in}}%
\pgfpathlineto{\pgfqpoint{2.265382in}{1.738295in}}%
\pgfpathlineto{\pgfqpoint{2.273682in}{1.739799in}}%
\pgfpathlineto{\pgfqpoint{2.281981in}{1.741254in}}%
\pgfpathlineto{\pgfqpoint{2.290281in}{1.742663in}}%
\pgfpathlineto{\pgfqpoint{2.298581in}{1.744025in}}%
\pgfpathlineto{\pgfqpoint{2.306881in}{1.745341in}}%
\pgfpathlineto{\pgfqpoint{2.315180in}{1.746614in}}%
\pgfpathlineto{\pgfqpoint{2.323480in}{1.747844in}}%
\pgfpathlineto{\pgfqpoint{2.331780in}{1.749032in}}%
\pgfpathlineto{\pgfqpoint{2.340079in}{1.750180in}}%
\pgfpathlineto{\pgfqpoint{2.348379in}{1.751287in}}%
\pgfpathlineto{\pgfqpoint{2.356679in}{1.752356in}}%
\pgfpathlineto{\pgfqpoint{2.364979in}{1.753387in}}%
\pgfpathlineto{\pgfqpoint{2.373278in}{1.754381in}}%
\pgfpathlineto{\pgfqpoint{2.381578in}{1.755340in}}%
\pgfpathlineto{\pgfqpoint{2.389878in}{1.756264in}}%
\pgfpathlineto{\pgfqpoint{2.398177in}{1.757154in}}%
\pgfpathlineto{\pgfqpoint{2.406477in}{1.758012in}}%
\pgfpathlineto{\pgfqpoint{2.414777in}{1.758837in}}%
\pgfpathlineto{\pgfqpoint{2.423077in}{1.759632in}}%
\pgfpathlineto{\pgfqpoint{2.431376in}{1.760396in}}%
\pgfpathlineto{\pgfqpoint{2.439676in}{1.761132in}}%
\pgfpathlineto{\pgfqpoint{2.447976in}{1.761839in}}%
\pgfpathlineto{\pgfqpoint{2.456276in}{1.762519in}}%
\pgfpathlineto{\pgfqpoint{2.464575in}{1.763173in}}%
\pgfpathlineto{\pgfqpoint{2.472875in}{1.763800in}}%
\pgfpathlineto{\pgfqpoint{2.481175in}{1.764403in}}%
\pgfpathlineto{\pgfqpoint{2.489474in}{1.764982in}}%
\pgfpathlineto{\pgfqpoint{2.497774in}{1.765537in}}%
\pgfpathlineto{\pgfqpoint{2.506074in}{1.766070in}}%
\pgfpathlineto{\pgfqpoint{2.514374in}{1.766581in}}%
\pgfpathlineto{\pgfqpoint{2.522673in}{1.767071in}}%
\pgfpathlineto{\pgfqpoint{2.530973in}{1.767541in}}%
\pgfpathlineto{\pgfqpoint{2.539273in}{1.767991in}}%
\pgfpathlineto{\pgfqpoint{2.547572in}{1.768422in}}%
\pgfpathlineto{\pgfqpoint{2.555872in}{1.768835in}}%
\pgfpathlineto{\pgfqpoint{2.564172in}{1.769230in}}%
\pgfpathlineto{\pgfqpoint{2.572472in}{1.769608in}}%
\pgfpathlineto{\pgfqpoint{2.580771in}{1.769970in}}%
\pgfpathlineto{\pgfqpoint{2.589071in}{1.770315in}}%
\pgfpathlineto{\pgfqpoint{2.597371in}{1.770646in}}%
\pgfpathlineto{\pgfqpoint{2.605670in}{1.770962in}}%
\pgfpathlineto{\pgfqpoint{2.613970in}{1.771264in}}%
\pgfpathlineto{\pgfqpoint{2.622270in}{1.771552in}}%
\pgfpathlineto{\pgfqpoint{2.630570in}{1.771828in}}%
\pgfpathlineto{\pgfqpoint{2.638869in}{1.772090in}}%
\pgfpathlineto{\pgfqpoint{2.647169in}{1.772341in}}%
\pgfpathlineto{\pgfqpoint{2.655469in}{1.772580in}}%
\pgfpathlineto{\pgfqpoint{2.663768in}{1.772808in}}%
\pgfpathlineto{\pgfqpoint{2.672068in}{1.773025in}}%
\pgfpathlineto{\pgfqpoint{2.680368in}{1.773232in}}%
\pgfpathlineto{\pgfqpoint{2.688668in}{1.773429in}}%
\pgfpathlineto{\pgfqpoint{2.696967in}{1.773617in}}%
\pgfpathlineto{\pgfqpoint{2.705267in}{1.773796in}}%
\pgfpathlineto{\pgfqpoint{2.713567in}{1.773966in}}%
\pgfpathlineto{\pgfqpoint{2.721867in}{1.774127in}}%
\pgfpathlineto{\pgfqpoint{2.730166in}{1.774281in}}%
\pgfpathlineto{\pgfqpoint{2.738466in}{1.774427in}}%
\pgfpathlineto{\pgfqpoint{2.746766in}{1.774566in}}%
\pgfpathlineto{\pgfqpoint{2.755065in}{1.774698in}}%
\pgfpathlineto{\pgfqpoint{2.763365in}{1.774824in}}%
\pgfpathlineto{\pgfqpoint{2.771665in}{1.774943in}}%
\pgfpathlineto{\pgfqpoint{2.779965in}{1.775056in}}%
\pgfpathlineto{\pgfqpoint{2.788264in}{1.775163in}}%
\pgfpathlineto{\pgfqpoint{2.796564in}{1.775265in}}%
\pgfpathlineto{\pgfqpoint{2.804864in}{1.775361in}}%
\pgfpathlineto{\pgfqpoint{2.813163in}{1.775452in}}%
\pgfpathlineto{\pgfqpoint{2.821463in}{1.775539in}}%
\pgfpathlineto{\pgfqpoint{2.829763in}{1.775621in}}%
\pgfpathlineto{\pgfqpoint{2.838063in}{1.775698in}}%
\pgfpathlineto{\pgfqpoint{2.846362in}{1.775772in}}%
\pgfpathlineto{\pgfqpoint{2.854662in}{1.775841in}}%
\pgfpathlineto{\pgfqpoint{2.862962in}{1.775907in}}%
\pgfpathlineto{\pgfqpoint{2.871261in}{1.775969in}}%
\pgfpathlineto{\pgfqpoint{2.879561in}{1.776028in}}%
\pgfpathlineto{\pgfqpoint{2.887861in}{1.776084in}}%
\pgfpathlineto{\pgfqpoint{2.896161in}{1.776136in}}%
\pgfpathlineto{\pgfqpoint{2.904460in}{1.776185in}}%
\pgfpathlineto{\pgfqpoint{2.912760in}{1.776232in}}%
\pgfpathlineto{\pgfqpoint{2.921060in}{1.776276in}}%
\pgfpathlineto{\pgfqpoint{2.921060in}{1.776276in}}%
\pgfpathlineto{\pgfqpoint{2.921060in}{1.776953in}}%
\pgfpathlineto{\pgfqpoint{2.912760in}{1.776953in}}%
\pgfpathlineto{\pgfqpoint{2.904460in}{1.776953in}}%
\pgfpathlineto{\pgfqpoint{2.896161in}{1.776953in}}%
\pgfpathlineto{\pgfqpoint{2.887861in}{1.776953in}}%
\pgfpathlineto{\pgfqpoint{2.879561in}{1.776953in}}%
\pgfpathlineto{\pgfqpoint{2.871261in}{1.776953in}}%
\pgfpathlineto{\pgfqpoint{2.862962in}{1.776953in}}%
\pgfpathlineto{\pgfqpoint{2.854662in}{1.776953in}}%
\pgfpathlineto{\pgfqpoint{2.846362in}{1.776953in}}%
\pgfpathlineto{\pgfqpoint{2.838063in}{1.776953in}}%
\pgfpathlineto{\pgfqpoint{2.829763in}{1.776953in}}%
\pgfpathlineto{\pgfqpoint{2.821463in}{1.776953in}}%
\pgfpathlineto{\pgfqpoint{2.813163in}{1.776953in}}%
\pgfpathlineto{\pgfqpoint{2.804864in}{1.776953in}}%
\pgfpathlineto{\pgfqpoint{2.796564in}{1.776953in}}%
\pgfpathlineto{\pgfqpoint{2.788264in}{1.776953in}}%
\pgfpathlineto{\pgfqpoint{2.779965in}{1.776953in}}%
\pgfpathlineto{\pgfqpoint{2.771665in}{1.776953in}}%
\pgfpathlineto{\pgfqpoint{2.763365in}{1.776953in}}%
\pgfpathlineto{\pgfqpoint{2.755065in}{1.776953in}}%
\pgfpathlineto{\pgfqpoint{2.746766in}{1.776953in}}%
\pgfpathlineto{\pgfqpoint{2.738466in}{1.776953in}}%
\pgfpathlineto{\pgfqpoint{2.730166in}{1.776953in}}%
\pgfpathlineto{\pgfqpoint{2.721867in}{1.776953in}}%
\pgfpathlineto{\pgfqpoint{2.713567in}{1.776953in}}%
\pgfpathlineto{\pgfqpoint{2.705267in}{1.776953in}}%
\pgfpathlineto{\pgfqpoint{2.696967in}{1.776953in}}%
\pgfpathlineto{\pgfqpoint{2.688668in}{1.776953in}}%
\pgfpathlineto{\pgfqpoint{2.680368in}{1.776953in}}%
\pgfpathlineto{\pgfqpoint{2.672068in}{1.776953in}}%
\pgfpathlineto{\pgfqpoint{2.663768in}{1.776953in}}%
\pgfpathlineto{\pgfqpoint{2.655469in}{1.776953in}}%
\pgfpathlineto{\pgfqpoint{2.647169in}{1.776953in}}%
\pgfpathlineto{\pgfqpoint{2.638869in}{1.776953in}}%
\pgfpathlineto{\pgfqpoint{2.630570in}{1.776953in}}%
\pgfpathlineto{\pgfqpoint{2.622270in}{1.776953in}}%
\pgfpathlineto{\pgfqpoint{2.613970in}{1.776953in}}%
\pgfpathlineto{\pgfqpoint{2.605670in}{1.776953in}}%
\pgfpathlineto{\pgfqpoint{2.597371in}{1.776953in}}%
\pgfpathlineto{\pgfqpoint{2.589071in}{1.776953in}}%
\pgfpathlineto{\pgfqpoint{2.580771in}{1.776953in}}%
\pgfpathlineto{\pgfqpoint{2.572472in}{1.776953in}}%
\pgfpathlineto{\pgfqpoint{2.564172in}{1.776953in}}%
\pgfpathlineto{\pgfqpoint{2.555872in}{1.776953in}}%
\pgfpathlineto{\pgfqpoint{2.547572in}{1.776953in}}%
\pgfpathlineto{\pgfqpoint{2.539273in}{1.776953in}}%
\pgfpathlineto{\pgfqpoint{2.530973in}{1.776953in}}%
\pgfpathlineto{\pgfqpoint{2.522673in}{1.776953in}}%
\pgfpathlineto{\pgfqpoint{2.514374in}{1.776953in}}%
\pgfpathlineto{\pgfqpoint{2.506074in}{1.776953in}}%
\pgfpathlineto{\pgfqpoint{2.497774in}{1.776953in}}%
\pgfpathlineto{\pgfqpoint{2.489474in}{1.776953in}}%
\pgfpathlineto{\pgfqpoint{2.481175in}{1.776953in}}%
\pgfpathlineto{\pgfqpoint{2.472875in}{1.776953in}}%
\pgfpathlineto{\pgfqpoint{2.464575in}{1.776953in}}%
\pgfpathlineto{\pgfqpoint{2.456276in}{1.776953in}}%
\pgfpathlineto{\pgfqpoint{2.447976in}{1.776953in}}%
\pgfpathlineto{\pgfqpoint{2.439676in}{1.776953in}}%
\pgfpathlineto{\pgfqpoint{2.431376in}{1.776953in}}%
\pgfpathlineto{\pgfqpoint{2.423077in}{1.776953in}}%
\pgfpathlineto{\pgfqpoint{2.414777in}{1.776953in}}%
\pgfpathlineto{\pgfqpoint{2.406477in}{1.776953in}}%
\pgfpathlineto{\pgfqpoint{2.398177in}{1.776953in}}%
\pgfpathlineto{\pgfqpoint{2.389878in}{1.776953in}}%
\pgfpathlineto{\pgfqpoint{2.381578in}{1.776953in}}%
\pgfpathlineto{\pgfqpoint{2.373278in}{1.776953in}}%
\pgfpathlineto{\pgfqpoint{2.364979in}{1.776953in}}%
\pgfpathlineto{\pgfqpoint{2.356679in}{1.776953in}}%
\pgfpathlineto{\pgfqpoint{2.348379in}{1.776953in}}%
\pgfpathlineto{\pgfqpoint{2.340079in}{1.776953in}}%
\pgfpathlineto{\pgfqpoint{2.331780in}{1.776953in}}%
\pgfpathlineto{\pgfqpoint{2.323480in}{1.776953in}}%
\pgfpathlineto{\pgfqpoint{2.315180in}{1.776953in}}%
\pgfpathlineto{\pgfqpoint{2.306881in}{1.776953in}}%
\pgfpathlineto{\pgfqpoint{2.298581in}{1.776953in}}%
\pgfpathlineto{\pgfqpoint{2.290281in}{1.776953in}}%
\pgfpathlineto{\pgfqpoint{2.281981in}{1.776953in}}%
\pgfpathlineto{\pgfqpoint{2.273682in}{1.776953in}}%
\pgfpathlineto{\pgfqpoint{2.265382in}{1.776953in}}%
\pgfpathlineto{\pgfqpoint{2.257082in}{1.776953in}}%
\pgfpathlineto{\pgfqpoint{2.248783in}{1.776953in}}%
\pgfpathlineto{\pgfqpoint{2.240483in}{1.776953in}}%
\pgfpathlineto{\pgfqpoint{2.232183in}{1.776953in}}%
\pgfpathlineto{\pgfqpoint{2.223883in}{1.776953in}}%
\pgfpathlineto{\pgfqpoint{2.215584in}{1.776953in}}%
\pgfpathlineto{\pgfqpoint{2.207284in}{1.776953in}}%
\pgfpathlineto{\pgfqpoint{2.198984in}{1.776953in}}%
\pgfpathlineto{\pgfqpoint{2.190685in}{1.776953in}}%
\pgfpathlineto{\pgfqpoint{2.182385in}{1.776953in}}%
\pgfpathlineto{\pgfqpoint{2.174085in}{1.776953in}}%
\pgfpathlineto{\pgfqpoint{2.165785in}{1.776953in}}%
\pgfpathlineto{\pgfqpoint{2.157486in}{1.776953in}}%
\pgfpathlineto{\pgfqpoint{2.149186in}{1.776953in}}%
\pgfpathlineto{\pgfqpoint{2.140886in}{1.776953in}}%
\pgfpathlineto{\pgfqpoint{2.132586in}{1.776953in}}%
\pgfpathlineto{\pgfqpoint{2.124287in}{1.776953in}}%
\pgfpathlineto{\pgfqpoint{2.115987in}{1.776953in}}%
\pgfpathlineto{\pgfqpoint{2.107687in}{1.776953in}}%
\pgfpathlineto{\pgfqpoint{2.099388in}{1.776953in}}%
\pgfpathlineto{\pgfqpoint{2.091088in}{1.776953in}}%
\pgfpathlineto{\pgfqpoint{2.082788in}{1.776953in}}%
\pgfpathlineto{\pgfqpoint{2.074488in}{1.776953in}}%
\pgfpathlineto{\pgfqpoint{2.066189in}{1.776953in}}%
\pgfpathlineto{\pgfqpoint{2.057889in}{1.776953in}}%
\pgfpathlineto{\pgfqpoint{2.049589in}{1.776953in}}%
\pgfpathlineto{\pgfqpoint{2.041290in}{1.776953in}}%
\pgfpathlineto{\pgfqpoint{2.032990in}{1.776953in}}%
\pgfpathlineto{\pgfqpoint{2.024690in}{1.776953in}}%
\pgfpathlineto{\pgfqpoint{2.016390in}{1.776953in}}%
\pgfpathlineto{\pgfqpoint{2.008091in}{1.776953in}}%
\pgfpathlineto{\pgfqpoint{1.999791in}{1.776953in}}%
\pgfpathlineto{\pgfqpoint{1.991491in}{1.776953in}}%
\pgfpathlineto{\pgfqpoint{1.983192in}{1.776953in}}%
\pgfpathlineto{\pgfqpoint{1.974892in}{1.776953in}}%
\pgfpathlineto{\pgfqpoint{1.966592in}{1.776953in}}%
\pgfpathlineto{\pgfqpoint{1.958292in}{1.776953in}}%
\pgfpathlineto{\pgfqpoint{1.949993in}{1.776953in}}%
\pgfpathlineto{\pgfqpoint{1.941693in}{1.776953in}}%
\pgfpathlineto{\pgfqpoint{1.933393in}{1.776953in}}%
\pgfpathlineto{\pgfqpoint{1.925093in}{1.776953in}}%
\pgfpathlineto{\pgfqpoint{1.916794in}{1.776953in}}%
\pgfpathlineto{\pgfqpoint{1.908494in}{1.776953in}}%
\pgfpathlineto{\pgfqpoint{1.900194in}{1.776953in}}%
\pgfpathlineto{\pgfqpoint{1.891895in}{1.776953in}}%
\pgfpathlineto{\pgfqpoint{1.883595in}{1.776953in}}%
\pgfpathlineto{\pgfqpoint{1.875295in}{1.776953in}}%
\pgfpathlineto{\pgfqpoint{1.866995in}{1.776953in}}%
\pgfpathlineto{\pgfqpoint{1.858696in}{1.776953in}}%
\pgfpathlineto{\pgfqpoint{1.850396in}{1.776953in}}%
\pgfpathlineto{\pgfqpoint{1.842096in}{1.776953in}}%
\pgfpathlineto{\pgfqpoint{1.833797in}{1.776953in}}%
\pgfpathlineto{\pgfqpoint{1.825497in}{1.776953in}}%
\pgfpathlineto{\pgfqpoint{1.817197in}{1.776953in}}%
\pgfpathlineto{\pgfqpoint{1.808897in}{1.776953in}}%
\pgfpathlineto{\pgfqpoint{1.800598in}{1.776953in}}%
\pgfpathlineto{\pgfqpoint{1.792298in}{1.776953in}}%
\pgfpathlineto{\pgfqpoint{1.783998in}{1.776953in}}%
\pgfpathlineto{\pgfqpoint{1.775699in}{1.776953in}}%
\pgfpathlineto{\pgfqpoint{1.767399in}{1.776953in}}%
\pgfpathlineto{\pgfqpoint{1.759099in}{1.776953in}}%
\pgfpathlineto{\pgfqpoint{1.750799in}{1.776953in}}%
\pgfpathlineto{\pgfqpoint{1.742500in}{1.776953in}}%
\pgfpathlineto{\pgfqpoint{1.734200in}{1.776953in}}%
\pgfpathlineto{\pgfqpoint{1.725900in}{1.776953in}}%
\pgfpathlineto{\pgfqpoint{1.717601in}{1.776953in}}%
\pgfpathlineto{\pgfqpoint{1.709301in}{1.776953in}}%
\pgfpathlineto{\pgfqpoint{1.701001in}{1.776953in}}%
\pgfpathlineto{\pgfqpoint{1.692701in}{1.776953in}}%
\pgfpathlineto{\pgfqpoint{1.684402in}{1.776953in}}%
\pgfpathlineto{\pgfqpoint{1.676102in}{1.776953in}}%
\pgfpathclose%
\pgfusepath{fill}%
\end{pgfscope}%
\begin{pgfscope}%
\pgfpathrectangle{\pgfqpoint{0.306648in}{0.403361in}}{\pgfqpoint{2.738907in}{1.439002in}} %
\pgfusepath{clip}%
\pgfsetbuttcap%
\pgfsetroundjoin%
\definecolor{currentfill}{rgb}{1.000000,0.588235,0.470588}%
\pgfsetfillcolor{currentfill}%
\pgfsetlinewidth{0.000000pt}%
\definecolor{currentstroke}{rgb}{0.000000,0.000000,0.000000}%
\pgfsetstrokecolor{currentstroke}%
\pgfsetdash{}{0pt}%
\pgfpathmoveto{\pgfqpoint{0.431144in}{0.527616in}}%
\pgfpathlineto{\pgfqpoint{0.431144in}{0.527616in}}%
\pgfpathlineto{\pgfqpoint{0.439444in}{0.529480in}}%
\pgfpathlineto{\pgfqpoint{0.447744in}{0.531393in}}%
\pgfpathlineto{\pgfqpoint{0.456043in}{0.533356in}}%
\pgfpathlineto{\pgfqpoint{0.464343in}{0.535369in}}%
\pgfpathlineto{\pgfqpoint{0.472643in}{0.537434in}}%
\pgfpathlineto{\pgfqpoint{0.480942in}{0.539550in}}%
\pgfpathlineto{\pgfqpoint{0.489242in}{0.541719in}}%
\pgfpathlineto{\pgfqpoint{0.497542in}{0.543943in}}%
\pgfpathlineto{\pgfqpoint{0.505842in}{0.546221in}}%
\pgfpathlineto{\pgfqpoint{0.514141in}{0.548554in}}%
\pgfpathlineto{\pgfqpoint{0.522441in}{0.550945in}}%
\pgfpathlineto{\pgfqpoint{0.530741in}{0.553392in}}%
\pgfpathlineto{\pgfqpoint{0.539040in}{0.555898in}}%
\pgfpathlineto{\pgfqpoint{0.547340in}{0.558463in}}%
\pgfpathlineto{\pgfqpoint{0.555640in}{0.561088in}}%
\pgfpathlineto{\pgfqpoint{0.563940in}{0.563774in}}%
\pgfpathlineto{\pgfqpoint{0.572239in}{0.566521in}}%
\pgfpathlineto{\pgfqpoint{0.580539in}{0.569331in}}%
\pgfpathlineto{\pgfqpoint{0.588839in}{0.572204in}}%
\pgfpathlineto{\pgfqpoint{0.597138in}{0.575142in}}%
\pgfpathlineto{\pgfqpoint{0.605438in}{0.578144in}}%
\pgfpathlineto{\pgfqpoint{0.613738in}{0.581212in}}%
\pgfpathlineto{\pgfqpoint{0.622038in}{0.584347in}}%
\pgfpathlineto{\pgfqpoint{0.630337in}{0.587549in}}%
\pgfpathlineto{\pgfqpoint{0.638637in}{0.590819in}}%
\pgfpathlineto{\pgfqpoint{0.646937in}{0.594158in}}%
\pgfpathlineto{\pgfqpoint{0.655237in}{0.597567in}}%
\pgfpathlineto{\pgfqpoint{0.663536in}{0.601045in}}%
\pgfpathlineto{\pgfqpoint{0.671836in}{0.604595in}}%
\pgfpathlineto{\pgfqpoint{0.680136in}{0.608217in}}%
\pgfpathlineto{\pgfqpoint{0.688435in}{0.611910in}}%
\pgfpathlineto{\pgfqpoint{0.696735in}{0.615677in}}%
\pgfpathlineto{\pgfqpoint{0.705035in}{0.619517in}}%
\pgfpathlineto{\pgfqpoint{0.713335in}{0.623431in}}%
\pgfpathlineto{\pgfqpoint{0.721634in}{0.627420in}}%
\pgfpathlineto{\pgfqpoint{0.729934in}{0.631485in}}%
\pgfpathlineto{\pgfqpoint{0.738234in}{0.635625in}}%
\pgfpathlineto{\pgfqpoint{0.746533in}{0.639841in}}%
\pgfpathlineto{\pgfqpoint{0.754833in}{0.644134in}}%
\pgfpathlineto{\pgfqpoint{0.763133in}{0.648505in}}%
\pgfpathlineto{\pgfqpoint{0.771433in}{0.652953in}}%
\pgfpathlineto{\pgfqpoint{0.779732in}{0.657478in}}%
\pgfpathlineto{\pgfqpoint{0.788032in}{0.662083in}}%
\pgfpathlineto{\pgfqpoint{0.796332in}{0.666766in}}%
\pgfpathlineto{\pgfqpoint{0.804631in}{0.671528in}}%
\pgfpathlineto{\pgfqpoint{0.812931in}{0.676369in}}%
\pgfpathlineto{\pgfqpoint{0.821231in}{0.681289in}}%
\pgfpathlineto{\pgfqpoint{0.829531in}{0.686289in}}%
\pgfpathlineto{\pgfqpoint{0.837830in}{0.691369in}}%
\pgfpathlineto{\pgfqpoint{0.846130in}{0.696529in}}%
\pgfpathlineto{\pgfqpoint{0.854430in}{0.701768in}}%
\pgfpathlineto{\pgfqpoint{0.862729in}{0.707088in}}%
\pgfpathlineto{\pgfqpoint{0.871029in}{0.712487in}}%
\pgfpathlineto{\pgfqpoint{0.879329in}{0.717967in}}%
\pgfpathlineto{\pgfqpoint{0.887629in}{0.723526in}}%
\pgfpathlineto{\pgfqpoint{0.895928in}{0.729165in}}%
\pgfpathlineto{\pgfqpoint{0.904228in}{0.734883in}}%
\pgfpathlineto{\pgfqpoint{0.912528in}{0.740681in}}%
\pgfpathlineto{\pgfqpoint{0.920828in}{0.746557in}}%
\pgfpathlineto{\pgfqpoint{0.929127in}{0.752513in}}%
\pgfpathlineto{\pgfqpoint{0.937427in}{0.758547in}}%
\pgfpathlineto{\pgfqpoint{0.945727in}{0.764658in}}%
\pgfpathlineto{\pgfqpoint{0.954026in}{0.770848in}}%
\pgfpathlineto{\pgfqpoint{0.962326in}{0.777114in}}%
\pgfpathlineto{\pgfqpoint{0.970626in}{0.783457in}}%
\pgfpathlineto{\pgfqpoint{0.978926in}{0.789876in}}%
\pgfpathlineto{\pgfqpoint{0.987225in}{0.796370in}}%
\pgfpathlineto{\pgfqpoint{0.995525in}{0.802938in}}%
\pgfpathlineto{\pgfqpoint{1.003825in}{0.809581in}}%
\pgfpathlineto{\pgfqpoint{1.012124in}{0.816297in}}%
\pgfpathlineto{\pgfqpoint{1.020424in}{0.823085in}}%
\pgfpathlineto{\pgfqpoint{1.028724in}{0.829944in}}%
\pgfpathlineto{\pgfqpoint{1.037024in}{0.836874in}}%
\pgfpathlineto{\pgfqpoint{1.045323in}{0.843873in}}%
\pgfpathlineto{\pgfqpoint{1.053623in}{0.850941in}}%
\pgfpathlineto{\pgfqpoint{1.061923in}{0.858077in}}%
\pgfpathlineto{\pgfqpoint{1.070222in}{0.865279in}}%
\pgfpathlineto{\pgfqpoint{1.078522in}{0.872546in}}%
\pgfpathlineto{\pgfqpoint{1.086822in}{0.879877in}}%
\pgfpathlineto{\pgfqpoint{1.095122in}{0.887270in}}%
\pgfpathlineto{\pgfqpoint{1.103421in}{0.894725in}}%
\pgfpathlineto{\pgfqpoint{1.111721in}{0.902240in}}%
\pgfpathlineto{\pgfqpoint{1.120021in}{0.909814in}}%
\pgfpathlineto{\pgfqpoint{1.128320in}{0.917445in}}%
\pgfpathlineto{\pgfqpoint{1.136620in}{0.925132in}}%
\pgfpathlineto{\pgfqpoint{1.144920in}{0.932873in}}%
\pgfpathlineto{\pgfqpoint{1.153220in}{0.940666in}}%
\pgfpathlineto{\pgfqpoint{1.161519in}{0.948511in}}%
\pgfpathlineto{\pgfqpoint{1.169819in}{0.956405in}}%
\pgfpathlineto{\pgfqpoint{1.178119in}{0.964347in}}%
\pgfpathlineto{\pgfqpoint{1.186419in}{0.972334in}}%
\pgfpathlineto{\pgfqpoint{1.194718in}{0.980366in}}%
\pgfpathlineto{\pgfqpoint{1.203018in}{0.988440in}}%
\pgfpathlineto{\pgfqpoint{1.211318in}{0.996554in}}%
\pgfpathlineto{\pgfqpoint{1.219617in}{1.004707in}}%
\pgfpathlineto{\pgfqpoint{1.227917in}{1.012897in}}%
\pgfpathlineto{\pgfqpoint{1.236217in}{1.021122in}}%
\pgfpathlineto{\pgfqpoint{1.244517in}{1.029379in}}%
\pgfpathlineto{\pgfqpoint{1.252816in}{1.037667in}}%
\pgfpathlineto{\pgfqpoint{1.261116in}{1.045983in}}%
\pgfpathlineto{\pgfqpoint{1.269416in}{1.054326in}}%
\pgfpathlineto{\pgfqpoint{1.277715in}{1.062694in}}%
\pgfpathlineto{\pgfqpoint{1.286015in}{1.071084in}}%
\pgfpathlineto{\pgfqpoint{1.294315in}{1.079494in}}%
\pgfpathlineto{\pgfqpoint{1.302615in}{1.087922in}}%
\pgfpathlineto{\pgfqpoint{1.310914in}{1.096366in}}%
\pgfpathlineto{\pgfqpoint{1.319214in}{1.104824in}}%
\pgfpathlineto{\pgfqpoint{1.327514in}{1.113293in}}%
\pgfpathlineto{\pgfqpoint{1.335813in}{1.121771in}}%
\pgfpathlineto{\pgfqpoint{1.344113in}{1.130256in}}%
\pgfpathlineto{\pgfqpoint{1.352413in}{1.138746in}}%
\pgfpathlineto{\pgfqpoint{1.360713in}{1.147238in}}%
\pgfpathlineto{\pgfqpoint{1.369012in}{1.155730in}}%
\pgfpathlineto{\pgfqpoint{1.377312in}{1.164220in}}%
\pgfpathlineto{\pgfqpoint{1.385612in}{1.172706in}}%
\pgfpathlineto{\pgfqpoint{1.393911in}{1.181185in}}%
\pgfpathlineto{\pgfqpoint{1.402211in}{1.189655in}}%
\pgfpathlineto{\pgfqpoint{1.410511in}{1.198113in}}%
\pgfpathlineto{\pgfqpoint{1.418811in}{1.206558in}}%
\pgfpathlineto{\pgfqpoint{1.427110in}{1.214987in}}%
\pgfpathlineto{\pgfqpoint{1.435410in}{1.223398in}}%
\pgfpathlineto{\pgfqpoint{1.443710in}{1.231788in}}%
\pgfpathlineto{\pgfqpoint{1.452010in}{1.240156in}}%
\pgfpathlineto{\pgfqpoint{1.460309in}{1.248499in}}%
\pgfpathlineto{\pgfqpoint{1.468609in}{1.256814in}}%
\pgfpathlineto{\pgfqpoint{1.476909in}{1.265100in}}%
\pgfpathlineto{\pgfqpoint{1.485208in}{1.273355in}}%
\pgfpathlineto{\pgfqpoint{1.493508in}{1.281576in}}%
\pgfpathlineto{\pgfqpoint{1.501808in}{1.289761in}}%
\pgfpathlineto{\pgfqpoint{1.510108in}{1.297909in}}%
\pgfpathlineto{\pgfqpoint{1.518407in}{1.306016in}}%
\pgfpathlineto{\pgfqpoint{1.526707in}{1.314081in}}%
\pgfpathlineto{\pgfqpoint{1.535007in}{1.322102in}}%
\pgfpathlineto{\pgfqpoint{1.543306in}{1.330077in}}%
\pgfpathlineto{\pgfqpoint{1.551606in}{1.338003in}}%
\pgfpathlineto{\pgfqpoint{1.559906in}{1.345880in}}%
\pgfpathlineto{\pgfqpoint{1.568206in}{1.353704in}}%
\pgfpathlineto{\pgfqpoint{1.576505in}{1.361475in}}%
\pgfpathlineto{\pgfqpoint{1.584805in}{1.369191in}}%
\pgfpathlineto{\pgfqpoint{1.593105in}{1.376848in}}%
\pgfpathlineto{\pgfqpoint{1.601404in}{1.384447in}}%
\pgfpathlineto{\pgfqpoint{1.609704in}{1.391985in}}%
\pgfpathlineto{\pgfqpoint{1.618004in}{1.399460in}}%
\pgfpathlineto{\pgfqpoint{1.626304in}{1.406871in}}%
\pgfpathlineto{\pgfqpoint{1.634603in}{1.414217in}}%
\pgfpathlineto{\pgfqpoint{1.642903in}{1.421495in}}%
\pgfpathlineto{\pgfqpoint{1.651203in}{1.428705in}}%
\pgfpathlineto{\pgfqpoint{1.659502in}{1.435844in}}%
\pgfpathlineto{\pgfqpoint{1.667802in}{1.442912in}}%
\pgfpathlineto{\pgfqpoint{1.676102in}{1.449908in}}%
\pgfpathlineto{\pgfqpoint{1.676102in}{1.449908in}}%
\pgfpathlineto{\pgfqpoint{1.676102in}{0.468770in}}%
\pgfpathlineto{\pgfqpoint{1.667802in}{0.468770in}}%
\pgfpathlineto{\pgfqpoint{1.659502in}{0.468770in}}%
\pgfpathlineto{\pgfqpoint{1.651203in}{0.468770in}}%
\pgfpathlineto{\pgfqpoint{1.642903in}{0.468770in}}%
\pgfpathlineto{\pgfqpoint{1.634603in}{0.468770in}}%
\pgfpathlineto{\pgfqpoint{1.626304in}{0.468770in}}%
\pgfpathlineto{\pgfqpoint{1.618004in}{0.468770in}}%
\pgfpathlineto{\pgfqpoint{1.609704in}{0.468770in}}%
\pgfpathlineto{\pgfqpoint{1.601404in}{0.468770in}}%
\pgfpathlineto{\pgfqpoint{1.593105in}{0.468770in}}%
\pgfpathlineto{\pgfqpoint{1.584805in}{0.468770in}}%
\pgfpathlineto{\pgfqpoint{1.576505in}{0.468770in}}%
\pgfpathlineto{\pgfqpoint{1.568206in}{0.468770in}}%
\pgfpathlineto{\pgfqpoint{1.559906in}{0.468770in}}%
\pgfpathlineto{\pgfqpoint{1.551606in}{0.468770in}}%
\pgfpathlineto{\pgfqpoint{1.543306in}{0.468770in}}%
\pgfpathlineto{\pgfqpoint{1.535007in}{0.468770in}}%
\pgfpathlineto{\pgfqpoint{1.526707in}{0.468770in}}%
\pgfpathlineto{\pgfqpoint{1.518407in}{0.468770in}}%
\pgfpathlineto{\pgfqpoint{1.510108in}{0.468770in}}%
\pgfpathlineto{\pgfqpoint{1.501808in}{0.468770in}}%
\pgfpathlineto{\pgfqpoint{1.493508in}{0.468770in}}%
\pgfpathlineto{\pgfqpoint{1.485208in}{0.468770in}}%
\pgfpathlineto{\pgfqpoint{1.476909in}{0.468770in}}%
\pgfpathlineto{\pgfqpoint{1.468609in}{0.468770in}}%
\pgfpathlineto{\pgfqpoint{1.460309in}{0.468770in}}%
\pgfpathlineto{\pgfqpoint{1.452010in}{0.468770in}}%
\pgfpathlineto{\pgfqpoint{1.443710in}{0.468770in}}%
\pgfpathlineto{\pgfqpoint{1.435410in}{0.468770in}}%
\pgfpathlineto{\pgfqpoint{1.427110in}{0.468770in}}%
\pgfpathlineto{\pgfqpoint{1.418811in}{0.468770in}}%
\pgfpathlineto{\pgfqpoint{1.410511in}{0.468770in}}%
\pgfpathlineto{\pgfqpoint{1.402211in}{0.468770in}}%
\pgfpathlineto{\pgfqpoint{1.393911in}{0.468770in}}%
\pgfpathlineto{\pgfqpoint{1.385612in}{0.468770in}}%
\pgfpathlineto{\pgfqpoint{1.377312in}{0.468770in}}%
\pgfpathlineto{\pgfqpoint{1.369012in}{0.468770in}}%
\pgfpathlineto{\pgfqpoint{1.360713in}{0.468770in}}%
\pgfpathlineto{\pgfqpoint{1.352413in}{0.468770in}}%
\pgfpathlineto{\pgfqpoint{1.344113in}{0.468770in}}%
\pgfpathlineto{\pgfqpoint{1.335813in}{0.468770in}}%
\pgfpathlineto{\pgfqpoint{1.327514in}{0.468770in}}%
\pgfpathlineto{\pgfqpoint{1.319214in}{0.468770in}}%
\pgfpathlineto{\pgfqpoint{1.310914in}{0.468770in}}%
\pgfpathlineto{\pgfqpoint{1.302615in}{0.468770in}}%
\pgfpathlineto{\pgfqpoint{1.294315in}{0.468770in}}%
\pgfpathlineto{\pgfqpoint{1.286015in}{0.468770in}}%
\pgfpathlineto{\pgfqpoint{1.277715in}{0.468770in}}%
\pgfpathlineto{\pgfqpoint{1.269416in}{0.468770in}}%
\pgfpathlineto{\pgfqpoint{1.261116in}{0.468770in}}%
\pgfpathlineto{\pgfqpoint{1.252816in}{0.468770in}}%
\pgfpathlineto{\pgfqpoint{1.244517in}{0.468770in}}%
\pgfpathlineto{\pgfqpoint{1.236217in}{0.468770in}}%
\pgfpathlineto{\pgfqpoint{1.227917in}{0.468770in}}%
\pgfpathlineto{\pgfqpoint{1.219617in}{0.468770in}}%
\pgfpathlineto{\pgfqpoint{1.211318in}{0.468770in}}%
\pgfpathlineto{\pgfqpoint{1.203018in}{0.468770in}}%
\pgfpathlineto{\pgfqpoint{1.194718in}{0.468770in}}%
\pgfpathlineto{\pgfqpoint{1.186419in}{0.468770in}}%
\pgfpathlineto{\pgfqpoint{1.178119in}{0.468770in}}%
\pgfpathlineto{\pgfqpoint{1.169819in}{0.468770in}}%
\pgfpathlineto{\pgfqpoint{1.161519in}{0.468770in}}%
\pgfpathlineto{\pgfqpoint{1.153220in}{0.468770in}}%
\pgfpathlineto{\pgfqpoint{1.144920in}{0.468770in}}%
\pgfpathlineto{\pgfqpoint{1.136620in}{0.468770in}}%
\pgfpathlineto{\pgfqpoint{1.128320in}{0.468770in}}%
\pgfpathlineto{\pgfqpoint{1.120021in}{0.468770in}}%
\pgfpathlineto{\pgfqpoint{1.111721in}{0.468770in}}%
\pgfpathlineto{\pgfqpoint{1.103421in}{0.468770in}}%
\pgfpathlineto{\pgfqpoint{1.095122in}{0.468770in}}%
\pgfpathlineto{\pgfqpoint{1.086822in}{0.468770in}}%
\pgfpathlineto{\pgfqpoint{1.078522in}{0.468770in}}%
\pgfpathlineto{\pgfqpoint{1.070222in}{0.468770in}}%
\pgfpathlineto{\pgfqpoint{1.061923in}{0.468770in}}%
\pgfpathlineto{\pgfqpoint{1.053623in}{0.468770in}}%
\pgfpathlineto{\pgfqpoint{1.045323in}{0.468770in}}%
\pgfpathlineto{\pgfqpoint{1.037024in}{0.468770in}}%
\pgfpathlineto{\pgfqpoint{1.028724in}{0.468770in}}%
\pgfpathlineto{\pgfqpoint{1.020424in}{0.468770in}}%
\pgfpathlineto{\pgfqpoint{1.012124in}{0.468770in}}%
\pgfpathlineto{\pgfqpoint{1.003825in}{0.468770in}}%
\pgfpathlineto{\pgfqpoint{0.995525in}{0.468770in}}%
\pgfpathlineto{\pgfqpoint{0.987225in}{0.468770in}}%
\pgfpathlineto{\pgfqpoint{0.978926in}{0.468770in}}%
\pgfpathlineto{\pgfqpoint{0.970626in}{0.468770in}}%
\pgfpathlineto{\pgfqpoint{0.962326in}{0.468770in}}%
\pgfpathlineto{\pgfqpoint{0.954026in}{0.468770in}}%
\pgfpathlineto{\pgfqpoint{0.945727in}{0.468770in}}%
\pgfpathlineto{\pgfqpoint{0.937427in}{0.468770in}}%
\pgfpathlineto{\pgfqpoint{0.929127in}{0.468770in}}%
\pgfpathlineto{\pgfqpoint{0.920828in}{0.468770in}}%
\pgfpathlineto{\pgfqpoint{0.912528in}{0.468770in}}%
\pgfpathlineto{\pgfqpoint{0.904228in}{0.468770in}}%
\pgfpathlineto{\pgfqpoint{0.895928in}{0.468770in}}%
\pgfpathlineto{\pgfqpoint{0.887629in}{0.468770in}}%
\pgfpathlineto{\pgfqpoint{0.879329in}{0.468770in}}%
\pgfpathlineto{\pgfqpoint{0.871029in}{0.468770in}}%
\pgfpathlineto{\pgfqpoint{0.862729in}{0.468770in}}%
\pgfpathlineto{\pgfqpoint{0.854430in}{0.468770in}}%
\pgfpathlineto{\pgfqpoint{0.846130in}{0.468770in}}%
\pgfpathlineto{\pgfqpoint{0.837830in}{0.468770in}}%
\pgfpathlineto{\pgfqpoint{0.829531in}{0.468770in}}%
\pgfpathlineto{\pgfqpoint{0.821231in}{0.468770in}}%
\pgfpathlineto{\pgfqpoint{0.812931in}{0.468770in}}%
\pgfpathlineto{\pgfqpoint{0.804631in}{0.468770in}}%
\pgfpathlineto{\pgfqpoint{0.796332in}{0.468770in}}%
\pgfpathlineto{\pgfqpoint{0.788032in}{0.468770in}}%
\pgfpathlineto{\pgfqpoint{0.779732in}{0.468770in}}%
\pgfpathlineto{\pgfqpoint{0.771433in}{0.468770in}}%
\pgfpathlineto{\pgfqpoint{0.763133in}{0.468770in}}%
\pgfpathlineto{\pgfqpoint{0.754833in}{0.468770in}}%
\pgfpathlineto{\pgfqpoint{0.746533in}{0.468770in}}%
\pgfpathlineto{\pgfqpoint{0.738234in}{0.468770in}}%
\pgfpathlineto{\pgfqpoint{0.729934in}{0.468770in}}%
\pgfpathlineto{\pgfqpoint{0.721634in}{0.468770in}}%
\pgfpathlineto{\pgfqpoint{0.713335in}{0.468770in}}%
\pgfpathlineto{\pgfqpoint{0.705035in}{0.468770in}}%
\pgfpathlineto{\pgfqpoint{0.696735in}{0.468770in}}%
\pgfpathlineto{\pgfqpoint{0.688435in}{0.468770in}}%
\pgfpathlineto{\pgfqpoint{0.680136in}{0.468770in}}%
\pgfpathlineto{\pgfqpoint{0.671836in}{0.468770in}}%
\pgfpathlineto{\pgfqpoint{0.663536in}{0.468770in}}%
\pgfpathlineto{\pgfqpoint{0.655237in}{0.468770in}}%
\pgfpathlineto{\pgfqpoint{0.646937in}{0.468770in}}%
\pgfpathlineto{\pgfqpoint{0.638637in}{0.468770in}}%
\pgfpathlineto{\pgfqpoint{0.630337in}{0.468770in}}%
\pgfpathlineto{\pgfqpoint{0.622038in}{0.468770in}}%
\pgfpathlineto{\pgfqpoint{0.613738in}{0.468770in}}%
\pgfpathlineto{\pgfqpoint{0.605438in}{0.468770in}}%
\pgfpathlineto{\pgfqpoint{0.597138in}{0.468770in}}%
\pgfpathlineto{\pgfqpoint{0.588839in}{0.468770in}}%
\pgfpathlineto{\pgfqpoint{0.580539in}{0.468770in}}%
\pgfpathlineto{\pgfqpoint{0.572239in}{0.468770in}}%
\pgfpathlineto{\pgfqpoint{0.563940in}{0.468770in}}%
\pgfpathlineto{\pgfqpoint{0.555640in}{0.468770in}}%
\pgfpathlineto{\pgfqpoint{0.547340in}{0.468770in}}%
\pgfpathlineto{\pgfqpoint{0.539040in}{0.468770in}}%
\pgfpathlineto{\pgfqpoint{0.530741in}{0.468770in}}%
\pgfpathlineto{\pgfqpoint{0.522441in}{0.468770in}}%
\pgfpathlineto{\pgfqpoint{0.514141in}{0.468770in}}%
\pgfpathlineto{\pgfqpoint{0.505842in}{0.468770in}}%
\pgfpathlineto{\pgfqpoint{0.497542in}{0.468770in}}%
\pgfpathlineto{\pgfqpoint{0.489242in}{0.468770in}}%
\pgfpathlineto{\pgfqpoint{0.480942in}{0.468770in}}%
\pgfpathlineto{\pgfqpoint{0.472643in}{0.468770in}}%
\pgfpathlineto{\pgfqpoint{0.464343in}{0.468770in}}%
\pgfpathlineto{\pgfqpoint{0.456043in}{0.468770in}}%
\pgfpathlineto{\pgfqpoint{0.447744in}{0.468770in}}%
\pgfpathlineto{\pgfqpoint{0.439444in}{0.468770in}}%
\pgfpathlineto{\pgfqpoint{0.431144in}{0.468770in}}%
\pgfpathclose%
\pgfusepath{fill}%
\end{pgfscope}%
\begin{pgfscope}%
\pgfsetbuttcap%
\pgfsetroundjoin%
\definecolor{currentfill}{rgb}{0.000000,0.000000,0.000000}%
\pgfsetfillcolor{currentfill}%
\pgfsetlinewidth{0.803000pt}%
\definecolor{currentstroke}{rgb}{0.000000,0.000000,0.000000}%
\pgfsetstrokecolor{currentstroke}%
\pgfsetdash{}{0pt}%
\pgfsys@defobject{currentmarker}{\pgfqpoint{0.000000in}{-0.048611in}}{\pgfqpoint{0.000000in}{0.000000in}}{%
\pgfpathmoveto{\pgfqpoint{0.000000in}{0.000000in}}%
\pgfpathlineto{\pgfqpoint{0.000000in}{-0.048611in}}%
\pgfusepath{stroke,fill}%
}%
\begin{pgfscope}%
\pgfsys@transformshift{1.676102in}{0.468770in}%
\pgfsys@useobject{currentmarker}{}%
\end{pgfscope}%
\end{pgfscope}%
\begin{pgfscope}%
\pgftext[x=1.676102in,y=0.371548in,,top]{\sffamily\fontsize{10.000000}{12.000000}\selectfont \(\displaystyle 0\)}%
\end{pgfscope}%
\begin{pgfscope}%
\pgftext[x=1.676102in,y=0.192659in,,top]{\sffamily\fontsize{10.000000}{12.000000}\selectfont \(\displaystyle z\)}%
\end{pgfscope}%
\begin{pgfscope}%
\pgfsetbuttcap%
\pgfsetroundjoin%
\definecolor{currentfill}{rgb}{0.000000,0.000000,0.000000}%
\pgfsetfillcolor{currentfill}%
\pgfsetlinewidth{0.803000pt}%
\definecolor{currentstroke}{rgb}{0.000000,0.000000,0.000000}%
\pgfsetstrokecolor{currentstroke}%
\pgfsetdash{}{0pt}%
\pgfsys@defobject{currentmarker}{\pgfqpoint{-0.048611in}{0.000000in}}{\pgfqpoint{0.000000in}{0.000000in}}{%
\pgfpathmoveto{\pgfqpoint{0.000000in}{0.000000in}}%
\pgfpathlineto{\pgfqpoint{-0.048611in}{0.000000in}}%
\pgfusepath{stroke,fill}%
}%
\begin{pgfscope}%
\pgfsys@transformshift{0.431144in}{0.468770in}%
\pgfsys@useobject{currentmarker}{}%
\end{pgfscope}%
\end{pgfscope}%
\begin{pgfscope}%
\pgftext[x=0.264477in,y=0.420576in,left,base]{\sffamily\fontsize{10.000000}{12.000000}\selectfont \(\displaystyle 0\)}%
\end{pgfscope}%
\begin{pgfscope}%
\pgfsetbuttcap%
\pgfsetroundjoin%
\definecolor{currentfill}{rgb}{0.000000,0.000000,0.000000}%
\pgfsetfillcolor{currentfill}%
\pgfsetlinewidth{0.803000pt}%
\definecolor{currentstroke}{rgb}{0.000000,0.000000,0.000000}%
\pgfsetstrokecolor{currentstroke}%
\pgfsetdash{}{0pt}%
\pgfsys@defobject{currentmarker}{\pgfqpoint{-0.048611in}{0.000000in}}{\pgfqpoint{0.000000in}{0.000000in}}{%
\pgfpathmoveto{\pgfqpoint{0.000000in}{0.000000in}}%
\pgfpathlineto{\pgfqpoint{-0.048611in}{0.000000in}}%
\pgfusepath{stroke,fill}%
}%
\begin{pgfscope}%
\pgfsys@transformshift{0.431144in}{1.776953in}%
\pgfsys@useobject{currentmarker}{}%
\end{pgfscope}%
\end{pgfscope}%
\begin{pgfscope}%
\pgftext[x=0.264477in,y=1.728759in,left,base]{\sffamily\fontsize{10.000000}{12.000000}\selectfont \(\displaystyle 1\)}%
\end{pgfscope}%
\begin{pgfscope}%
\pgftext[x=0.208922in,y=1.122862in,,bottom,rotate=90.000000]{\sffamily\fontsize{10.000000}{12.000000}\selectfont \(\displaystyle 1 - y(1 -F(z))\)}%
\end{pgfscope}%
\begin{pgfscope}%
\pgfpathrectangle{\pgfqpoint{0.306648in}{0.403361in}}{\pgfqpoint{2.738907in}{1.439002in}} %
\pgfusepath{clip}%
\pgfsetrectcap%
\pgfsetroundjoin%
\pgfsetlinewidth{1.204500pt}%
\definecolor{currentstroke}{rgb}{0.000000,0.000000,0.000000}%
\pgfsetstrokecolor{currentstroke}%
\pgfsetdash{}{0pt}%
\pgfpathmoveto{\pgfqpoint{0.431144in}{0.527616in}}%
\pgfpathlineto{\pgfqpoint{0.489242in}{0.541719in}}%
\pgfpathlineto{\pgfqpoint{0.539040in}{0.555898in}}%
\pgfpathlineto{\pgfqpoint{0.588839in}{0.572204in}}%
\pgfpathlineto{\pgfqpoint{0.638637in}{0.590819in}}%
\pgfpathlineto{\pgfqpoint{0.688435in}{0.611910in}}%
\pgfpathlineto{\pgfqpoint{0.738234in}{0.635625in}}%
\pgfpathlineto{\pgfqpoint{0.788032in}{0.662083in}}%
\pgfpathlineto{\pgfqpoint{0.837830in}{0.691369in}}%
\pgfpathlineto{\pgfqpoint{0.887629in}{0.723526in}}%
\pgfpathlineto{\pgfqpoint{0.937427in}{0.758547in}}%
\pgfpathlineto{\pgfqpoint{0.987225in}{0.796370in}}%
\pgfpathlineto{\pgfqpoint{1.037024in}{0.836874in}}%
\pgfpathlineto{\pgfqpoint{1.095122in}{0.887270in}}%
\pgfpathlineto{\pgfqpoint{1.153220in}{0.940666in}}%
\pgfpathlineto{\pgfqpoint{1.219617in}{1.004707in}}%
\pgfpathlineto{\pgfqpoint{1.302615in}{1.087922in}}%
\pgfpathlineto{\pgfqpoint{1.518407in}{1.306016in}}%
\pgfpathlineto{\pgfqpoint{1.584805in}{1.369191in}}%
\pgfpathlineto{\pgfqpoint{1.642903in}{1.421495in}}%
\pgfpathlineto{\pgfqpoint{1.692701in}{1.463675in}}%
\pgfpathlineto{\pgfqpoint{1.742500in}{1.503110in}}%
\pgfpathlineto{\pgfqpoint{1.792298in}{1.539593in}}%
\pgfpathlineto{\pgfqpoint{1.842096in}{1.572986in}}%
\pgfpathlineto{\pgfqpoint{1.883595in}{1.598404in}}%
\pgfpathlineto{\pgfqpoint{1.925093in}{1.621627in}}%
\pgfpathlineto{\pgfqpoint{1.966592in}{1.642686in}}%
\pgfpathlineto{\pgfqpoint{2.008091in}{1.661636in}}%
\pgfpathlineto{\pgfqpoint{2.057889in}{1.681708in}}%
\pgfpathlineto{\pgfqpoint{2.107687in}{1.699037in}}%
\pgfpathlineto{\pgfqpoint{2.157486in}{1.713831in}}%
\pgfpathlineto{\pgfqpoint{2.207284in}{1.726319in}}%
\pgfpathlineto{\pgfqpoint{2.257082in}{1.736742in}}%
\pgfpathlineto{\pgfqpoint{2.315180in}{1.746614in}}%
\pgfpathlineto{\pgfqpoint{2.381578in}{1.755340in}}%
\pgfpathlineto{\pgfqpoint{2.447976in}{1.761839in}}%
\pgfpathlineto{\pgfqpoint{2.530973in}{1.767541in}}%
\pgfpathlineto{\pgfqpoint{2.622270in}{1.771552in}}%
\pgfpathlineto{\pgfqpoint{2.738466in}{1.774427in}}%
\pgfpathlineto{\pgfqpoint{2.904460in}{1.776185in}}%
\pgfpathlineto{\pgfqpoint{2.921060in}{1.776276in}}%
\pgfpathlineto{\pgfqpoint{2.921060in}{1.776276in}}%
\pgfusepath{stroke}%
\end{pgfscope}%
\begin{pgfscope}%
\pgfpathrectangle{\pgfqpoint{0.306648in}{0.403361in}}{\pgfqpoint{2.738907in}{1.439002in}} %
\pgfusepath{clip}%
\pgfsetbuttcap%
\pgfsetroundjoin%
\pgfsetlinewidth{1.204500pt}%
\definecolor{currentstroke}{rgb}{0.000000,0.000000,0.000000}%
\pgfsetstrokecolor{currentstroke}%
\pgfsetdash{{4.440000pt}{1.920000pt}}{0.000000pt}%
\pgfpathmoveto{\pgfqpoint{0.431144in}{0.498531in}}%
\pgfpathlineto{\pgfqpoint{0.505842in}{0.508086in}}%
\pgfpathlineto{\pgfqpoint{0.572239in}{0.518594in}}%
\pgfpathlineto{\pgfqpoint{0.638637in}{0.531289in}}%
\pgfpathlineto{\pgfqpoint{0.705035in}{0.546450in}}%
\pgfpathlineto{\pgfqpoint{0.763133in}{0.561956in}}%
\pgfpathlineto{\pgfqpoint{0.821231in}{0.579736in}}%
\pgfpathlineto{\pgfqpoint{0.879329in}{0.599945in}}%
\pgfpathlineto{\pgfqpoint{0.937427in}{0.622716in}}%
\pgfpathlineto{\pgfqpoint{0.995525in}{0.648153in}}%
\pgfpathlineto{\pgfqpoint{1.053623in}{0.676320in}}%
\pgfpathlineto{\pgfqpoint{1.111721in}{0.707241in}}%
\pgfpathlineto{\pgfqpoint{1.169819in}{0.740890in}}%
\pgfpathlineto{\pgfqpoint{1.227917in}{0.777191in}}%
\pgfpathlineto{\pgfqpoint{1.286015in}{0.816013in}}%
\pgfpathlineto{\pgfqpoint{1.352413in}{0.863229in}}%
\pgfpathlineto{\pgfqpoint{1.418811in}{0.913135in}}%
\pgfpathlineto{\pgfqpoint{1.493508in}{0.971941in}}%
\pgfpathlineto{\pgfqpoint{1.593105in}{1.053482in}}%
\pgfpathlineto{\pgfqpoint{1.883595in}{1.293656in}}%
\pgfpathlineto{\pgfqpoint{1.958292in}{1.351592in}}%
\pgfpathlineto{\pgfqpoint{2.024690in}{1.400537in}}%
\pgfpathlineto{\pgfqpoint{2.091088in}{1.446647in}}%
\pgfpathlineto{\pgfqpoint{2.149186in}{1.484408in}}%
\pgfpathlineto{\pgfqpoint{2.207284in}{1.519585in}}%
\pgfpathlineto{\pgfqpoint{2.265382in}{1.552071in}}%
\pgfpathlineto{\pgfqpoint{2.323480in}{1.581812in}}%
\pgfpathlineto{\pgfqpoint{2.381578in}{1.608803in}}%
\pgfpathlineto{\pgfqpoint{2.439676in}{1.633087in}}%
\pgfpathlineto{\pgfqpoint{2.497774in}{1.654746in}}%
\pgfpathlineto{\pgfqpoint{2.555872in}{1.673896in}}%
\pgfpathlineto{\pgfqpoint{2.613970in}{1.690681in}}%
\pgfpathlineto{\pgfqpoint{2.672068in}{1.705266in}}%
\pgfpathlineto{\pgfqpoint{2.738466in}{1.719469in}}%
\pgfpathlineto{\pgfqpoint{2.804864in}{1.731311in}}%
\pgfpathlineto{\pgfqpoint{2.879561in}{1.742159in}}%
\pgfpathlineto{\pgfqpoint{2.921060in}{1.747192in}}%
\pgfpathlineto{\pgfqpoint{2.921060in}{1.747192in}}%
\pgfusepath{stroke}%
\end{pgfscope}%
\begin{pgfscope}%
\pgfsetrectcap%
\pgfsetmiterjoin%
\pgfsetlinewidth{0.803000pt}%
\definecolor{currentstroke}{rgb}{0.000000,0.000000,0.000000}%
\pgfsetstrokecolor{currentstroke}%
\pgfsetdash{}{0pt}%
\pgfpathmoveto{\pgfqpoint{0.431144in}{0.468770in}}%
\pgfpathlineto{\pgfqpoint{0.431144in}{1.776953in}}%
\pgfusepath{stroke}%
\end{pgfscope}%
\begin{pgfscope}%
\pgfsetrectcap%
\pgfsetmiterjoin%
\pgfsetlinewidth{0.803000pt}%
\definecolor{currentstroke}{rgb}{0.000000,0.000000,0.000000}%
\pgfsetstrokecolor{currentstroke}%
\pgfsetdash{}{0pt}%
\pgfpathmoveto{\pgfqpoint{2.921060in}{0.468770in}}%
\pgfpathlineto{\pgfqpoint{2.921060in}{1.776953in}}%
\pgfusepath{stroke}%
\end{pgfscope}%
\begin{pgfscope}%
\pgfsetrectcap%
\pgfsetmiterjoin%
\pgfsetlinewidth{0.803000pt}%
\definecolor{currentstroke}{rgb}{0.000000,0.000000,0.000000}%
\pgfsetstrokecolor{currentstroke}%
\pgfsetdash{}{0pt}%
\pgfpathmoveto{\pgfqpoint{0.431144in}{0.468770in}}%
\pgfpathlineto{\pgfqpoint{2.921060in}{0.468770in}}%
\pgfusepath{stroke}%
\end{pgfscope}%
\begin{pgfscope}%
\pgfsetrectcap%
\pgfsetmiterjoin%
\pgfsetlinewidth{0.803000pt}%
\definecolor{currentstroke}{rgb}{0.000000,0.000000,0.000000}%
\pgfsetstrokecolor{currentstroke}%
\pgfsetdash{}{0pt}%
\pgfpathmoveto{\pgfqpoint{0.431144in}{1.776953in}}%
\pgfpathlineto{\pgfqpoint{2.921060in}{1.776953in}}%
\pgfusepath{stroke}%
\end{pgfscope}%
\end{pgfpicture}%
\makeatother%
\endgroup%

%% file: figures/beta_bounded.pgf
%% Creator: Matplotlib, PGF backend
%%
%% To include the figure in your LaTeX document, write
%%   \input{<filename>.pgf}
%%
%% Make sure the required packages are loaded in your preamble
%%   \usepackage{pgf}
%%
%% Figures using additional raster images can only be included by \input if
%% they are in the same directory as the main LaTeX file. For loading figures
%% from other directories you can use the `import` package
%%   \usepackage{import}
%% and then include the figures with
%%   \import{<path to file>}{<filename>.pgf}
%%
%% Matplotlib used the following preamble
%%   \usepackage{fontspec}
%%
\begingroup%
\makeatletter%
\begin{pgfpicture}%
\pgfpathrectangle{\pgfpointorigin}{\pgfqpoint{2.310000in}{2.310000in}}%
\pgfusepath{use as bounding box, clip}%
\begin{pgfscope}%
\pgfsetbuttcap%
\pgfsetmiterjoin%
\pgfsetlinewidth{0.000000pt}%
\definecolor{currentstroke}{rgb}{1.000000,1.000000,1.000000}%
\pgfsetstrokecolor{currentstroke}%
\pgfsetstrokeopacity{0.000000}%
\pgfsetdash{}{0pt}%
\pgfpathmoveto{\pgfqpoint{0.000000in}{0.000000in}}%
\pgfpathlineto{\pgfqpoint{2.310000in}{0.000000in}}%
\pgfpathlineto{\pgfqpoint{2.310000in}{2.310000in}}%
\pgfpathlineto{\pgfqpoint{0.000000in}{2.310000in}}%
\pgfpathclose%
\pgfusepath{}%
\end{pgfscope}%
\begin{pgfscope}%
\pgfsetbuttcap%
\pgfsetmiterjoin%
\definecolor{currentfill}{rgb}{1.000000,1.000000,1.000000}%
\pgfsetfillcolor{currentfill}%
\pgfsetlinewidth{0.000000pt}%
\definecolor{currentstroke}{rgb}{0.000000,0.000000,0.000000}%
\pgfsetstrokecolor{currentstroke}%
\pgfsetstrokeopacity{0.000000}%
\pgfsetdash{}{0pt}%
\pgfpathmoveto{\pgfqpoint{0.346175in}{0.386274in}}%
\pgfpathlineto{\pgfqpoint{2.205556in}{0.386274in}}%
\pgfpathlineto{\pgfqpoint{2.205556in}{2.205556in}}%
\pgfpathlineto{\pgfqpoint{0.346175in}{2.205556in}}%
\pgfpathclose%
\pgfusepath{fill}%
\end{pgfscope}%
\begin{pgfscope}%
\pgfpathrectangle{\pgfqpoint{0.346175in}{0.386274in}}{\pgfqpoint{1.859380in}{1.819281in}} %
\pgfusepath{clip}%
\pgfsetbuttcap%
\pgfsetroundjoin%
\definecolor{currentfill}{rgb}{0.376471,0.482353,0.545098}%
\pgfsetfillcolor{currentfill}%
\pgfsetfillopacity{0.500000}%
\pgfsetlinewidth{0.000000pt}%
\definecolor{currentstroke}{rgb}{0.000000,0.000000,0.000000}%
\pgfsetstrokecolor{currentstroke}%
\pgfsetdash{}{0pt}%
\pgfpathmoveto{\pgfqpoint{1.557590in}{1.204032in}}%
\pgfpathlineto{\pgfqpoint{1.557590in}{1.204032in}}%
\pgfpathlineto{\pgfqpoint{2.121038in}{1.204032in}}%
\pgfpathlineto{\pgfqpoint{2.121038in}{1.204032in}}%
\pgfpathlineto{\pgfqpoint{2.121038in}{0.468969in}}%
\pgfpathlineto{\pgfqpoint{1.557590in}{0.468969in}}%
\pgfpathclose%
\pgfusepath{fill}%
\end{pgfscope}%
\begin{pgfscope}%
\pgfsetbuttcap%
\pgfsetroundjoin%
\definecolor{currentfill}{rgb}{0.000000,0.000000,0.000000}%
\pgfsetfillcolor{currentfill}%
\pgfsetlinewidth{0.803000pt}%
\definecolor{currentstroke}{rgb}{0.000000,0.000000,0.000000}%
\pgfsetstrokecolor{currentstroke}%
\pgfsetdash{}{0pt}%
\pgfsys@defobject{currentmarker}{\pgfqpoint{0.000000in}{-0.048611in}}{\pgfqpoint{0.000000in}{0.000000in}}{%
\pgfpathmoveto{\pgfqpoint{0.000000in}{0.000000in}}%
\pgfpathlineto{\pgfqpoint{0.000000in}{-0.048611in}}%
\pgfusepath{stroke,fill}%
}%
\begin{pgfscope}%
\pgfsys@transformshift{0.430692in}{0.468969in}%
\pgfsys@useobject{currentmarker}{}%
\end{pgfscope}%
\end{pgfscope}%
\begin{pgfscope}%
\pgftext[x=0.430692in,y=0.371747in,,top]{\sffamily\fontsize{10.000000}{12.000000}\selectfont \(\displaystyle 0\)}%
\end{pgfscope}%
\begin{pgfscope}%
\pgfsetbuttcap%
\pgfsetroundjoin%
\definecolor{currentfill}{rgb}{0.000000,0.000000,0.000000}%
\pgfsetfillcolor{currentfill}%
\pgfsetlinewidth{0.803000pt}%
\definecolor{currentstroke}{rgb}{0.000000,0.000000,0.000000}%
\pgfsetstrokecolor{currentstroke}%
\pgfsetdash{}{0pt}%
\pgfsys@defobject{currentmarker}{\pgfqpoint{0.000000in}{-0.048611in}}{\pgfqpoint{0.000000in}{0.000000in}}{%
\pgfpathmoveto{\pgfqpoint{0.000000in}{0.000000in}}%
\pgfpathlineto{\pgfqpoint{0.000000in}{-0.048611in}}%
\pgfusepath{stroke,fill}%
}%
\begin{pgfscope}%
\pgfsys@transformshift{2.121038in}{0.468969in}%
\pgfsys@useobject{currentmarker}{}%
\end{pgfscope}%
\end{pgfscope}%
\begin{pgfscope}%
\pgftext[x=2.121038in,y=0.371747in,,top]{\sffamily\fontsize{10.000000}{12.000000}\selectfont \(\displaystyle 1\)}%
\end{pgfscope}%
\begin{pgfscope}%
\pgftext[x=1.275865in,y=0.192858in,,top]{\sffamily\fontsize{10.000000}{12.000000}\selectfont \(\displaystyle x\)}%
\end{pgfscope}%
\begin{pgfscope}%
\pgfsetbuttcap%
\pgfsetroundjoin%
\definecolor{currentfill}{rgb}{0.000000,0.000000,0.000000}%
\pgfsetfillcolor{currentfill}%
\pgfsetlinewidth{0.803000pt}%
\definecolor{currentstroke}{rgb}{0.000000,0.000000,0.000000}%
\pgfsetstrokecolor{currentstroke}%
\pgfsetdash{}{0pt}%
\pgfsys@defobject{currentmarker}{\pgfqpoint{-0.048611in}{0.000000in}}{\pgfqpoint{0.000000in}{0.000000in}}{%
\pgfpathmoveto{\pgfqpoint{0.000000in}{0.000000in}}%
\pgfpathlineto{\pgfqpoint{-0.048611in}{0.000000in}}%
\pgfusepath{stroke,fill}%
}%
\begin{pgfscope}%
\pgfsys@transformshift{0.430692in}{0.468969in}%
\pgfsys@useobject{currentmarker}{}%
\end{pgfscope}%
\end{pgfscope}%
\begin{pgfscope}%
\pgftext[x=0.264026in,y=0.420774in,left,base]{\sffamily\fontsize{10.000000}{12.000000}\selectfont \(\displaystyle 0\)}%
\end{pgfscope}%
\begin{pgfscope}%
\pgfsetbuttcap%
\pgfsetroundjoin%
\definecolor{currentfill}{rgb}{0.000000,0.000000,0.000000}%
\pgfsetfillcolor{currentfill}%
\pgfsetlinewidth{0.803000pt}%
\definecolor{currentstroke}{rgb}{0.000000,0.000000,0.000000}%
\pgfsetstrokecolor{currentstroke}%
\pgfsetdash{}{0pt}%
\pgfsys@defobject{currentmarker}{\pgfqpoint{-0.048611in}{0.000000in}}{\pgfqpoint{0.000000in}{0.000000in}}{%
\pgfpathmoveto{\pgfqpoint{0.000000in}{0.000000in}}%
\pgfpathlineto{\pgfqpoint{-0.048611in}{0.000000in}}%
\pgfusepath{stroke,fill}%
}%
\begin{pgfscope}%
\pgfsys@transformshift{0.430692in}{2.122861in}%
\pgfsys@useobject{currentmarker}{}%
\end{pgfscope}%
\end{pgfscope}%
\begin{pgfscope}%
\pgftext[x=0.264026in,y=2.074667in,left,base]{\sffamily\fontsize{10.000000}{12.000000}\selectfont \(\displaystyle 1\)}%
\end{pgfscope}%
\begin{pgfscope}%
\pgftext[x=0.208470in,y=1.295915in,,bottom,rotate=90.000000]{\sffamily\fontsize{10.000000}{12.000000}\selectfont \(\displaystyle y(x)\)}%
\end{pgfscope}%
\begin{pgfscope}%
\pgfpathrectangle{\pgfqpoint{0.346175in}{0.386274in}}{\pgfqpoint{1.859380in}{1.819281in}} %
\pgfusepath{clip}%
\pgfsetbuttcap%
\pgfsetroundjoin%
\pgfsetlinewidth{1.204500pt}%
\definecolor{currentstroke}{rgb}{0.000000,0.000000,0.000000}%
\pgfsetstrokecolor{currentstroke}%
\pgfsetdash{{4.440000pt}{1.920000pt}}{0.000000pt}%
\pgfpathmoveto{\pgfqpoint{0.430692in}{0.468969in}}%
\pgfpathlineto{\pgfqpoint{2.121038in}{2.122861in}}%
\pgfusepath{stroke}%
\end{pgfscope}%
\begin{pgfscope}%
\pgfpathrectangle{\pgfqpoint{0.346175in}{0.386274in}}{\pgfqpoint{1.859380in}{1.819281in}} %
\pgfusepath{clip}%
\pgfsetrectcap%
\pgfsetroundjoin%
\pgfsetlinewidth{1.204500pt}%
\definecolor{currentstroke}{rgb}{0.376471,0.482353,0.545098}%
\pgfsetstrokecolor{currentstroke}%
\pgfsetdash{}{0pt}%
\pgfpathmoveto{\pgfqpoint{0.430692in}{0.468969in}}%
\pgfpathlineto{\pgfqpoint{0.447596in}{0.469134in}}%
\pgfpathlineto{\pgfqpoint{0.464499in}{0.469630in}}%
\pgfpathlineto{\pgfqpoint{0.481403in}{0.470457in}}%
\pgfpathlineto{\pgfqpoint{0.498306in}{0.471615in}}%
\pgfpathlineto{\pgfqpoint{0.515210in}{0.473104in}}%
\pgfpathlineto{\pgfqpoint{0.532113in}{0.474923in}}%
\pgfpathlineto{\pgfqpoint{0.549017in}{0.477073in}}%
\pgfpathlineto{\pgfqpoint{0.565920in}{0.479554in}}%
\pgfpathlineto{\pgfqpoint{0.582824in}{0.482365in}}%
\pgfpathlineto{\pgfqpoint{0.599727in}{0.485508in}}%
\pgfpathlineto{\pgfqpoint{0.616630in}{0.488981in}}%
\pgfpathlineto{\pgfqpoint{0.633534in}{0.492785in}}%
\pgfpathlineto{\pgfqpoint{0.650437in}{0.496920in}}%
\pgfpathlineto{\pgfqpoint{0.667341in}{0.501385in}}%
\pgfpathlineto{\pgfqpoint{0.684244in}{0.506181in}}%
\pgfpathlineto{\pgfqpoint{0.701148in}{0.511308in}}%
\pgfpathlineto{\pgfqpoint{0.718051in}{0.516766in}}%
\pgfpathlineto{\pgfqpoint{0.734955in}{0.522555in}}%
\pgfpathlineto{\pgfqpoint{0.751858in}{0.528674in}}%
\pgfpathlineto{\pgfqpoint{0.768762in}{0.535124in}}%
\pgfpathlineto{\pgfqpoint{0.785665in}{0.541905in}}%
\pgfpathlineto{\pgfqpoint{0.802569in}{0.549017in}}%
\pgfpathlineto{\pgfqpoint{0.819472in}{0.556460in}}%
\pgfpathlineto{\pgfqpoint{0.836375in}{0.564233in}}%
\pgfpathlineto{\pgfqpoint{0.853279in}{0.572337in}}%
\pgfpathlineto{\pgfqpoint{0.870182in}{0.580772in}}%
\pgfpathlineto{\pgfqpoint{0.887086in}{0.589538in}}%
\pgfpathlineto{\pgfqpoint{0.903989in}{0.598634in}}%
\pgfpathlineto{\pgfqpoint{0.920893in}{0.608061in}}%
\pgfpathlineto{\pgfqpoint{0.937796in}{0.617819in}}%
\pgfpathlineto{\pgfqpoint{0.954700in}{0.627908in}}%
\pgfpathlineto{\pgfqpoint{0.971603in}{0.638327in}}%
\pgfpathlineto{\pgfqpoint{0.988507in}{0.649078in}}%
\pgfpathlineto{\pgfqpoint{1.005410in}{0.660159in}}%
\pgfpathlineto{\pgfqpoint{1.022313in}{0.671571in}}%
\pgfpathlineto{\pgfqpoint{1.039217in}{0.683313in}}%
\pgfpathlineto{\pgfqpoint{1.056120in}{0.695387in}}%
\pgfpathlineto{\pgfqpoint{1.073024in}{0.707791in}}%
\pgfpathlineto{\pgfqpoint{1.089927in}{0.720526in}}%
\pgfpathlineto{\pgfqpoint{1.106831in}{0.733592in}}%
\pgfpathlineto{\pgfqpoint{1.123734in}{0.746988in}}%
\pgfpathlineto{\pgfqpoint{1.140638in}{0.760715in}}%
\pgfpathlineto{\pgfqpoint{1.157541in}{0.774773in}}%
\pgfpathlineto{\pgfqpoint{1.174445in}{0.789162in}}%
\pgfpathlineto{\pgfqpoint{1.191348in}{0.803882in}}%
\pgfpathlineto{\pgfqpoint{1.208252in}{0.818932in}}%
\pgfpathlineto{\pgfqpoint{1.225155in}{0.834314in}}%
\pgfpathlineto{\pgfqpoint{1.242058in}{0.850026in}}%
\pgfpathlineto{\pgfqpoint{1.258962in}{0.866068in}}%
\pgfpathlineto{\pgfqpoint{1.275865in}{0.882442in}}%
\pgfpathlineto{\pgfqpoint{1.292769in}{0.899146in}}%
\pgfpathlineto{\pgfqpoint{1.309672in}{0.916181in}}%
\pgfpathlineto{\pgfqpoint{1.326576in}{0.933547in}}%
\pgfpathlineto{\pgfqpoint{1.343479in}{0.951244in}}%
\pgfpathlineto{\pgfqpoint{1.360383in}{0.969271in}}%
\pgfpathlineto{\pgfqpoint{1.377286in}{0.987629in}}%
\pgfpathlineto{\pgfqpoint{1.394190in}{1.006318in}}%
\pgfpathlineto{\pgfqpoint{1.411093in}{1.025338in}}%
\pgfpathlineto{\pgfqpoint{1.427996in}{1.044689in}}%
\pgfpathlineto{\pgfqpoint{1.444900in}{1.064370in}}%
\pgfpathlineto{\pgfqpoint{1.461803in}{1.084382in}}%
\pgfpathlineto{\pgfqpoint{1.478707in}{1.104725in}}%
\pgfpathlineto{\pgfqpoint{1.495610in}{1.125399in}}%
\pgfpathlineto{\pgfqpoint{1.512514in}{1.146403in}}%
\pgfpathlineto{\pgfqpoint{1.529417in}{1.167738in}}%
\pgfpathlineto{\pgfqpoint{1.546321in}{1.189404in}}%
\pgfpathlineto{\pgfqpoint{1.563224in}{1.211401in}}%
\pgfpathlineto{\pgfqpoint{1.580128in}{1.233729in}}%
\pgfpathlineto{\pgfqpoint{1.597031in}{1.256387in}}%
\pgfpathlineto{\pgfqpoint{1.613935in}{1.279376in}}%
\pgfpathlineto{\pgfqpoint{1.630838in}{1.302696in}}%
\pgfpathlineto{\pgfqpoint{1.647741in}{1.326346in}}%
\pgfpathlineto{\pgfqpoint{1.664645in}{1.350328in}}%
\pgfpathlineto{\pgfqpoint{1.681548in}{1.374640in}}%
\pgfpathlineto{\pgfqpoint{1.698452in}{1.399283in}}%
\pgfpathlineto{\pgfqpoint{1.715355in}{1.424257in}}%
\pgfpathlineto{\pgfqpoint{1.732259in}{1.449561in}}%
\pgfpathlineto{\pgfqpoint{1.749162in}{1.475197in}}%
\pgfpathlineto{\pgfqpoint{1.766066in}{1.501163in}}%
\pgfpathlineto{\pgfqpoint{1.782969in}{1.527460in}}%
\pgfpathlineto{\pgfqpoint{1.799873in}{1.554087in}}%
\pgfpathlineto{\pgfqpoint{1.816776in}{1.581046in}}%
\pgfpathlineto{\pgfqpoint{1.833679in}{1.608335in}}%
\pgfpathlineto{\pgfqpoint{1.850583in}{1.635955in}}%
\pgfpathlineto{\pgfqpoint{1.867486in}{1.663906in}}%
\pgfpathlineto{\pgfqpoint{1.884390in}{1.692187in}}%
\pgfpathlineto{\pgfqpoint{1.901293in}{1.720800in}}%
\pgfpathlineto{\pgfqpoint{1.918197in}{1.749743in}}%
\pgfpathlineto{\pgfqpoint{1.935100in}{1.779017in}}%
\pgfpathlineto{\pgfqpoint{1.952004in}{1.808621in}}%
\pgfpathlineto{\pgfqpoint{1.968907in}{1.838557in}}%
\pgfpathlineto{\pgfqpoint{1.985811in}{1.868823in}}%
\pgfpathlineto{\pgfqpoint{2.002714in}{1.899420in}}%
\pgfpathlineto{\pgfqpoint{2.019618in}{1.930348in}}%
\pgfpathlineto{\pgfqpoint{2.036521in}{1.961606in}}%
\pgfpathlineto{\pgfqpoint{2.053424in}{1.993196in}}%
\pgfpathlineto{\pgfqpoint{2.070328in}{2.025116in}}%
\pgfpathlineto{\pgfqpoint{2.087231in}{2.057367in}}%
\pgfpathlineto{\pgfqpoint{2.104135in}{2.089948in}}%
\pgfpathlineto{\pgfqpoint{2.121038in}{2.122861in}}%
\pgfusepath{stroke}%
\end{pgfscope}%
\begin{pgfscope}%
\pgfsetrectcap%
\pgfsetmiterjoin%
\pgfsetlinewidth{0.803000pt}%
\definecolor{currentstroke}{rgb}{0.000000,0.000000,0.000000}%
\pgfsetstrokecolor{currentstroke}%
\pgfsetdash{}{0pt}%
\pgfpathmoveto{\pgfqpoint{0.430692in}{0.468969in}}%
\pgfpathlineto{\pgfqpoint{0.430692in}{2.122861in}}%
\pgfusepath{stroke}%
\end{pgfscope}%
\begin{pgfscope}%
\pgfsetrectcap%
\pgfsetmiterjoin%
\pgfsetlinewidth{0.803000pt}%
\definecolor{currentstroke}{rgb}{0.000000,0.000000,0.000000}%
\pgfsetstrokecolor{currentstroke}%
\pgfsetdash{}{0pt}%
\pgfpathmoveto{\pgfqpoint{2.121038in}{0.468969in}}%
\pgfpathlineto{\pgfqpoint{2.121038in}{2.122861in}}%
\pgfusepath{stroke}%
\end{pgfscope}%
\begin{pgfscope}%
\pgfsetrectcap%
\pgfsetmiterjoin%
\pgfsetlinewidth{0.803000pt}%
\definecolor{currentstroke}{rgb}{0.000000,0.000000,0.000000}%
\pgfsetstrokecolor{currentstroke}%
\pgfsetdash{}{0pt}%
\pgfpathmoveto{\pgfqpoint{0.430692in}{0.468969in}}%
\pgfpathlineto{\pgfqpoint{2.121038in}{0.468969in}}%
\pgfusepath{stroke}%
\end{pgfscope}%
\begin{pgfscope}%
\pgfsetrectcap%
\pgfsetmiterjoin%
\pgfsetlinewidth{0.803000pt}%
\definecolor{currentstroke}{rgb}{0.000000,0.000000,0.000000}%
\pgfsetstrokecolor{currentstroke}%
\pgfsetdash{}{0pt}%
\pgfpathmoveto{\pgfqpoint{0.430692in}{2.122861in}}%
\pgfpathlineto{\pgfqpoint{2.121038in}{2.122861in}}%
\pgfusepath{stroke}%
\end{pgfscope}%
\begin{pgfscope}%
\pgftext[x=1.839314in,y=0.836500in,,base]{\sffamily\fontsize{10.000000}{12.000000}\selectfont \(\displaystyle \beta\)}%
\end{pgfscope}%
\end{pgfpicture}%
\makeatother%
\endgroup%

%% file: one-shot.tex
\section{Single-Shot Revenue Maximization}
\label{sec:single-shot}

We begin by considering the ``single-shot'' setting in which the seller
wants to sell a single item to the buyer. The buyer's value $v$ for
the item is drawn from a known distribution $F$. When the buyer is
risk-neutral, it is well known that the optimal mechanism is
deterministic, and in particular is a posted-price mechanism. Observe that a
risk-averse buyer obtains the same utility from a posted-price as a
risk-neutral one, so over the class of posted-price mechanisms, the
optimal one remains the same regardless of the buyer's risk
attitude. We will refer to the optimal posted price as the Myerson
price, and to the corresponding mechanism as Myerson's mechanism. We
denote the revenue obtained by Myerson's mechanism as
\[\myer(F) = \max_p\, \{p(1-F(p))\}.\]

Given that a risk-averse buyer derives less utility from a randomized outcome
than a risk-neutral buyer does, one might conclude that the optimal mechanism
for a risk-averse buyer continues to be deterministic.  Perhaps surprisingly,
this is not the case, as our next example shows. The seller exploits the
buyer's risk aversion to price higher allocations superlinearly.

\begin{example}
  Suppose $v \sim U[0,1]$ and $\rprof(x)=x^2$. The optimal
  deterministic mechanism offers a price of $1/2$ and earns revenue of
  $1/4$ in expectation.

  Suppose we offer an additional option: the binary lottery $(1/2,
  3/8)$ allocates the item to the buyer with probability $1/2$ and if
  the item is allocated, charges the price $3/8$.

  The buyer chooses the deterministic option if $(v-1/2) \geq
  (v-3/8)(1/2)^2$; that is, if $v \geq \frac{13}{24}$. Otherwise, he
  chooses the second option as long as $v \geq 3/8$. The expected
  revenue is therefore
    \[
        \half\left(1-\frac{13}{24}\right)(1)
        + \frac38\left(\frac{13}{24} - \frac38\right)\left(\half\right)
        = \frac{25}{96} > 1/4.
    \]
\end{example}
% \vspace{.3cm}
% In fact, we observe that a mechanism with many menu options can allow 
% us to extract far more revenue from risk-averse buyers. \bmnote{what does this
% refer to?}

\subsection{Incentive Compatible Mechanisms.}

Any single-shot mechanism can be described by the allocation it makes
and the prices it charges to the buyer as a function of the buyer's
value. Let $X(v)$ and $P(v)$ denote these functions, with
$X(v)\in\{0,1\}$ and $P(v)\in \R_{\geq 0}$. Observe that $X(v)$ and
$P(v)$ are random variables and may be correlated. Then, the buyer's
risk-averse utility from the mechanism's outcome is given by
$\rae{vX(v)-P(v)}$. We say that a mechanism with allocation and
pricing functions $(X,P)$ is incentive compatible (IC) for a buyer with
weighting function $\wt$ if for all possible values $v, v'$ of the
buyer, it holds that
\begin{align*}
  \rae{vX(v)-P(v)} &\ge \rae{vX(v')-P(v')}.
\end{align*}

It is without loss of generality to express an incentive compatible mechanism
in the form of a menu, $\menu$, with each menu option, a.k.a. {\em lottery},
corresponding to a particular correlated random (allocation, payment) pair, $(X,P)$. Then,
the allocation and payment of a buyer with value $v$ and weighting function
$\wt$ is given by the utility-maximizing menu option:\footnote{We assume that
  any ties are broken in favor of menu options with a higher expected price.}  
\[
    (X_{\wt}(v),P_{\wt}(v)) = \argmax_{(X,P)\in\menu} \rae{vX-P}.
\]
The revenue of the mechanism is therefore given by
\[
    \rev_{\wt,F}(\menu) = \expect[v\sim F]{\expect{P_{\wt}(v)}},
\] 
where $F$ denotes the distribution from which $v$ is drawn.

Let $\opt(\wt, F)$ denote the optimal revenue achievable by an
incentive compatible mechanism from selling an item to a buyer with weighting
function $\wt$ and value drawn from $F$: 
\[
    \opt(\wt, F) = \max_{\menu} \rev_{\wt,F}(\menu).
\]
We will drop $\wt$ from the above definitions
when it is clear from the context.

\subsubsection*{Utility functions and binary lotteries.}

Lotteries in a mechanism may have many different outcomes---they may charge a
random price when the item is not allocated, and a different random price when
the item is allocated. When a buyer purchases such a lottery, his risk-averse
utility as a function of his value depends on which of the outcomes of the
lottery bring him negative utility and which ones bring positive utility.

\begin{example}
\label{ex:concave-utility}
    Consider a lottery with three outcomes and a buyer with
    weighting function $\wt(x)=x^2$. With probability $1/2$, $X=P=0$; with
    probability $1/4$, $X=1$ and $P=1$; with probability $1/4$, $X=1$ and
    $P=2$. The risk-averse utility of this lottery is plotted on the left in
    Figure~\ref{fig:concave-utility}. At worst, a buyer pays 2 and has $v=0$, 
    hence the buyer has a base value of $-2$ and measures increments from
    there.  A buyer with value $v\in [0,1)$ gets a negative utility from both
    of the second and third outcomes. His risk-averse utility is
    \[
        -2 + (v)\wt(1) + (1)\wt(3/4) + (1-v)\wt(1/2)
                %= v (1-\wt(1/2)) + \wt(3/4) + \wt(1/2) - 2 
                = \frac 34 v - \frac {19}{16}.
    \] 
    A buyer with value $v\in [1,2)$ gets negative utility from the third
    outcome alone. His risk-averse utility is
    \[
        -2 + (v) \wt(1) + (2-v)\wt(3/4) + (v-1)\wt(1/4)
                %= v(1-y(3/4)+y(1/4)) - y(1/4) - 2(1-y(3/4)) 
                = \frac 12 v - \frac{15}{16}.
    \] 
    Finally, a buyer with value $v\ge 2$ gets positive utility from every
    outcome. His risk-averse utility is
    \[
        (v-2)\wt(1/2) + (1)\wt(1/4) = \frac 14 v - \frac{7}{16}.
    \]

    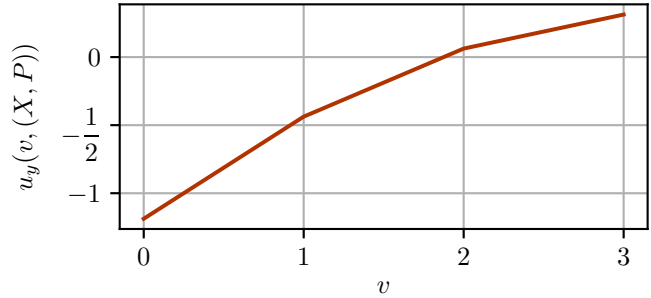
\begin{figure}[htbp]
    \begin{center}
        \scalebox{1}{\input{figures/three_outcome_ex.pgf}} \qquad
        \caption{The risk-averse utility as a function of $v$ for a buyer
        evaluating the lottery described in Example~\ref{ex:concave-utility}.
        Notice that the utility is a concave function of $v$.}
        \label{fig:concave-utility}
    \end{center}
    \end{figure}

\end{example}

In general, the utility $\lrau{v}$ that a buyer derives from a lottery $(X,P)$
is a concave function.  See Appendix~\ref{sec:app-proofs} for a proof.

\begin{lemma}
      \label{lem:concaveUtil}
      For any $\wt$ and lottery $\lotteryp$, $\lrau{v}$ is a concave
      function of $v$. The slope of this function lies between $1 - \wt(1-x)$
      and $\wt(x)$, where $x = \prob{\alloc = 1}$.
\end{lemma}

\begin{figure}[htbp]
\begin{center}
    %\scalebox{.9}{\input{figures/concaveUtil.pgf}} \qquad
    \scalebox{1}{\input{figures/bin_lot.pgf}}
    % \caption{{\em Left:} The risk-averse utility as a function of $v$ for a
    % buyer facing the lottery described in Example~\ref{ex:concave-utility}.
    % {\em Center:} The effect of increasing risk aversion on utility for a
    % fixed lottery. The top line corresponds to the utility of a
    % risk-neutral buyer; the lower lines correspond to increasing risk
    % aversion. {\em Right:} The utility curve of a binary lottery in which,
    % with probability $x$, the buyer receives the item and pays a fixed
    % price of $p$ or else receives nothing and pays nothing.}
    \caption{The utility curve for a buyer with weighting function $y$
    evaluating a binary lottery. The lottery allocates with probability $x$ and
    charges $p$ only when the item is allocated. Notice that the curve is
    linear above zero; the slope here is $y(x)$.}
    \label{fig:bin_lot}
\end{center}
\end{figure}
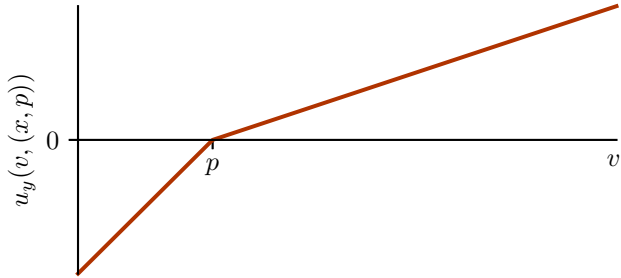

We will make extensive use of a particularly simple lottery which has just two
outcomes: with some probability the buyer is allocated the item and charged a
deterministic price; with the remaining probability, both the allocation and
price are $0$.  We refer to such a lottery as a {\em binary lottery}, and
denote it by the pair $(x, p)$ where $x\in [0,1]$ is the probability of
allocation and $p$ is the price charged upon allocation.  The buyer's utility
function for a binary lottery has a convenient form---it is linear for $v\ge
p$. In particular, when $v > p$, $\lrau{v}{x, p} = \wt(x)(v-p)$. See
Figure~\ref{fig:bin_lot}.

\subsection{Optimal Mechanisms.}
\label{sec:optOneShot}

When the buyer is risk-neutral, the utility that the buyer receives as a
function of his value in any IC mechanism is a concave function. This property
allows for convenient analysis of optimal mechanisms. For a risk-averse buyer,
this is no longer necessarily true. For example, if the mechanism offers a
single lottery with more than two outcomes, as in
Example~\ref{ex:concave-utility}, the buyer's utility function is concave on
its support. In general, the buyer's utility function is the maximum over
concave functions.

We now study properties of revenue optimal mechanisms for a single
buyer. We will show that revenue optimal mechanisms can be described
by a menu composed of binary lotteries, and always induce a convex
utility curve.  We also observe that payment of such a mechanism is 
explicitly determined by the allocation, and that optimal revenue 
weakly increases with risk aversion.
%is non-decreasing in increased aversion to risk.

\begin{theorem}
\label{thm:single-shot-opt}
  For any revenue-optimal IC mechanism $(\allocs, \paymts)$ in the
  single-shot setting, the buyer's utility function
  $\lrau{v}{\lottery[v]}$ is convex and nondecreasing. Furthermore,
  there exists an optimal ex-post IR mechanism that can be described
  as a menu of binary lotteries.
\end{theorem}

\begin{figure*}[htbp]
    \begin{center}
        \scalebox{1}{\input{figures/cvx_env_a.pgf}}~
        \scalebox{1}{\input{figures/cvx_env_b.pgf}}
        \caption{{\em Left:} The utility curve for a menu of lotteries as a
        function of the buyer's value is the pointwise maximum of the utility
        curves of the individual lotteries.  {\em Right:} The lower convex
        envelope of the utility curve corresponds to a menu of binary lotteries
        which, by Theorem~\ref{thm:single-shot-opt}, obtains at least as much
        revenue as the original menu.}
        \label{fig:cvx_env}
    \end{center}
\end{figure*}
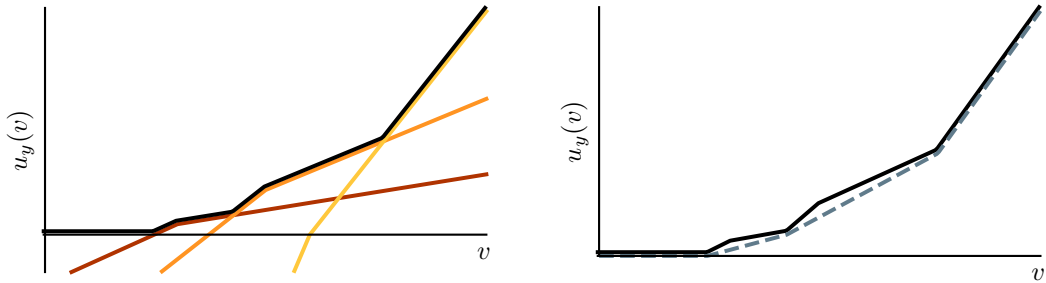

The proof of this theorem 
is defered to Appendix~\ref{sec:app-proofs}. 
% can be found in the full version of the paper.
At a high-level, our proof proceeds as follows. We start with an arbitrary IC
mechanism, and consider the buyer's utility function induced by this mechanism.
We then take the lower convex envelope of this function (see
Figure~\ref{fig:cvx_env}). We show that every point on this curve can be
supported by the utility curve of a binary lottery which has an expected
payment as least that of the menu option it replaces in the original mechanism.
At points where the lower convex envelope is strictly below the original
utility curve, the new mechanism obtains strictly more revenue.

%%
%% We may want to add the figure 4 here for the camera ready? I am not sure.
%%

%  See Figure~\ref{fig:cvx_env} in the appendix.%\kgnote{I put the figure in the appendix; not sure if you wanted it here instead.}

\subsubsection*{Payment Identity.} Consider a mechanism that offers the
buyer a menu of binary lotteries. The utility of a menu option $(x,p)$
for a buyer with value $v$ is $y(x)(v - p)$. This form permits a
standard payment identity analysis, and gives the following identity
for any IC mechanism:
\begin{align}
  \label{eq:payment-id} y(x(v))p(v) = y(x(v))v - \int _0 ^v y(x(z)) \, dz.
\end{align}
Unfortunately, unlike the risk-neutral setting, this payment identity
does not lead to an expression for the mechanism's revenue that is
linear in the allocation function, and so does not allow a
Myerson-type theorem characterizing optimal mechanisms.

\subsection{Optimal Revenue Approaches Social Welfare.}

As we observed earlier, the optimal revenue in general exceeds
Myerson's revenue when the buyer is risk-averse. But to what extent?
We now show that as long as the buyer is sufficiently risk-averse, the
seller can extract nearly the entire expected value of the buyer as
revenue, {\em regardless of the buyer's value distribution}. This
stands in contrast to the risk-neutral setting where for some
distributions (e.g. the equal revenue distribution), the
revenue-welfare gap can be unbounded. More generally,
Lemma~\ref{lem:rev-risk-monotonicity} in the following subsection
shows that the revenue of {\em every} IC mechanism composed of
binary lotteries increases weakly with increasing aversion to risk.

\begin{theorem}
\label{thm:extract-sw}
  For every $\vareps>0$ and $H>1$, if the buyer's weighting function
  satisfies $\wt(1-\vareps)\le 2^{-H/\vareps}$, there exists a
  mechanism that for any value distribution $F$ supported over $[1,H]$
  obtains revenue at least $1-O(\vareps)$ times the buyer's expected
  value $\expect[v \sim F]{v}$.
\end{theorem}
\begin{proof}
Fix the parameters $\vareps$, $H$, and $\wt$, as specified in the
theorem statement. Consider a mechanism that has $H/\vareps$ menu
options of the form $(x_i, y_i)=(y^{-1}(2^{-(i-1)}), H-i \vareps)$.

We first claim that for any value $v$, among all of the menu options
corresponding to prices $ H-i\vareps\le v-\vareps$, the buyer prefers the
option with the maximum price. This follows from observing that $2^{-(i-1)}(v - (H-i \vareps)) \geq 2^{-(i'-1)}(v - (H-i'
  \vareps))$, or equivalently, $2^{i'-i} \geq 1 + ((i'-i) \vareps)/(v - (H-i\vareps))$.
% Then a buyer with value $v$ has more utility for the option $H-i \vareps := \max _j H-j \vareps < v$ than any option with a cheaper payment $H - i' \vareps$:
% \begin{align*}
%   2^{-(i-1)}(v - (H-i \vareps)) & \geq 2^{-(i'-1)}(v - (H-i'
%   \vareps))\\
%   \intertext{or equivalently,}
%   2^{i'-i} & \geq 1 + \frac{(i'-i) \vareps}{v - (H-i\vareps)}
% \end{align*}
The last inequality follows from noting that $v - (H-i \vareps) \geq
\vareps$, and therefore, $1 + \frac{(i'-i) \vareps}{v -
  (H-i\vareps)}\le 1+ i'-i$ which is at most $2^{i'-i}$ for $i' -i \geq 1$.

Note that any menu option with a positive payment has allocation at least $y^{-1}(2^{-(H/\vareps - 1)})$.  Then 
\begin{align*}
\rev(\mech) &= \int _0 ^H f(v) x(v) p(v) dv \\ &\geq y^{-1}(2^{-(H/\vareps - 1)}) \int _0 ^H f(v) (v-2\vareps) dv.
\end{align*}
Then $\rev(\mech) \geq y^{-1}(2^{-(H/\vareps - 1)}) (\expect{v}-
2\vareps)$ where $\expect{v}$ is the social welfare. The theorem now
follows by recalling that $y^{-1}(2^{-(H/\vareps - 1)}) \ge 1-\vareps$.
\end{proof}

We note that the extreme risk aversion required by the statement of
Theorem~\ref{thm:extract-sw} is unrealistic; we address this in
Section~\ref{sec:rr-myerson}.

% It is also clear that $y^{-1}(\cdot)$ is strictly increasing as risk-aversion increases.  Let $z(\cdot)$ be the uncertainty weighting function of a more risk averse profile than $y(\cdot)$, where $z(x) \leq y(x)$ for all $x$, and of course $z(x)$ is strictly increasing in $x$.  Then for any $k$,
% $$z(y^{-1}(k)) \leq y(y^{-1}(k)) = k = z(z^{-1}(k))$$
% hence $y^{-1}(k) < z^{-1}(k)$.

%Then, as risk-aversion increases, $y^{-1}(2^{-(H/\vareps - 1)})$ approaches $1$, and hence revenue approaches social welfare. 

\subsection{Risk-Robust Approximations to Revenue.}
\label{sec:risk-robust-oneshot}

We now turn to the problem of designing near-optimal mechanisms without
detailed knowledge of the buyer's weighting function. As discussed in the
introduction, it is unreasonable to expect the seller to possess precise
information about the buyer's risk attitude. We therefore ask whether it is
possible to design a mechanism that simultaneously approximates the optimal
revenue for any risk attitude. Theorem~\ref{thm:extract-sw} suggests that this
is challenging: for some value distributions supported over $[1,H]$, as we vary
the buyer's weighting function, the optimal revenue can vary by a factor as
large as $\Theta(\log H)$.

Formally, we say that a mechanism defined by menu $\menu$ achieves
{\em an $\alpha$ risk-robust approximation} to revenue for value
distribution $F$ over class $\profiles$ of weighting functions, if for all
$\wt\in\profiles$,
\[\rev_{\wt,F}(\menu)\ge \alpha\, \opt(\wt, F).\]

Achieving a risk-robust approximation to revenue requires
understanding how the revenue of a mechanism changes as the buyer's
risk attitude changes. Ideally, since the optimal revenue tends to
increase as the buyer gets more and more risk averse, we would require
that the revenue of our robust mechanism also increases in
tandem. We show that this is indeed true for mechanisms composed of
binary lotteries under a certain assumption about how risk aversion
increases.

\subsubsection{An $O(\log \log H)$ risk-robust approximation to
  revenue.}

Consider a family of weighting functions $\wtfam$. We say that
$\wtfam$ is {\em non-crossing} if for all pairs of functions $\wt_1$
and $\wt_2$ in $\wtfam$, for all $x_1, x_2\in [0,1]$, $\wt_1(x_1)\ge
\wt_2(x_1)$ implies $\wt_1(x_2)\ge \wt_2(x_2)$. In other words, one
function always lies above the other -- they never cross. We express
this relationship between the functions as $\wt_1\ge \wt_2$. We say
that the family $\wtfam$ is {\em monotone} if for all pairs of
functions $\wt_1\ge\wt_2$ in $\wtfam$, $\wt_2(x)/\wt_1(x)$ is monotone
non-decreasing in $x$. In other words, $\wt_2$ is relatively more
risk-averse at small probabilities than at large probabilities.

\begin{lemma} \label{lem:rev-risk-monotonicity}
    Let $\wtfam$ be a monotone non-crossing family of weighting
    functions, and let $\wt_1\ge \wt_2$ be any two weighting functions in
    $\wtfam$. Then, for any IC mechanism $\mech$ composed of binary
    lotteries, we have $\rev_{\wt_2,F}(\mech) \geq \rev_{\wt_1, F}(\mech)$.
% $y$ and $z$ be two risk profiles such that $y(x) \geq z(x)$ for all $x
%     \in [0,1]$. Fix an IC mechanism $\mech = (x,p)$. If $z(x)/y(x)$ is increasing in
%     $x$, then $\rev(z,\mech,F) \geq \rev(y,\mech,F)$. That is, revenue is monotone in
%     increasing risk aversion, subject to monotonicity in the ratio of the risk
%     profiles.
\end{lemma}

This lemma follows by observing that a buyer with weighting function $\wt_2$
selects a menu option in $\mech$ that is no cheaper than the menu
option a buyer with the same value but weighting function $\wt_1$ selects. We show in Appendix~\ref{sec:app-monotone}  that the monotonicity property of the family $ \wtfam$ is necessary to obtain this revenue monotonicity.
%A formal proof is given in Appendix~\ref{sec:app-proofs}. We also show in the appendix  that the monotonicity property of the family $ \wtfam$
%is necessary to obtain this revenue monotonicity.

\begin{proof}
    Fix a value $v$. We abuse notation and write $\wt_1(v)$ to mean
    $\wt_1(x(v))$, and similarly with $\wt_2$ and bid $b$. By
    incentive compatibility of $\mech$, $\wt_1(v)(v-p(v)) \geq
    \wt_1(b)(v-p(b))$ for all $b$.  By assumption and monotonicity of $x(v)$ in
    $v$, for any $b < v$, $\wt_2(v)/\wt_1(v) \geq \wt_2(b)/\wt_1(b)$.
    Multiplying these inequalities gives
    \[
        \frac{\wt_2(v)}{\wt_1(v)}\wt_1(v) (v-p(v))
                \geq \frac{\wt_2(b)}{\wt_1(b)}\wt_1(b)(v-p(b)),
    \]
    or equivalently
    \[
        \wt_2(v)(v-p(v)) \geq \wt_2(b)(v-p(b)).
    \]
    Therefore, a buyer $v$ will not underreport his value, and so the revenue
    of $\mech$ can only increase under $\wt_2$.
\end{proof}

An implication of this lemma is that if the optimal revenues under
weighting functions $\wt_1$ and $\wt_2$ are close, then a mechanism
that is approximately optimal for $\wt_1$ continues to be
approximately optimal for $\wt_2$. This observation allows us to
develop a mechanism that obtains a risk-robust $O(\log \log H)$
approximation to revenue over monotone non-crossing families of
weighting functions. Observe that this approximation factor is
exponentially smaller than the range of optimal revenues for the
different risk attitudes.

Our mechanism chooses $\log \log H$ representative weighting functions
from the given family, and then picks a random one of the optimal
menus from each family. We note that this is only an existential and
not a computationally efficient result. 
%See Appendix~\ref{sec:app-proofs} for a proof.

\begin{theorem}
\label{thm:log-log}
Let $\wtfam$ be a monotone non-crossing family of weighting
functions and let $F$ be a value distribution supported on
$[0,H]$. Then there exists a mechanism $\mech$ that for any weighting
function in $\wtfam$ achieves an $O(\log\log H)$ approximation to
revenue. Formally, for all $\wt\in\wtfam$,
\[
\rev_{\wt,F}(\mech)\ge \Omega\left( \frac 1{\log\log H} \right)
\opt(\wt, F).
\]
    % Assume a family of revenue-monotone risk profiles parameterized by $a$.  For
    % any $F$ supported on $[0,H]$,
    % \[
    %     \min_\mech \max_{a} \frac{\opt(a, F)}{\rev(\mech, a, F)} \leq    %\rev(M_a^*, a, F)
    %         e \log\log H,
    % \]
    % %where $M_a^*$ is the revenue-optimal mechanism for $a$ and $F$.
    % where $\opt(a,F)$ is the revenue of the revenue-optimal mechanism for $a$ and $F$.
\end{theorem}

\begin{proof}
Since $\wtfam$ is a non-crossing family of functions, the relation
$\ge$ defines a total ordering over the functions. We say that $\wt_1$
is larger than $\wt_2$ if $\wt_1\ge\wt_2$. 

Let $n$ be a constant to be determined later. For $i\in \{0,\cdots,
n\}$, let $k_i = \expect{v}/(\log H)^{i/n}$. Observe that $k_0$ is the
buyer's expected value and $k_n$ is a lower bound on the revenue of
Myerson's mechanism. Therefore, for all $\wt\in\wtfam$, we have
$k_n\le \opt(\wt, F) \le k_0$.

Let $\wtfam_i = \{\wt\in\wtfam : k_i\le\opt(\wt,F)< k_{i-1}\}$, and
let $\wt_i$ be the largest (i.e. least risk-averse) weighting function
in $\wtfam_i$. Define $\mech_i$ to be the revenue-optimal mechanism
for $\wt_i$, that is, $\rev_{\wt_i, F}(\mech_i) = \opt(\wt_i,F)$. 

We now claim that a mechanism that randomly chooses one of the
mechanisms $\mech_i$ to offer to the agent achieves the desired risk
robust approximation. 

Consider some $\wt\in\wtfam$ and suppose that this weighting function
belongs to the set $\wtfam_i$. With probability $1/n$, we choose to
run the mechanism $\mech_i$. Now we observe:
\begin{align*}
  \opt(\wt,F) & \le (\log H)^{1/n}\opt(\wt_i,F) \\
  & = (\log H)^{1/n}\rev_{\wt_i, F}(\mech_i) & \text{(by 
    definition of $\mech_i$)}\\
  & \le (\log H)^{1/n}\rev_{\wt, F}(\mech_i) & \text{(by
    Lemma~\ref{lem:rev-risk-monotonicity})}\\
\end{align*}
Therefore, we get an approximation factor of $n (\log H)^{1/n}$, which is
minimized at $n=\log\log H$.
\end{proof}

\subsubsection{Risk-robust approximation via Myerson's mechanism.}
\label{sec:rr-myerson}

Theorem~\ref{thm:log-log} is unsatisfying for two reasons. One,
finding the mechanism that achieves the revenue guarantee in the
theorem appears challenging. Second, the theorem works only for
certain families of weighting functions, and not for arbitrary
ones. 

We now consider risk-robust approximation from a different
viewpoint. We observe that obtaining the high revenue guaranteed by
Theorem~\ref{thm:extract-sw} requires the buyer to heavily discount
any probabilities that are bounded away from 1. Such extreme risk
aversion is unrealistic. We therefore focus on weighting functions
that map some probabilities bounded away from 1 (i.e. $x =
1-\Theta(1)$) to weights bounded away from 0 (i.e. $\wt(x) =
\Theta(1)$). In other words, the weighting function is $\beta$-bounded
for some $\beta=\Theta(1)$ (Definition~\ref{def:boundedness}). We show
that for such weighting functions, Myerson's mechanism already
achieves an approximation to the optimal revenue. Of course, Myerson's
mechanism is defined independently of the buyer's risk attitude. This
therefore implies a risk-robust approximation.

\begin{theorem}
\label{thm:risk-robust-myer}
  For any $\beta$-bounded convex weighting function $\wt$, and any
  value distribution $F$, we have 
  \[\myer(F) \ge \beta \, \opt(\wt,F).\]
\end{theorem}

To understand the intuition behind this theorem, consider a $\beta$-bounded
convex weighting function $\wt$, and let $\mech=(x(v),p(v))$ denote any
mechanism composed of binary lotteries. We now make two observations. First,
this mechanism cannot obtain too much revenue from low-probability allocations.
Intuitively, this is because if most of its revenue came from buyers who
purchase the low-probability allocations, it could extract much more revenue
from these buyers by selling to them with higher probability. Second, when the
allocation probability is large, the buyer faces less risk, and so the
mechanism behaves nearly like the optimal risk-neutral mechanism. We now
formalize this intuition.

\begin{lemma}
    \label{lem:small-x-bound}
    Let $\mech=(x(v),p(v))$ denote an IC mechanism composed of binary
    lotteries for value distribution $F$ and weighting function
    $\wt$. Then, for any type $t$,
    \[
        \int_0^t f(v)x(v)p(v)dv \leq x(t) \opt(\wt, F).
    \]
\end{lemma}
\begin{proof}
    We examine an alternate mechanism that increases the probability of
    allocation to types below $t$ by a factor of $\frac{1}{x(t)}$.  Consider
    the alternate mechanism $\hat{\mech}$ with allocation rule $\hat{x}(b) =
    \frac{x(b)}{x(t)} $ if $b<t$, and $1$ otherwise, and payment rule
    $\hat{p}(b) = p(b)$. Note that this mechanism is not truthful; in
    particular, an agent with value $v$ may choose a menu option $(\hat{x}(b),
    \hat{p}(b))$ for $b\ne v$. We will show, however, that an agent will never
    deviate to a bid below his true value, and will continue to pay a price at
    least as high as in $\mech$.

    For convenience, we abuse notation and write $\wt(x(v))$ as $\wt(v)$ and
    $\wt(\hat{x}(v))$ as $\hat{\wt}(v)$.  Because $\mech$ is
    incentive compatible, for any $v$ and $w < v$, we have that  $u(v,v) =
    (v-p(v))\rprof(v) \geq (v-p(w))\rprof(w) = u(v,w)$. The weighting function
    $\wt$ is positive, increasing, and convex as a function of $x$, and so
    $\frac{\hat{y}(v)}{\rprof(v)} = \frac{\wt(x(v)/x(t))}{\wt(x(v))}$ is
    nondecreasing in $v$ for $v\le t$.  Therefore
    \begin{align*}
        \frac{\hat{\rprof}(w)}{\hat{\rprof}(v)}(v-p(w)) &\leq
                \frac{\rprof(w)}{\rprof(v)}(v-p(w)) 
            \leq (v-p(v)),
    \end{align*}
    so $\hat{u}(v,w) = \hat{\rprof}(w)(v-p(w)) \leq \hat{\rprof}(v)(v-p(v)) =
    \hat{u}(v,v)$, and the buyer will not underreport. Let $b(v)$ be the
    optimal bid for buyer $v$ in mechanism $\hat{\mech}$; then
    \begin{align*}
        \rev(\hat{\mech}) &\ge \int_0^tf(v)\hat{x}(b(v))p(b(v))dv \\
        &\geq \frac{1}{x(t)}\int_0^tf(v)x(v)p(v)dv,
    \end{align*}
    and the lemma follows.
%   The result now follows from the assumption that $\mech$ was optimal.
\end{proof}

\begin{lemma}
    \label{lem:large-x-bound}
    %If $y(x) = xe^{-a(1/x-1)}$ and $t = \inf\{v : x(v) \geq 1/c\}$ for some
    %$k > 0$, then
    Let $\mech=(x(v),p(v))$ denote an IC mechanism composed of binary
    lotteries for value distribution $F$ and weighting function
    $\wt$. Then, for any type $t$,
   \[
        \int_t^\infty f(v)x(v)p(v)dv \leq \frac{1}{y(t)}\myer(F)
    \]
   % \[
   %      \int_t^\infty f(v)x(v)p(v)dv \leq \frac{x(t)}{y(t)}\myer(F)
   %  \]
\end{lemma}

\begin{proof}
    Let $\tilde{p}(v) = p(v)y(x(v))$. Then the payment identity in
    Equation~\eqref{eq:payment-id} implies that the mechanism
    $(\wt(x(v)),\tilde{p}(v))$ is IC for a risk-neutral buyer. % Note that,
    % because $\wt$ is convex and $\wt(0) = 0$, it must be that $\wt(v) \geq
    % x(v)\frac{\wt(t)}{x(t)}$ for any $v \geq t$, and therefore 
    Therefore, noting that $x(v)\le 1$ and $\wt(v)\ge\wt(t)$ for all
    $v\ge t$, we have
    \begin{align*}
        \int_t^\infty f(v)x(v)p(v)dv &=
                \int_t^\infty f(v)\frac{x(v)}{\wt(v)}\tilde{p}(v)dv \\
            &\leq \frac{1}{\wt(t)}\int_t^\infty f(v)\tilde{p}(v)dv \\
        &\leq \frac{1}{\wt(t)} \int_0^\infty f(v)\tilde{p}(v)dv \\
            &\leq \frac{1}{\wt(t)} \myer(F).
    \end{align*}
\end{proof}
% \bmnote{Note that we can drop the convexity assumption and use only that
% $\wt(t) \leq \wt(v)$ to get a bound with $x(t)$ replaced by 1, which suffices
% for the following proof.}

We can now combine the two lemmas into a proof for
Theorem~\ref{thm:risk-robust-myer}.

\begin{proof}{Theorem~\ref{thm:risk-robust-myer}}
  Let $\mech = (x(v), p(v))$ be an optimal mechanism for value
  distribution $F$ and weighting function $\wt$. Let $t$ be any type
  at which $(1-x(t))y(x(t)) \ge \beta$. By the definition of
  $\beta$-boundedness, such a type exists.

  By Lemma~\ref{lem:small-x-bound}, we have
    \begin{align*}
      \opt &\leq \left(\frac{1}{1-x(t)}\right) \int_t^\infty
      f(v)x(v)p(v)dv,
      \intertext{which, using Lemma~\ref{lem:large-x-bound}, gives}
      &%\leq \left(\frac{1}{1-x(t)}\right) \frac{x(t)}{y(t)} \myer(F) 
      \leq \frac{1}{(1-x(t))y(t)} \myer(F) \le \frac{1}{\beta}\myer(F).
    \end{align*}
    % Note that $\max _x (1-x) y(x)$ is the size of the largest rectangle under the curve $y(\cdot)$, 
    % which by assumption, is at least $1/c$.  Hence $\frac{1}{(1-x(t))y(t)} \leq c$, which gives the desired result.
\end{proof}

%% file: figures/three_outcome_ex.pgf
%% Creator: Matplotlib, PGF backend
%%
%% To include the figure in your LaTeX document, write
%%   \input{<filename>.pgf}
%%
%% Make sure the required packages are loaded in your preamble
%%   \usepackage{pgf}
%%
%% Figures using additional raster images can only be included by \input if
%% they are in the same directory as the main LaTeX file. For loading figures
%% from other directories you can use the `import` package
%%   \usepackage{import}
%% and then include the figures with
%%   \import{<path to file>}{<filename>.pgf}
%%
%% Matplotlib used the following preamble
%%   \usepackage{fontspec}
%%
\begingroup%
\makeatletter%
\begin{pgfpicture}%
\pgfpathrectangle{\pgfpointorigin}{\pgfqpoint{3.500000in}{1.750000in}}%
\pgfusepath{use as bounding box, clip}%
\begin{pgfscope}%
\pgfsetbuttcap%
\pgfsetmiterjoin%
\pgfsetlinewidth{0.000000pt}%
\definecolor{currentstroke}{rgb}{1.000000,1.000000,1.000000}%
\pgfsetstrokecolor{currentstroke}%
\pgfsetstrokeopacity{0.000000}%
\pgfsetdash{}{0pt}%
\pgfpathmoveto{\pgfqpoint{0.000000in}{0.000000in}}%
\pgfpathlineto{\pgfqpoint{3.500000in}{0.000000in}}%
\pgfpathlineto{\pgfqpoint{3.500000in}{1.750000in}}%
\pgfpathlineto{\pgfqpoint{0.000000in}{1.750000in}}%
\pgfpathclose%
\pgfusepath{}%
\end{pgfscope}%
\begin{pgfscope}%
\pgfsetbuttcap%
\pgfsetmiterjoin%
\definecolor{currentfill}{rgb}{1.000000,1.000000,1.000000}%
\pgfsetfillcolor{currentfill}%
\pgfsetlinewidth{0.000000pt}%
\definecolor{currentstroke}{rgb}{0.000000,0.000000,0.000000}%
\pgfsetstrokecolor{currentstroke}%
\pgfsetstrokeopacity{0.000000}%
\pgfsetdash{}{0pt}%
\pgfpathmoveto{\pgfqpoint{0.632485in}{0.468889in}}%
\pgfpathlineto{\pgfqpoint{3.395556in}{0.468889in}}%
\pgfpathlineto{\pgfqpoint{3.395556in}{1.645556in}}%
\pgfpathlineto{\pgfqpoint{0.632485in}{1.645556in}}%
\pgfpathclose%
\pgfusepath{fill}%
\end{pgfscope}%
\begin{pgfscope}%
\pgfpathrectangle{\pgfqpoint{0.632485in}{0.468889in}}{\pgfqpoint{2.763071in}{1.176667in}} %
\pgfusepath{clip}%
\pgfsetrectcap%
\pgfsetroundjoin%
\pgfsetlinewidth{0.803000pt}%
\definecolor{currentstroke}{rgb}{0.690196,0.690196,0.690196}%
\pgfsetstrokecolor{currentstroke}%
\pgfsetdash{}{0pt}%
\pgfpathmoveto{\pgfqpoint{0.758079in}{0.468889in}}%
\pgfpathlineto{\pgfqpoint{0.758079in}{1.645556in}}%
\pgfusepath{stroke}%
\end{pgfscope}%
\begin{pgfscope}%
\pgfsetbuttcap%
\pgfsetroundjoin%
\definecolor{currentfill}{rgb}{0.000000,0.000000,0.000000}%
\pgfsetfillcolor{currentfill}%
\pgfsetlinewidth{0.803000pt}%
\definecolor{currentstroke}{rgb}{0.000000,0.000000,0.000000}%
\pgfsetstrokecolor{currentstroke}%
\pgfsetdash{}{0pt}%
\pgfsys@defobject{currentmarker}{\pgfqpoint{0.000000in}{-0.048611in}}{\pgfqpoint{0.000000in}{0.000000in}}{%
\pgfpathmoveto{\pgfqpoint{0.000000in}{0.000000in}}%
\pgfpathlineto{\pgfqpoint{0.000000in}{-0.048611in}}%
\pgfusepath{stroke,fill}%
}%
\begin{pgfscope}%
\pgfsys@transformshift{0.758079in}{0.468889in}%
\pgfsys@useobject{currentmarker}{}%
\end{pgfscope}%
\end{pgfscope}%
\begin{pgfscope}%
\pgftext[x=0.758079in,y=0.371666in,,top]{\sffamily\fontsize{10.000000}{12.000000}\selectfont \(\displaystyle 0\)}%
\end{pgfscope}%
\begin{pgfscope}%
\pgfpathrectangle{\pgfqpoint{0.632485in}{0.468889in}}{\pgfqpoint{2.763071in}{1.176667in}} %
\pgfusepath{clip}%
\pgfsetrectcap%
\pgfsetroundjoin%
\pgfsetlinewidth{0.803000pt}%
\definecolor{currentstroke}{rgb}{0.690196,0.690196,0.690196}%
\pgfsetstrokecolor{currentstroke}%
\pgfsetdash{}{0pt}%
\pgfpathmoveto{\pgfqpoint{1.595373in}{0.468889in}}%
\pgfpathlineto{\pgfqpoint{1.595373in}{1.645556in}}%
\pgfusepath{stroke}%
\end{pgfscope}%
\begin{pgfscope}%
\pgfsetbuttcap%
\pgfsetroundjoin%
\definecolor{currentfill}{rgb}{0.000000,0.000000,0.000000}%
\pgfsetfillcolor{currentfill}%
\pgfsetlinewidth{0.803000pt}%
\definecolor{currentstroke}{rgb}{0.000000,0.000000,0.000000}%
\pgfsetstrokecolor{currentstroke}%
\pgfsetdash{}{0pt}%
\pgfsys@defobject{currentmarker}{\pgfqpoint{0.000000in}{-0.048611in}}{\pgfqpoint{0.000000in}{0.000000in}}{%
\pgfpathmoveto{\pgfqpoint{0.000000in}{0.000000in}}%
\pgfpathlineto{\pgfqpoint{0.000000in}{-0.048611in}}%
\pgfusepath{stroke,fill}%
}%
\begin{pgfscope}%
\pgfsys@transformshift{1.595373in}{0.468889in}%
\pgfsys@useobject{currentmarker}{}%
\end{pgfscope}%
\end{pgfscope}%
\begin{pgfscope}%
\pgftext[x=1.595373in,y=0.371666in,,top]{\sffamily\fontsize{10.000000}{12.000000}\selectfont \(\displaystyle 1\)}%
\end{pgfscope}%
\begin{pgfscope}%
\pgfpathrectangle{\pgfqpoint{0.632485in}{0.468889in}}{\pgfqpoint{2.763071in}{1.176667in}} %
\pgfusepath{clip}%
\pgfsetrectcap%
\pgfsetroundjoin%
\pgfsetlinewidth{0.803000pt}%
\definecolor{currentstroke}{rgb}{0.690196,0.690196,0.690196}%
\pgfsetstrokecolor{currentstroke}%
\pgfsetdash{}{0pt}%
\pgfpathmoveto{\pgfqpoint{2.432667in}{0.468889in}}%
\pgfpathlineto{\pgfqpoint{2.432667in}{1.645556in}}%
\pgfusepath{stroke}%
\end{pgfscope}%
\begin{pgfscope}%
\pgfsetbuttcap%
\pgfsetroundjoin%
\definecolor{currentfill}{rgb}{0.000000,0.000000,0.000000}%
\pgfsetfillcolor{currentfill}%
\pgfsetlinewidth{0.803000pt}%
\definecolor{currentstroke}{rgb}{0.000000,0.000000,0.000000}%
\pgfsetstrokecolor{currentstroke}%
\pgfsetdash{}{0pt}%
\pgfsys@defobject{currentmarker}{\pgfqpoint{0.000000in}{-0.048611in}}{\pgfqpoint{0.000000in}{0.000000in}}{%
\pgfpathmoveto{\pgfqpoint{0.000000in}{0.000000in}}%
\pgfpathlineto{\pgfqpoint{0.000000in}{-0.048611in}}%
\pgfusepath{stroke,fill}%
}%
\begin{pgfscope}%
\pgfsys@transformshift{2.432667in}{0.468889in}%
\pgfsys@useobject{currentmarker}{}%
\end{pgfscope}%
\end{pgfscope}%
\begin{pgfscope}%
\pgftext[x=2.432667in,y=0.371666in,,top]{\sffamily\fontsize{10.000000}{12.000000}\selectfont \(\displaystyle 2\)}%
\end{pgfscope}%
\begin{pgfscope}%
\pgfpathrectangle{\pgfqpoint{0.632485in}{0.468889in}}{\pgfqpoint{2.763071in}{1.176667in}} %
\pgfusepath{clip}%
\pgfsetrectcap%
\pgfsetroundjoin%
\pgfsetlinewidth{0.803000pt}%
\definecolor{currentstroke}{rgb}{0.690196,0.690196,0.690196}%
\pgfsetstrokecolor{currentstroke}%
\pgfsetdash{}{0pt}%
\pgfpathmoveto{\pgfqpoint{3.269961in}{0.468889in}}%
\pgfpathlineto{\pgfqpoint{3.269961in}{1.645556in}}%
\pgfusepath{stroke}%
\end{pgfscope}%
\begin{pgfscope}%
\pgfsetbuttcap%
\pgfsetroundjoin%
\definecolor{currentfill}{rgb}{0.000000,0.000000,0.000000}%
\pgfsetfillcolor{currentfill}%
\pgfsetlinewidth{0.803000pt}%
\definecolor{currentstroke}{rgb}{0.000000,0.000000,0.000000}%
\pgfsetstrokecolor{currentstroke}%
\pgfsetdash{}{0pt}%
\pgfsys@defobject{currentmarker}{\pgfqpoint{0.000000in}{-0.048611in}}{\pgfqpoint{0.000000in}{0.000000in}}{%
\pgfpathmoveto{\pgfqpoint{0.000000in}{0.000000in}}%
\pgfpathlineto{\pgfqpoint{0.000000in}{-0.048611in}}%
\pgfusepath{stroke,fill}%
}%
\begin{pgfscope}%
\pgfsys@transformshift{3.269961in}{0.468889in}%
\pgfsys@useobject{currentmarker}{}%
\end{pgfscope}%
\end{pgfscope}%
\begin{pgfscope}%
\pgftext[x=3.269961in,y=0.371666in,,top]{\sffamily\fontsize{10.000000}{12.000000}\selectfont \(\displaystyle 3\)}%
\end{pgfscope}%
\begin{pgfscope}%
\pgftext[x=2.014020in,y=0.192778in,,top]{\sffamily\fontsize{10.000000}{12.000000}\selectfont \(\displaystyle v\)}%
\end{pgfscope}%
\begin{pgfscope}%
\pgfpathrectangle{\pgfqpoint{0.632485in}{0.468889in}}{\pgfqpoint{2.763071in}{1.176667in}} %
\pgfusepath{clip}%
\pgfsetrectcap%
\pgfsetroundjoin%
\pgfsetlinewidth{0.803000pt}%
\definecolor{currentstroke}{rgb}{0.690196,0.690196,0.690196}%
\pgfsetstrokecolor{currentstroke}%
\pgfsetdash{}{0pt}%
\pgfpathmoveto{\pgfqpoint{0.632485in}{0.656086in}}%
\pgfpathlineto{\pgfqpoint{3.395556in}{0.656086in}}%
\pgfusepath{stroke}%
\end{pgfscope}%
\begin{pgfscope}%
\pgfsetbuttcap%
\pgfsetroundjoin%
\definecolor{currentfill}{rgb}{0.000000,0.000000,0.000000}%
\pgfsetfillcolor{currentfill}%
\pgfsetlinewidth{0.803000pt}%
\definecolor{currentstroke}{rgb}{0.000000,0.000000,0.000000}%
\pgfsetstrokecolor{currentstroke}%
\pgfsetdash{}{0pt}%
\pgfsys@defobject{currentmarker}{\pgfqpoint{-0.048611in}{0.000000in}}{\pgfqpoint{0.000000in}{0.000000in}}{%
\pgfpathmoveto{\pgfqpoint{0.000000in}{0.000000in}}%
\pgfpathlineto{\pgfqpoint{-0.048611in}{0.000000in}}%
\pgfusepath{stroke,fill}%
}%
\begin{pgfscope}%
\pgfsys@transformshift{0.632485in}{0.656086in}%
\pgfsys@useobject{currentmarker}{}%
\end{pgfscope}%
\end{pgfscope}%
\begin{pgfscope}%
\pgftext[x=0.357793in,y=0.607891in,left,base]{\sffamily\fontsize{10.000000}{12.000000}\selectfont \(\displaystyle -1\)}%
\end{pgfscope}%
\begin{pgfscope}%
\pgfpathrectangle{\pgfqpoint{0.632485in}{0.468889in}}{\pgfqpoint{2.763071in}{1.176667in}} %
\pgfusepath{clip}%
\pgfsetrectcap%
\pgfsetroundjoin%
\pgfsetlinewidth{0.803000pt}%
\definecolor{currentstroke}{rgb}{0.690196,0.690196,0.690196}%
\pgfsetstrokecolor{currentstroke}%
\pgfsetdash{}{0pt}%
\pgfpathmoveto{\pgfqpoint{0.632485in}{1.012651in}}%
\pgfpathlineto{\pgfqpoint{3.395556in}{1.012651in}}%
\pgfusepath{stroke}%
\end{pgfscope}%
\begin{pgfscope}%
\pgfsetbuttcap%
\pgfsetroundjoin%
\definecolor{currentfill}{rgb}{0.000000,0.000000,0.000000}%
\pgfsetfillcolor{currentfill}%
\pgfsetlinewidth{0.803000pt}%
\definecolor{currentstroke}{rgb}{0.000000,0.000000,0.000000}%
\pgfsetstrokecolor{currentstroke}%
\pgfsetdash{}{0pt}%
\pgfsys@defobject{currentmarker}{\pgfqpoint{-0.048611in}{0.000000in}}{\pgfqpoint{0.000000in}{0.000000in}}{%
\pgfpathmoveto{\pgfqpoint{0.000000in}{0.000000in}}%
\pgfpathlineto{\pgfqpoint{-0.048611in}{0.000000in}}%
\pgfusepath{stroke,fill}%
}%
\begin{pgfscope}%
\pgfsys@transformshift{0.632485in}{1.012651in}%
\pgfsys@useobject{currentmarker}{}%
\end{pgfscope}%
\end{pgfscope}%
\begin{pgfscope}%
\pgftext[x=0.324459in,y=0.920919in,left,base]{\sffamily\fontsize{10.000000}{12.000000}\selectfont \(\displaystyle -\frac{1}{2}\)}%
\end{pgfscope}%
\begin{pgfscope}%
\pgfpathrectangle{\pgfqpoint{0.632485in}{0.468889in}}{\pgfqpoint{2.763071in}{1.176667in}} %
\pgfusepath{clip}%
\pgfsetrectcap%
\pgfsetroundjoin%
\pgfsetlinewidth{0.803000pt}%
\definecolor{currentstroke}{rgb}{0.690196,0.690196,0.690196}%
\pgfsetstrokecolor{currentstroke}%
\pgfsetdash{}{0pt}%
\pgfpathmoveto{\pgfqpoint{0.632485in}{1.369217in}}%
\pgfpathlineto{\pgfqpoint{3.395556in}{1.369217in}}%
\pgfusepath{stroke}%
\end{pgfscope}%
\begin{pgfscope}%
\pgfsetbuttcap%
\pgfsetroundjoin%
\definecolor{currentfill}{rgb}{0.000000,0.000000,0.000000}%
\pgfsetfillcolor{currentfill}%
\pgfsetlinewidth{0.803000pt}%
\definecolor{currentstroke}{rgb}{0.000000,0.000000,0.000000}%
\pgfsetstrokecolor{currentstroke}%
\pgfsetdash{}{0pt}%
\pgfsys@defobject{currentmarker}{\pgfqpoint{-0.048611in}{0.000000in}}{\pgfqpoint{0.000000in}{0.000000in}}{%
\pgfpathmoveto{\pgfqpoint{0.000000in}{0.000000in}}%
\pgfpathlineto{\pgfqpoint{-0.048611in}{0.000000in}}%
\pgfusepath{stroke,fill}%
}%
\begin{pgfscope}%
\pgfsys@transformshift{0.632485in}{1.369217in}%
\pgfsys@useobject{currentmarker}{}%
\end{pgfscope}%
\end{pgfscope}%
\begin{pgfscope}%
\pgftext[x=0.465818in,y=1.321023in,left,base]{\sffamily\fontsize{10.000000}{12.000000}\selectfont \(\displaystyle 0\)}%
\end{pgfscope}%
\begin{pgfscope}%
\pgftext[x=0.213348in,y=1.057222in,,bottom,rotate=90.000000]{\sffamily\fontsize{10.000000}{12.000000}\selectfont \(\displaystyle u_y(v, (X, P))\)}%
\end{pgfscope}%
\begin{pgfscope}%
\pgfpathrectangle{\pgfqpoint{0.632485in}{0.468889in}}{\pgfqpoint{2.763071in}{1.176667in}} %
\pgfusepath{clip}%
\pgfsetrectcap%
\pgfsetroundjoin%
\pgfsetlinewidth{1.505625pt}%
\definecolor{currentstroke}{rgb}{0.690196,0.200000,0.000000}%
\pgfsetstrokecolor{currentstroke}%
\pgfsetdash{}{0pt}%
\pgfpathmoveto{\pgfqpoint{0.758079in}{0.522373in}}%
\pgfpathlineto{\pgfqpoint{0.783451in}{0.538581in}}%
\pgfpathlineto{\pgfqpoint{0.808824in}{0.554789in}}%
\pgfpathlineto{\pgfqpoint{0.834196in}{0.570996in}}%
\pgfpathlineto{\pgfqpoint{0.859569in}{0.587204in}}%
\pgfpathlineto{\pgfqpoint{0.884941in}{0.603411in}}%
\pgfpathlineto{\pgfqpoint{0.910314in}{0.619619in}}%
\pgfpathlineto{\pgfqpoint{0.935687in}{0.635826in}}%
\pgfpathlineto{\pgfqpoint{0.961059in}{0.652034in}}%
\pgfpathlineto{\pgfqpoint{0.986432in}{0.668241in}}%
\pgfpathlineto{\pgfqpoint{1.011804in}{0.684449in}}%
\pgfpathlineto{\pgfqpoint{1.037177in}{0.700656in}}%
\pgfpathlineto{\pgfqpoint{1.062549in}{0.716864in}}%
\pgfpathlineto{\pgfqpoint{1.087922in}{0.733071in}}%
\pgfpathlineto{\pgfqpoint{1.113294in}{0.749279in}}%
\pgfpathlineto{\pgfqpoint{1.138667in}{0.765486in}}%
\pgfpathlineto{\pgfqpoint{1.164040in}{0.781694in}}%
\pgfpathlineto{\pgfqpoint{1.189412in}{0.797902in}}%
\pgfpathlineto{\pgfqpoint{1.214785in}{0.814109in}}%
\pgfpathlineto{\pgfqpoint{1.240157in}{0.830317in}}%
\pgfpathlineto{\pgfqpoint{1.265530in}{0.846524in}}%
\pgfpathlineto{\pgfqpoint{1.290902in}{0.862732in}}%
\pgfpathlineto{\pgfqpoint{1.316275in}{0.878939in}}%
\pgfpathlineto{\pgfqpoint{1.341647in}{0.895147in}}%
\pgfpathlineto{\pgfqpoint{1.367020in}{0.911354in}}%
\pgfpathlineto{\pgfqpoint{1.392393in}{0.927562in}}%
\pgfpathlineto{\pgfqpoint{1.417765in}{0.943769in}}%
\pgfpathlineto{\pgfqpoint{1.443138in}{0.959977in}}%
\pgfpathlineto{\pgfqpoint{1.468510in}{0.976184in}}%
\pgfpathlineto{\pgfqpoint{1.493883in}{0.992392in}}%
\pgfpathlineto{\pgfqpoint{1.519255in}{1.008599in}}%
\pgfpathlineto{\pgfqpoint{1.544628in}{1.024807in}}%
\pgfpathlineto{\pgfqpoint{1.570000in}{1.041015in}}%
\pgfpathlineto{\pgfqpoint{1.595373in}{1.057222in}}%
\pgfpathlineto{\pgfqpoint{1.620746in}{1.068027in}}%
\pgfpathlineto{\pgfqpoint{1.646118in}{1.078832in}}%
\pgfpathlineto{\pgfqpoint{1.671491in}{1.089637in}}%
\pgfpathlineto{\pgfqpoint{1.696863in}{1.100442in}}%
\pgfpathlineto{\pgfqpoint{1.722236in}{1.111247in}}%
\pgfpathlineto{\pgfqpoint{1.747608in}{1.122052in}}%
\pgfpathlineto{\pgfqpoint{1.772981in}{1.132857in}}%
\pgfpathlineto{\pgfqpoint{1.798353in}{1.143662in}}%
\pgfpathlineto{\pgfqpoint{1.823726in}{1.154467in}}%
\pgfpathlineto{\pgfqpoint{1.849098in}{1.165272in}}%
\pgfpathlineto{\pgfqpoint{1.874471in}{1.176077in}}%
\pgfpathlineto{\pgfqpoint{1.899844in}{1.186882in}}%
\pgfpathlineto{\pgfqpoint{1.925216in}{1.197687in}}%
\pgfpathlineto{\pgfqpoint{1.950589in}{1.208492in}}%
\pgfpathlineto{\pgfqpoint{1.975961in}{1.219297in}}%
\pgfpathlineto{\pgfqpoint{2.001334in}{1.230102in}}%
\pgfpathlineto{\pgfqpoint{2.026706in}{1.240907in}}%
\pgfpathlineto{\pgfqpoint{2.052079in}{1.251712in}}%
\pgfpathlineto{\pgfqpoint{2.077451in}{1.262518in}}%
\pgfpathlineto{\pgfqpoint{2.102824in}{1.273323in}}%
\pgfpathlineto{\pgfqpoint{2.128197in}{1.284128in}}%
\pgfpathlineto{\pgfqpoint{2.153569in}{1.294933in}}%
\pgfpathlineto{\pgfqpoint{2.178942in}{1.305738in}}%
\pgfpathlineto{\pgfqpoint{2.204314in}{1.316543in}}%
\pgfpathlineto{\pgfqpoint{2.229687in}{1.327348in}}%
\pgfpathlineto{\pgfqpoint{2.255059in}{1.338153in}}%
\pgfpathlineto{\pgfqpoint{2.280432in}{1.348958in}}%
\pgfpathlineto{\pgfqpoint{2.305804in}{1.359763in}}%
\pgfpathlineto{\pgfqpoint{2.331177in}{1.370568in}}%
\pgfpathlineto{\pgfqpoint{2.356550in}{1.381373in}}%
\pgfpathlineto{\pgfqpoint{2.381922in}{1.392178in}}%
\pgfpathlineto{\pgfqpoint{2.407295in}{1.402983in}}%
\pgfpathlineto{\pgfqpoint{2.432667in}{1.413788in}}%
\pgfpathlineto{\pgfqpoint{2.458040in}{1.419190in}}%
\pgfpathlineto{\pgfqpoint{2.483412in}{1.424593in}}%
\pgfpathlineto{\pgfqpoint{2.508785in}{1.429995in}}%
\pgfpathlineto{\pgfqpoint{2.534157in}{1.435398in}}%
\pgfpathlineto{\pgfqpoint{2.559530in}{1.440800in}}%
\pgfpathlineto{\pgfqpoint{2.584903in}{1.446203in}}%
\pgfpathlineto{\pgfqpoint{2.610275in}{1.451605in}}%
\pgfpathlineto{\pgfqpoint{2.635648in}{1.457008in}}%
\pgfpathlineto{\pgfqpoint{2.661020in}{1.462410in}}%
\pgfpathlineto{\pgfqpoint{2.686393in}{1.467813in}}%
\pgfpathlineto{\pgfqpoint{2.711765in}{1.473215in}}%
\pgfpathlineto{\pgfqpoint{2.737138in}{1.478618in}}%
\pgfpathlineto{\pgfqpoint{2.762510in}{1.484020in}}%
\pgfpathlineto{\pgfqpoint{2.787883in}{1.489423in}}%
\pgfpathlineto{\pgfqpoint{2.813255in}{1.494825in}}%
\pgfpathlineto{\pgfqpoint{2.838628in}{1.500228in}}%
\pgfpathlineto{\pgfqpoint{2.864001in}{1.505631in}}%
\pgfpathlineto{\pgfqpoint{2.889373in}{1.511033in}}%
\pgfpathlineto{\pgfqpoint{2.914746in}{1.516436in}}%
\pgfpathlineto{\pgfqpoint{2.940118in}{1.521838in}}%
\pgfpathlineto{\pgfqpoint{2.965491in}{1.527241in}}%
\pgfpathlineto{\pgfqpoint{2.990863in}{1.532643in}}%
\pgfpathlineto{\pgfqpoint{3.016236in}{1.538046in}}%
\pgfpathlineto{\pgfqpoint{3.041608in}{1.543448in}}%
\pgfpathlineto{\pgfqpoint{3.066981in}{1.548851in}}%
\pgfpathlineto{\pgfqpoint{3.092354in}{1.554253in}}%
\pgfpathlineto{\pgfqpoint{3.117726in}{1.559656in}}%
\pgfpathlineto{\pgfqpoint{3.143099in}{1.565058in}}%
\pgfpathlineto{\pgfqpoint{3.168471in}{1.570461in}}%
\pgfpathlineto{\pgfqpoint{3.193844in}{1.575863in}}%
\pgfpathlineto{\pgfqpoint{3.219216in}{1.581266in}}%
\pgfpathlineto{\pgfqpoint{3.244589in}{1.586668in}}%
\pgfpathlineto{\pgfqpoint{3.269961in}{1.592071in}}%
\pgfusepath{stroke}%
\end{pgfscope}%
\begin{pgfscope}%
\pgfsetrectcap%
\pgfsetmiterjoin%
\pgfsetlinewidth{0.803000pt}%
\definecolor{currentstroke}{rgb}{0.000000,0.000000,0.000000}%
\pgfsetstrokecolor{currentstroke}%
\pgfsetdash{}{0pt}%
\pgfpathmoveto{\pgfqpoint{0.632485in}{0.468889in}}%
\pgfpathlineto{\pgfqpoint{0.632485in}{1.645556in}}%
\pgfusepath{stroke}%
\end{pgfscope}%
\begin{pgfscope}%
\pgfsetrectcap%
\pgfsetmiterjoin%
\pgfsetlinewidth{0.803000pt}%
\definecolor{currentstroke}{rgb}{0.000000,0.000000,0.000000}%
\pgfsetstrokecolor{currentstroke}%
\pgfsetdash{}{0pt}%
\pgfpathmoveto{\pgfqpoint{3.395556in}{0.468889in}}%
\pgfpathlineto{\pgfqpoint{3.395556in}{1.645556in}}%
\pgfusepath{stroke}%
\end{pgfscope}%
\begin{pgfscope}%
\pgfsetrectcap%
\pgfsetmiterjoin%
\pgfsetlinewidth{0.803000pt}%
\definecolor{currentstroke}{rgb}{0.000000,0.000000,0.000000}%
\pgfsetstrokecolor{currentstroke}%
\pgfsetdash{}{0pt}%
\pgfpathmoveto{\pgfqpoint{0.632485in}{0.468889in}}%
\pgfpathlineto{\pgfqpoint{3.395556in}{0.468889in}}%
\pgfusepath{stroke}%
\end{pgfscope}%
\begin{pgfscope}%
\pgfsetrectcap%
\pgfsetmiterjoin%
\pgfsetlinewidth{0.803000pt}%
\definecolor{currentstroke}{rgb}{0.000000,0.000000,0.000000}%
\pgfsetstrokecolor{currentstroke}%
\pgfsetdash{}{0pt}%
\pgfpathmoveto{\pgfqpoint{0.632485in}{1.645556in}}%
\pgfpathlineto{\pgfqpoint{3.395556in}{1.645556in}}%
\pgfusepath{stroke}%
\end{pgfscope}%
\end{pgfpicture}%
\makeatother%
\endgroup%

%% file: figures/bin_lot.pgf
%% Creator: Matplotlib, PGF backend
%%
%% To include the figure in your LaTeX document, write
%%   \input{<filename>.pgf}
%%
%% Make sure the required packages are loaded in your preamble
%%   \usepackage{pgf}
%%
%% Figures using additional raster images can only be included by \input if
%% they are in the same directory as the main LaTeX file. For loading figures
%% from other directories you can use the `import` package
%%   \usepackage{import}
%% and then include the figures with
%%   \import{<path to file>}{<filename>.pgf}
%%
%% Matplotlib used the following preamble
%%   \usepackage{fontspec}
%%
\begingroup%
\makeatletter%
\begin{pgfpicture}%
\pgfpathrectangle{\pgfpointorigin}{\pgfqpoint{3.500000in}{1.750000in}}%
\pgfusepath{use as bounding box, clip}%
\begin{pgfscope}%
\pgfsetbuttcap%
\pgfsetmiterjoin%
\pgfsetlinewidth{0.000000pt}%
\definecolor{currentstroke}{rgb}{1.000000,1.000000,1.000000}%
\pgfsetstrokecolor{currentstroke}%
\pgfsetstrokeopacity{0.000000}%
\pgfsetdash{}{0pt}%
\pgfpathmoveto{\pgfqpoint{0.000000in}{0.000000in}}%
\pgfpathlineto{\pgfqpoint{3.500000in}{0.000000in}}%
\pgfpathlineto{\pgfqpoint{3.500000in}{1.750000in}}%
\pgfpathlineto{\pgfqpoint{0.000000in}{1.750000in}}%
\pgfpathclose%
\pgfusepath{}%
\end{pgfscope}%
\begin{pgfscope}%
\pgfsetbuttcap%
\pgfsetmiterjoin%
\definecolor{currentfill}{rgb}{1.000000,1.000000,1.000000}%
\pgfsetfillcolor{currentfill}%
\pgfsetlinewidth{0.000000pt}%
\definecolor{currentstroke}{rgb}{0.000000,0.000000,0.000000}%
\pgfsetstrokecolor{currentstroke}%
\pgfsetstrokeopacity{0.000000}%
\pgfsetdash{}{0pt}%
\pgfpathmoveto{\pgfqpoint{0.295422in}{0.104444in}}%
\pgfpathlineto{\pgfqpoint{3.395556in}{0.104444in}}%
\pgfpathlineto{\pgfqpoint{3.395556in}{1.645556in}}%
\pgfpathlineto{\pgfqpoint{0.295422in}{1.645556in}}%
\pgfpathclose%
\pgfusepath{fill}%
\end{pgfscope}%
\begin{pgfscope}%
\pgfsetbuttcap%
\pgfsetroundjoin%
\definecolor{currentfill}{rgb}{0.000000,0.000000,0.000000}%
\pgfsetfillcolor{currentfill}%
\pgfsetlinewidth{0.803000pt}%
\definecolor{currentstroke}{rgb}{0.000000,0.000000,0.000000}%
\pgfsetstrokecolor{currentstroke}%
\pgfsetdash{}{0pt}%
\pgfsys@defobject{currentmarker}{\pgfqpoint{0.000000in}{-0.048611in}}{\pgfqpoint{0.000000in}{0.000000in}}{%
\pgfpathmoveto{\pgfqpoint{0.000000in}{0.000000in}}%
\pgfpathlineto{\pgfqpoint{0.000000in}{-0.048611in}}%
\pgfusepath{stroke,fill}%
}%
\begin{pgfscope}%
\pgfsys@transformshift{1.140913in}{0.875000in}%
\pgfsys@useobject{currentmarker}{}%
\end{pgfscope}%
\end{pgfscope}%
\begin{pgfscope}%
\pgftext[x=1.140913in,y=0.777778in,,top]{\sffamily\fontsize{10.000000}{12.000000}\selectfont \(\displaystyle p\)}%
\end{pgfscope}%
\begin{pgfscope}%
\pgftext[x=3.240549in,y=0.797944in,,top]{\sffamily\fontsize{10.000000}{12.000000}\selectfont \(\displaystyle v\)}%
\end{pgfscope}%
\begin{pgfscope}%
\pgfsetbuttcap%
\pgfsetroundjoin%
\definecolor{currentfill}{rgb}{0.000000,0.000000,0.000000}%
\pgfsetfillcolor{currentfill}%
\pgfsetlinewidth{0.803000pt}%
\definecolor{currentstroke}{rgb}{0.000000,0.000000,0.000000}%
\pgfsetstrokecolor{currentstroke}%
\pgfsetdash{}{0pt}%
\pgfsys@defobject{currentmarker}{\pgfqpoint{-0.048611in}{0.000000in}}{\pgfqpoint{0.000000in}{0.000000in}}{%
\pgfpathmoveto{\pgfqpoint{0.000000in}{0.000000in}}%
\pgfpathlineto{\pgfqpoint{-0.048611in}{0.000000in}}%
\pgfusepath{stroke,fill}%
}%
\begin{pgfscope}%
\pgfsys@transformshift{0.436337in}{0.875000in}%
\pgfsys@useobject{currentmarker}{}%
\end{pgfscope}%
\end{pgfscope}%
\begin{pgfscope}%
\pgftext[x=0.269670in,y=0.826806in,left,base]{\sffamily\fontsize{10.000000}{12.000000}\selectfont \(\displaystyle 0\)}%
\end{pgfscope}%
\begin{pgfscope}%
\pgftext[x=0.214114in,y=0.875000in,,bottom,rotate=90.000000]{\sffamily\fontsize{10.000000}{12.000000}\selectfont \(\displaystyle u_y(v, (x, p))\)}%
\end{pgfscope}%
\begin{pgfscope}%
\pgfpathrectangle{\pgfqpoint{0.295422in}{0.104444in}}{\pgfqpoint{3.100134in}{1.541111in}} %
\pgfusepath{clip}%
\pgfsetrectcap%
\pgfsetroundjoin%
\pgfsetlinewidth{1.505625pt}%
\definecolor{currentstroke}{rgb}{0.690196,0.200000,0.000000}%
\pgfsetstrokecolor{currentstroke}%
\pgfsetdash{}{0pt}%
\pgfpathmoveto{\pgfqpoint{0.436337in}{0.174495in}}%
\pgfpathlineto{\pgfqpoint{1.140913in}{0.875000in}}%
\pgfpathlineto{\pgfqpoint{3.254640in}{1.575505in}}%
\pgfpathlineto{\pgfqpoint{3.254640in}{1.575505in}}%
\pgfusepath{stroke}%
\end{pgfscope}%
\begin{pgfscope}%
\pgfsetrectcap%
\pgfsetmiterjoin%
\pgfsetlinewidth{0.803000pt}%
\definecolor{currentstroke}{rgb}{0.000000,0.000000,0.000000}%
\pgfsetstrokecolor{currentstroke}%
\pgfsetdash{}{0pt}%
\pgfpathmoveto{\pgfqpoint{0.436337in}{0.174495in}}%
\pgfpathlineto{\pgfqpoint{0.436337in}{1.575505in}}%
\pgfusepath{stroke}%
\end{pgfscope}%
\begin{pgfscope}%
\pgfsetrectcap%
\pgfsetmiterjoin%
\pgfsetlinewidth{0.000000pt}%
\definecolor{currentstroke}{rgb}{0.000000,0.000000,0.000000}%
\pgfsetstrokecolor{currentstroke}%
\pgfsetstrokeopacity{0.000000}%
\pgfsetdash{}{0pt}%
\pgfpathmoveto{\pgfqpoint{3.395556in}{0.104444in}}%
\pgfpathlineto{\pgfqpoint{3.395556in}{1.645556in}}%
\pgfusepath{}%
\end{pgfscope}%
\begin{pgfscope}%
\pgfsetrectcap%
\pgfsetmiterjoin%
\pgfsetlinewidth{0.803000pt}%
\definecolor{currentstroke}{rgb}{0.000000,0.000000,0.000000}%
\pgfsetstrokecolor{currentstroke}%
\pgfsetdash{}{0pt}%
\pgfpathmoveto{\pgfqpoint{0.436337in}{0.875000in}}%
\pgfpathlineto{\pgfqpoint{3.254640in}{0.875000in}}%
\pgfusepath{stroke}%
\end{pgfscope}%
\begin{pgfscope}%
\pgfsetrectcap%
\pgfsetmiterjoin%
\pgfsetlinewidth{0.000000pt}%
\definecolor{currentstroke}{rgb}{0.000000,0.000000,0.000000}%
\pgfsetstrokecolor{currentstroke}%
\pgfsetstrokeopacity{0.000000}%
\pgfsetdash{}{0pt}%
\pgfpathmoveto{\pgfqpoint{0.295422in}{1.645556in}}%
\pgfpathlineto{\pgfqpoint{3.395556in}{1.645556in}}%
\pgfusepath{}%
\end{pgfscope}%
\end{pgfpicture}%
\makeatother%
\endgroup%

%% file: figures/cvx_env_a.pgf
%% Creator: Matplotlib, PGF backend
%%
%% To include the figure in your LaTeX document, write
%%   \input{<filename>.pgf}
%%
%% Make sure the required packages are loaded in your preamble
%%   \usepackage{pgf}
%%
%% Figures using additional raster images can only be included by \input if
%% they are in the same directory as the main LaTeX file. For loading figures
%% from other directories you can use the `import` package
%%   \usepackage{import}
%% and then include the figures with
%%   \import{<path to file>}{<filename>.pgf}
%%
%% Matplotlib used the following preamble
%%   \usepackage{fontspec}
%%
\begingroup%
\makeatletter%
\begin{pgfpicture}%
\pgfpathrectangle{\pgfpointorigin}{\pgfqpoint{2.800000in}{1.730495in}}%
\pgfusepath{use as bounding box, clip}%
\begin{pgfscope}%
\pgfsetbuttcap%
\pgfsetmiterjoin%
\pgfsetlinewidth{0.000000pt}%
\definecolor{currentstroke}{rgb}{1.000000,1.000000,1.000000}%
\pgfsetstrokecolor{currentstroke}%
\pgfsetstrokeopacity{0.000000}%
\pgfsetdash{}{0pt}%
\pgfpathmoveto{\pgfqpoint{0.000000in}{0.000000in}}%
\pgfpathlineto{\pgfqpoint{2.800000in}{0.000000in}}%
\pgfpathlineto{\pgfqpoint{2.800000in}{1.730495in}}%
\pgfpathlineto{\pgfqpoint{0.000000in}{1.730495in}}%
\pgfpathclose%
\pgfusepath{}%
\end{pgfscope}%
\begin{pgfscope}%
\pgfsetbuttcap%
\pgfsetmiterjoin%
\definecolor{currentfill}{rgb}{1.000000,1.000000,1.000000}%
\pgfsetfillcolor{currentfill}%
\pgfsetlinewidth{0.000000pt}%
\definecolor{currentstroke}{rgb}{0.000000,0.000000,0.000000}%
\pgfsetstrokecolor{currentstroke}%
\pgfsetstrokeopacity{0.000000}%
\pgfsetdash{}{0pt}%
\pgfpathmoveto{\pgfqpoint{0.146967in}{0.104444in}}%
\pgfpathlineto{\pgfqpoint{2.695556in}{0.104444in}}%
\pgfpathlineto{\pgfqpoint{2.695556in}{1.626051in}}%
\pgfpathlineto{\pgfqpoint{0.146967in}{1.626051in}}%
\pgfpathclose%
\pgfusepath{fill}%
\end{pgfscope}%
\begin{pgfscope}%
\pgftext[x=2.568126in,y=0.302253in,,top]{\sffamily\fontsize{10.000000}{12.000000}\selectfont \(\displaystyle v\)}%
\end{pgfscope}%
\begin{pgfscope}%
\pgftext[x=0.214187in,y=0.865248in,,bottom,rotate=90.000000]{\sffamily\fontsize{10.000000}{12.000000}\selectfont \(\displaystyle u_y(v)\)}%
\end{pgfscope}%
\begin{pgfscope}%
\pgfpathrectangle{\pgfqpoint{0.146967in}{0.104444in}}{\pgfqpoint{2.548588in}{1.521606in}} %
\pgfusepath{clip}%
\pgfsetrectcap%
\pgfsetroundjoin%
\pgfsetlinewidth{1.505625pt}%
\definecolor{currentstroke}{rgb}{0.690196,0.200000,0.000000}%
\pgfsetstrokecolor{currentstroke}%
\pgfsetdash{}{0pt}%
\pgfpathmoveto{\pgfqpoint{0.408340in}{0.173608in}}%
\pgfpathlineto{\pgfqpoint{0.962733in}{0.424232in}}%
\pgfpathlineto{\pgfqpoint{2.579711in}{0.685298in}}%
\pgfpathlineto{\pgfqpoint{2.579711in}{0.685298in}}%
\pgfusepath{stroke}%
\end{pgfscope}%
\begin{pgfscope}%
\pgfpathrectangle{\pgfqpoint{0.146967in}{0.104444in}}{\pgfqpoint{2.548588in}{1.521606in}} %
\pgfusepath{clip}%
\pgfsetrectcap%
\pgfsetroundjoin%
\pgfsetlinewidth{1.505625pt}%
\definecolor{currentstroke}{rgb}{1.000000,0.580392,0.133333}%
\pgfsetstrokecolor{currentstroke}%
\pgfsetdash{}{0pt}%
\pgfpathmoveto{\pgfqpoint{0.881884in}{0.177025in}}%
\pgfpathlineto{\pgfqpoint{1.424726in}{0.603323in}}%
\pgfpathlineto{\pgfqpoint{2.579711in}{1.080702in}}%
\pgfpathlineto{\pgfqpoint{2.579711in}{1.080702in}}%
\pgfusepath{stroke}%
\end{pgfscope}%
\begin{pgfscope}%
\pgfpathrectangle{\pgfqpoint{0.146967in}{0.104444in}}{\pgfqpoint{2.548588in}{1.521606in}} %
\pgfusepath{clip}%
\pgfsetrectcap%
\pgfsetroundjoin%
\pgfsetlinewidth{1.505625pt}%
\definecolor{currentstroke}{rgb}{1.000000,0.784314,0.235294}%
\pgfsetstrokecolor{currentstroke}%
\pgfsetdash{}{0pt}%
\pgfpathmoveto{\pgfqpoint{1.574874in}{0.179352in}}%
\pgfpathlineto{\pgfqpoint{1.586424in}{0.206503in}}%
\pgfpathlineto{\pgfqpoint{1.597974in}{0.233654in}}%
\pgfpathlineto{\pgfqpoint{1.609524in}{0.260805in}}%
\pgfpathlineto{\pgfqpoint{1.621074in}{0.287955in}}%
\pgfpathlineto{\pgfqpoint{1.632624in}{0.315106in}}%
\pgfpathlineto{\pgfqpoint{1.644173in}{0.342257in}}%
\pgfpathlineto{\pgfqpoint{1.655723in}{0.369408in}}%
\pgfpathlineto{\pgfqpoint{1.667273in}{0.384028in}}%
\pgfpathlineto{\pgfqpoint{1.678823in}{0.398648in}}%
\pgfpathlineto{\pgfqpoint{1.690373in}{0.413267in}}%
\pgfpathlineto{\pgfqpoint{1.701923in}{0.427887in}}%
\pgfpathlineto{\pgfqpoint{1.713472in}{0.442507in}}%
\pgfpathlineto{\pgfqpoint{1.725022in}{0.457126in}}%
\pgfpathlineto{\pgfqpoint{1.736572in}{0.471746in}}%
\pgfpathlineto{\pgfqpoint{1.748122in}{0.486366in}}%
\pgfpathlineto{\pgfqpoint{1.759672in}{0.500986in}}%
\pgfpathlineto{\pgfqpoint{1.771222in}{0.515605in}}%
\pgfpathlineto{\pgfqpoint{1.782771in}{0.530225in}}%
\pgfpathlineto{\pgfqpoint{1.794321in}{0.544845in}}%
\pgfpathlineto{\pgfqpoint{1.805871in}{0.559464in}}%
\pgfpathlineto{\pgfqpoint{1.817421in}{0.574084in}}%
\pgfpathlineto{\pgfqpoint{1.828971in}{0.588704in}}%
\pgfpathlineto{\pgfqpoint{1.840521in}{0.603323in}}%
\pgfpathlineto{\pgfqpoint{1.852071in}{0.617943in}}%
\pgfpathlineto{\pgfqpoint{1.863620in}{0.632563in}}%
\pgfpathlineto{\pgfqpoint{1.875170in}{0.647183in}}%
\pgfpathlineto{\pgfqpoint{1.886720in}{0.661802in}}%
\pgfpathlineto{\pgfqpoint{1.898270in}{0.676422in}}%
\pgfpathlineto{\pgfqpoint{1.909820in}{0.691042in}}%
\pgfpathlineto{\pgfqpoint{1.921370in}{0.705661in}}%
\pgfpathlineto{\pgfqpoint{1.932919in}{0.720281in}}%
\pgfpathlineto{\pgfqpoint{1.944469in}{0.734901in}}%
\pgfpathlineto{\pgfqpoint{1.956019in}{0.749521in}}%
\pgfpathlineto{\pgfqpoint{1.967569in}{0.764140in}}%
\pgfpathlineto{\pgfqpoint{1.979119in}{0.778760in}}%
\pgfpathlineto{\pgfqpoint{1.990669in}{0.793380in}}%
\pgfpathlineto{\pgfqpoint{2.002219in}{0.807999in}}%
\pgfpathlineto{\pgfqpoint{2.013768in}{0.822619in}}%
\pgfpathlineto{\pgfqpoint{2.025318in}{0.837239in}}%
\pgfpathlineto{\pgfqpoint{2.036868in}{0.851859in}}%
\pgfpathlineto{\pgfqpoint{2.048418in}{0.866478in}}%
\pgfpathlineto{\pgfqpoint{2.059968in}{0.881098in}}%
\pgfpathlineto{\pgfqpoint{2.071518in}{0.895718in}}%
\pgfpathlineto{\pgfqpoint{2.083067in}{0.910337in}}%
\pgfpathlineto{\pgfqpoint{2.094617in}{0.924957in}}%
\pgfpathlineto{\pgfqpoint{2.106167in}{0.939577in}}%
\pgfpathlineto{\pgfqpoint{2.117717in}{0.954197in}}%
\pgfpathlineto{\pgfqpoint{2.129267in}{0.968816in}}%
\pgfpathlineto{\pgfqpoint{2.140817in}{0.983436in}}%
\pgfpathlineto{\pgfqpoint{2.152366in}{0.998056in}}%
\pgfpathlineto{\pgfqpoint{2.163916in}{1.012675in}}%
\pgfpathlineto{\pgfqpoint{2.175466in}{1.027295in}}%
\pgfpathlineto{\pgfqpoint{2.187016in}{1.041915in}}%
\pgfpathlineto{\pgfqpoint{2.198566in}{1.056535in}}%
\pgfpathlineto{\pgfqpoint{2.210116in}{1.071154in}}%
\pgfpathlineto{\pgfqpoint{2.221666in}{1.085774in}}%
\pgfpathlineto{\pgfqpoint{2.233215in}{1.100394in}}%
\pgfpathlineto{\pgfqpoint{2.244765in}{1.115013in}}%
\pgfpathlineto{\pgfqpoint{2.256315in}{1.129633in}}%
\pgfpathlineto{\pgfqpoint{2.267865in}{1.144253in}}%
\pgfpathlineto{\pgfqpoint{2.279415in}{1.158873in}}%
\pgfpathlineto{\pgfqpoint{2.290965in}{1.173492in}}%
\pgfpathlineto{\pgfqpoint{2.302514in}{1.188112in}}%
\pgfpathlineto{\pgfqpoint{2.314064in}{1.202732in}}%
\pgfpathlineto{\pgfqpoint{2.325614in}{1.217351in}}%
\pgfpathlineto{\pgfqpoint{2.337164in}{1.231971in}}%
\pgfpathlineto{\pgfqpoint{2.348714in}{1.246591in}}%
\pgfpathlineto{\pgfqpoint{2.360264in}{1.261211in}}%
\pgfpathlineto{\pgfqpoint{2.371813in}{1.275830in}}%
\pgfpathlineto{\pgfqpoint{2.383363in}{1.290450in}}%
\pgfpathlineto{\pgfqpoint{2.394913in}{1.305070in}}%
\pgfpathlineto{\pgfqpoint{2.406463in}{1.319689in}}%
\pgfpathlineto{\pgfqpoint{2.418013in}{1.334309in}}%
\pgfpathlineto{\pgfqpoint{2.429563in}{1.348929in}}%
\pgfpathlineto{\pgfqpoint{2.441113in}{1.363549in}}%
\pgfpathlineto{\pgfqpoint{2.452662in}{1.378168in}}%
\pgfpathlineto{\pgfqpoint{2.464212in}{1.392788in}}%
\pgfpathlineto{\pgfqpoint{2.475762in}{1.407408in}}%
\pgfpathlineto{\pgfqpoint{2.487312in}{1.422027in}}%
\pgfpathlineto{\pgfqpoint{2.498862in}{1.436647in}}%
\pgfpathlineto{\pgfqpoint{2.510412in}{1.451267in}}%
\pgfpathlineto{\pgfqpoint{2.521961in}{1.465887in}}%
\pgfpathlineto{\pgfqpoint{2.533511in}{1.480506in}}%
\pgfpathlineto{\pgfqpoint{2.545061in}{1.495126in}}%
\pgfpathlineto{\pgfqpoint{2.556611in}{1.509746in}}%
\pgfpathlineto{\pgfqpoint{2.568161in}{1.524365in}}%
\pgfpathlineto{\pgfqpoint{2.579711in}{1.538985in}}%
\pgfusepath{stroke}%
\end{pgfscope}%
\begin{pgfscope}%
\pgfpathrectangle{\pgfqpoint{0.146967in}{0.104444in}}{\pgfqpoint{2.548588in}{1.521606in}} %
\pgfusepath{clip}%
\pgfsetrectcap%
\pgfsetroundjoin%
\pgfsetlinewidth{1.505625pt}%
\definecolor{currentstroke}{rgb}{0.000000,0.000000,0.000000}%
\pgfsetstrokecolor{currentstroke}%
\pgfsetdash{}{0pt}%
\pgfpathmoveto{\pgfqpoint{0.262812in}{0.387310in}}%
\pgfpathlineto{\pgfqpoint{0.828754in}{0.387310in}}%
\pgfpathlineto{\pgfqpoint{0.840304in}{0.389920in}}%
\pgfpathlineto{\pgfqpoint{0.955803in}{0.442134in}}%
\pgfpathlineto{\pgfqpoint{1.244549in}{0.488753in}}%
\pgfpathlineto{\pgfqpoint{1.256099in}{0.494243in}}%
\pgfpathlineto{\pgfqpoint{1.417796in}{0.621225in}}%
\pgfpathlineto{\pgfqpoint{2.029938in}{0.874236in}}%
\pgfpathlineto{\pgfqpoint{2.041488in}{0.884380in}}%
\pgfpathlineto{\pgfqpoint{2.572781in}{1.556887in}}%
\pgfpathlineto{\pgfqpoint{2.572781in}{1.556887in}}%
\pgfusepath{stroke}%
\end{pgfscope}%
\begin{pgfscope}%
\pgfsetrectcap%
\pgfsetmiterjoin%
\pgfsetlinewidth{0.803000pt}%
\definecolor{currentstroke}{rgb}{0.000000,0.000000,0.000000}%
\pgfsetstrokecolor{currentstroke}%
\pgfsetdash{}{0pt}%
\pgfpathmoveto{\pgfqpoint{0.269742in}{0.173608in}}%
\pgfpathlineto{\pgfqpoint{0.269742in}{1.556887in}}%
\pgfusepath{stroke}%
\end{pgfscope}%
\begin{pgfscope}%
\pgfsetrectcap%
\pgfsetmiterjoin%
\pgfsetlinewidth{0.000000pt}%
\definecolor{currentstroke}{rgb}{0.000000,0.000000,0.000000}%
\pgfsetstrokecolor{currentstroke}%
\pgfsetstrokeopacity{0.000000}%
\pgfsetdash{}{0pt}%
\pgfpathmoveto{\pgfqpoint{2.695556in}{0.104444in}}%
\pgfpathlineto{\pgfqpoint{2.695556in}{1.626051in}}%
\pgfusepath{}%
\end{pgfscope}%
\begin{pgfscope}%
\pgfsetrectcap%
\pgfsetmiterjoin%
\pgfsetlinewidth{0.803000pt}%
\definecolor{currentstroke}{rgb}{0.000000,0.000000,0.000000}%
\pgfsetstrokecolor{currentstroke}%
\pgfsetdash{}{0pt}%
\pgfpathmoveto{\pgfqpoint{0.262812in}{0.369408in}}%
\pgfpathlineto{\pgfqpoint{2.579711in}{0.369408in}}%
\pgfusepath{stroke}%
\end{pgfscope}%
\begin{pgfscope}%
\pgfsetrectcap%
\pgfsetmiterjoin%
\pgfsetlinewidth{0.000000pt}%
\definecolor{currentstroke}{rgb}{0.000000,0.000000,0.000000}%
\pgfsetstrokecolor{currentstroke}%
\pgfsetstrokeopacity{0.000000}%
\pgfsetdash{}{0pt}%
\pgfpathmoveto{\pgfqpoint{0.146967in}{1.626051in}}%
\pgfpathlineto{\pgfqpoint{2.695556in}{1.626051in}}%
\pgfusepath{}%
\end{pgfscope}%
\end{pgfpicture}%
\makeatother%
\endgroup%

%% file: figures/cvx_env_b.pgf
%% Creator: Matplotlib, PGF backend
%%
%% To include the figure in your LaTeX document, write
%%   \input{<filename>.pgf}
%%
%% Make sure the required packages are loaded in your preamble
%%   \usepackage{pgf}
%%
%% Figures using additional raster images can only be included by \input if
%% they are in the same directory as the main LaTeX file. For loading figures
%% from other directories you can use the `import` package
%%   \usepackage{import}
%% and then include the figures with
%%   \import{<path to file>}{<filename>.pgf}
%%
%% Matplotlib used the following preamble
%%   \usepackage{fontspec}
%%
\begingroup%
\makeatletter%
\begin{pgfpicture}%
\pgfpathrectangle{\pgfpointorigin}{\pgfqpoint{2.800000in}{1.730495in}}%
\pgfusepath{use as bounding box, clip}%
\begin{pgfscope}%
\pgfsetbuttcap%
\pgfsetmiterjoin%
\pgfsetlinewidth{0.000000pt}%
\definecolor{currentstroke}{rgb}{1.000000,1.000000,1.000000}%
\pgfsetstrokecolor{currentstroke}%
\pgfsetstrokeopacity{0.000000}%
\pgfsetdash{}{0pt}%
\pgfpathmoveto{\pgfqpoint{0.000000in}{0.000000in}}%
\pgfpathlineto{\pgfqpoint{2.800000in}{0.000000in}}%
\pgfpathlineto{\pgfqpoint{2.800000in}{1.730495in}}%
\pgfpathlineto{\pgfqpoint{0.000000in}{1.730495in}}%
\pgfpathclose%
\pgfusepath{}%
\end{pgfscope}%
\begin{pgfscope}%
\pgfsetbuttcap%
\pgfsetmiterjoin%
\definecolor{currentfill}{rgb}{1.000000,1.000000,1.000000}%
\pgfsetfillcolor{currentfill}%
\pgfsetlinewidth{0.000000pt}%
\definecolor{currentstroke}{rgb}{0.000000,0.000000,0.000000}%
\pgfsetstrokecolor{currentstroke}%
\pgfsetstrokeopacity{0.000000}%
\pgfsetdash{}{0pt}%
\pgfpathmoveto{\pgfqpoint{0.146967in}{0.192778in}}%
\pgfpathlineto{\pgfqpoint{2.695556in}{0.192778in}}%
\pgfpathlineto{\pgfqpoint{2.695556in}{1.626051in}}%
\pgfpathlineto{\pgfqpoint{0.146967in}{1.626051in}}%
\pgfpathclose%
\pgfusepath{fill}%
\end{pgfscope}%
\begin{pgfscope}%
\pgftext[x=2.568126in,y=0.192778in,,top]{\sffamily\fontsize{10.000000}{12.000000}\selectfont \(\displaystyle v\)}%
\end{pgfscope}%
\begin{pgfscope}%
\pgftext[x=0.214187in,y=0.909414in,,bottom,rotate=90.000000]{\sffamily\fontsize{10.000000}{12.000000}\selectfont \(\displaystyle u_y(v)\)}%
\end{pgfscope}%
\begin{pgfscope}%
\pgfpathrectangle{\pgfqpoint{0.146967in}{0.192778in}}{\pgfqpoint{2.548588in}{1.433273in}} %
\pgfusepath{clip}%
\pgfsetrectcap%
\pgfsetroundjoin%
\pgfsetlinewidth{1.505625pt}%
\definecolor{currentstroke}{rgb}{0.000000,0.000000,0.000000}%
\pgfsetstrokecolor{currentstroke}%
\pgfsetdash{}{0pt}%
\pgfpathmoveto{\pgfqpoint{0.262812in}{0.277569in}}%
\pgfpathlineto{\pgfqpoint{0.828754in}{0.277569in}}%
\pgfpathlineto{\pgfqpoint{0.840304in}{0.280434in}}%
\pgfpathlineto{\pgfqpoint{0.955803in}{0.337725in}}%
\pgfpathlineto{\pgfqpoint{1.244549in}{0.388879in}}%
\pgfpathlineto{\pgfqpoint{1.256099in}{0.394903in}}%
\pgfpathlineto{\pgfqpoint{1.417796in}{0.534236in}}%
\pgfpathlineto{\pgfqpoint{2.029938in}{0.811855in}}%
\pgfpathlineto{\pgfqpoint{2.041488in}{0.822986in}}%
\pgfpathlineto{\pgfqpoint{2.572781in}{1.560902in}}%
\pgfpathlineto{\pgfqpoint{2.572781in}{1.560902in}}%
\pgfusepath{stroke}%
\end{pgfscope}%
\begin{pgfscope}%
\pgfpathrectangle{\pgfqpoint{0.146967in}{0.192778in}}{\pgfqpoint{2.548588in}{1.433273in}} %
\pgfusepath{clip}%
\pgfsetbuttcap%
\pgfsetroundjoin%
\pgfsetlinewidth{1.505625pt}%
\definecolor{currentstroke}{rgb}{0.376471,0.482353,0.545098}%
\pgfsetstrokecolor{currentstroke}%
\pgfsetdash{{5.550000pt}{2.400000pt}}{0.000000pt}%
\pgfpathmoveto{\pgfqpoint{0.269742in}{0.257926in}}%
\pgfpathlineto{\pgfqpoint{0.835684in}{0.257926in}}%
\pgfpathlineto{\pgfqpoint{0.847234in}{0.260791in}}%
\pgfpathlineto{\pgfqpoint{1.251479in}{0.369236in}}%
\pgfpathlineto{\pgfqpoint{1.263029in}{0.375260in}}%
\pgfpathlineto{\pgfqpoint{2.036868in}{0.792212in}}%
\pgfpathlineto{\pgfqpoint{2.048418in}{0.803343in}}%
\pgfpathlineto{\pgfqpoint{2.579711in}{1.541259in}}%
\pgfusepath{stroke}%
\end{pgfscope}%
\begin{pgfscope}%
\pgfsetrectcap%
\pgfsetmiterjoin%
\pgfsetlinewidth{0.803000pt}%
\definecolor{currentstroke}{rgb}{0.000000,0.000000,0.000000}%
\pgfsetstrokecolor{currentstroke}%
\pgfsetdash{}{0pt}%
\pgfpathmoveto{\pgfqpoint{0.269742in}{0.257926in}}%
\pgfpathlineto{\pgfqpoint{0.269742in}{1.560902in}}%
\pgfusepath{stroke}%
\end{pgfscope}%
\begin{pgfscope}%
\pgfsetrectcap%
\pgfsetmiterjoin%
\pgfsetlinewidth{0.000000pt}%
\definecolor{currentstroke}{rgb}{0.000000,0.000000,0.000000}%
\pgfsetstrokecolor{currentstroke}%
\pgfsetstrokeopacity{0.000000}%
\pgfsetdash{}{0pt}%
\pgfpathmoveto{\pgfqpoint{2.695556in}{0.192778in}}%
\pgfpathlineto{\pgfqpoint{2.695556in}{1.626051in}}%
\pgfusepath{}%
\end{pgfscope}%
\begin{pgfscope}%
\pgfsetrectcap%
\pgfsetmiterjoin%
\pgfsetlinewidth{0.803000pt}%
\definecolor{currentstroke}{rgb}{0.000000,0.000000,0.000000}%
\pgfsetstrokecolor{currentstroke}%
\pgfsetdash{}{0pt}%
\pgfpathmoveto{\pgfqpoint{0.262812in}{0.257926in}}%
\pgfpathlineto{\pgfqpoint{2.579711in}{0.257926in}}%
\pgfusepath{stroke}%
\end{pgfscope}%
\begin{pgfscope}%
\pgfsetrectcap%
\pgfsetmiterjoin%
\pgfsetlinewidth{0.000000pt}%
\definecolor{currentstroke}{rgb}{0.000000,0.000000,0.000000}%
\pgfsetstrokecolor{currentstroke}%
\pgfsetstrokeopacity{0.000000}%
\pgfsetdash{}{0pt}%
\pgfpathmoveto{\pgfqpoint{0.146967in}{1.626051in}}%
\pgfpathlineto{\pgfqpoint{2.695556in}{1.626051in}}%
\pgfusepath{}%
\end{pgfscope}%
\end{pgfpicture}%
\makeatother%
\endgroup%

%% file: dynamic.tex
\newcolumntype{P}[1]{>{\centering\arraybackslash}p{#1}} %% horizontal centering
\newcolumntype{M}[1]{>{\centering\arraybackslash}m{#1}} %% vertical centering

\section{Two-Stage Revenue Maximization}

We now turn to a setting in which the seller has two items to sell to the buyer
in succession. We will denote the buyer's values for the two items by $v_1$ and
$v_2$. The values are drawn from independent distributions with c.d.f.s $F_1$
and $F_2$, and the buyer's value for receiving both of the items is $v_1+v_2$.
The mechanism proceeds in two stages. In the first stage, the seller announces
a mechanism by which to sell the first item and the buyer reveals $v_1$. At
this time, neither the buyer nor the seller knows the buyer's value for the
second item.  The buyer must report his value to the first-stage mechanism
before learning $v_2$. At the start of the second stage, the seller announces a
second mechanism, which may depend on the buyer's first-stage decision, by
which to sell the second item.

\paragraph{Properties of two-stage mechanisms.} A two-stage
mechanism is {\em incentive-compatible} if for any report $v_1$
during the first stage, the second-stage mechanism is incentive-compatible 
with respect to reporting $v_2$, and the combination of the first- and
second-stage mechanisms is incentive-compatible with respect to
reporting $v_1$. We further impose the constraint of {\em ex-post IR} which
states that the price charged to the buyer in either stage cannot
exceed the ex-post value obtained by the buyer in that stage.

Recall that $\opt(\wt, F_1)$ and $\opt(\wt, F_2)$ denote the optimal
revenue that the seller can obtain by selling items 1 and 2
respectively via independent mechanisms (such as posting a fixed price
in each stage). We will denote by $\opt(\wt, F)$ the optimal revenue
achievable by an incentive-compatible ex-post IR mechanism for the two
stages combined, where $F=F_1\times F_2$ denotes the joint
distribution over $(v_1, v_2)$. 

Of course, $\opt(\wt, F)\ge \opt(\wt, F_1)+\opt(\wt, F_2)$, but in
fact, the former can be much larger than the latter sum. Observe that
the second-stage mechanism can depend on the buyer's first-stage
report. This gives the seller some flexibility in extracting more
revenue in the first stage. \citet{ADH16} show, in particular, that for
risk-neutral buyers the seller can charge a premium on the first stage
in exchange for more utility in the second stage, which in some settings
allows the seller to extract almost the entire second-stage social
welfare as revenue. We will show that a similar result is achievable
under our model of risk aversion.

A two-stage mechanism can be described without loss of generality as a menu of
options with each option being a three-tuple of random variables $(X,P,M)$,
where $X$ is an indicator variable representing the allocation of item 1 to the
agent, $P$ is the price to be paid in stage one, and $M$ is an
incentive-compatible mechanism for the second stage. Let $U(v_2,M)$ be a random
variable denoting the ex-post utility that the agent obtains from mechanism $M$
in stage two. Then, the buyer's risk-averse utility from the menu option
$(X,P,M)$ in stage one is given by 
\[
    \rae{v_1X - P + U(v_2,M)},
\]
where the weighted expectation is taken over the randomness in the mechanisms
as well as the randomness in the agent's second-stage value. In the first
stage, the agent chooses the menu option that maximizes his risk-averse
utility.

Observe that once again $X, P,$ and $M$ can be arbitrarily correlated. In
particular, the contribution of the second-stage mechanism $M$ to the buyer's
risk-averse utility in stage one can depend not only on the chosen menu option
and his expected $v_2$, but also on his actual value $v_1$. This makes it
challenging to reason about the choices of the agent and account for the
contribution of the agent's second-stage utility to the first-stage revenue, as
we see in the following example.

\begin{example} %\kgnote{Someone please read over this.}
    Consider a menu option that with probability $x$ allocates the first item to 
    the bidder and charges him $p$, and then always gives the second item away
    for free.  Then with probability $1-F_2(v_2)$, the buyer gets utility of at least
    $v_2$ from the second item.  Suppose $v_2 \sim U\{1, 2\}$. Let
    $\wt(x)=x^2$ for all $x$. We will
    compute the utility of the buyer from this menu option at
    different first-stage values. Observe that although the menu
    option, in particular the second-stage mechanism, stays the same,
    the contribution of the first- and second-stage mechanisms to the
    buyer's utility vary as $v_1$ varies.

    Case 1: $v-p \geq 2$. Getting the first item is worth more than any
    second-stage utility alone.  Then the buyer's utility is
    \begin{align*}
      & y(1)1 + y((1-x)/2) (2-1) \\
      & \quad \quad + y(x)(v-p - 2) + y(x/2)(2) \\
      & = x^2(v-p) + 1 + \frac{(1-x)^2}{4} - \frac{3x^2}{2}.
%      & 1 + \frac{(1-x)^2}{4} - \frac{3x^2}{2}.$$ % = 1/4(4 + (1-x)^2 - 6x^2)
    \end{align*}
    % The gain in first-stage utility due to the mechanism is:
   
    Case 2: $v-p \in (1,2)$. Getting the second item when his value is high is
    worth more than just getting the first item.  His utility from this
    mechanism is
    \begin{align*}
        &y(1)1 + y((1-x)/2) (v-p-1) + y(1/2)(2-(v-p)) + y(x/2)(v-p) \\
      &\qquad = \frac{x(x-1)}{2}(v-p) + \frac{3}{2} - \frac{(1-x)^2}{4}.% - \frac{2(1-x)(v-p)}{2}.
    \end{align*}
    In the case where $p=1$, $x=1/2$, $y=x^2$, and $v$ is 4 in the first case
    and 2.5 in the second, we get that the first case has utility $12/16 +
    11/16 = 23/16$ and the second case has utility $- 3/16+
    23/16 = 5/4$.
\end{example}

We focus on a simple and practical class of mechanisms, namely posted-price
mechanisms, that in addition to $\opt(\wt, F_1)+\opt(\wt, F_2)$ can in some
cases obtain an additional $\rae{v_2}$ in revenue, matching results known for
the risk-neutral setting.

\subsection{Posted-price mechanisms and their revenue properties.}
A two-stage posted-price mechanism is specified by a menu, where each
menu option is a pair of prices $(p_1, p_2)$. If the buyer selects
this menu option, he is offered item 1 at a price of $p_1$ and
promised item 2 at a price of $p_2$. Observe that the buyer makes this
choice knowing $v_1$ but not knowing $v_2$. The buyer would
potentially be willing to pay a higher price for item 1 if in return
he is promised a lower price for item 2. Accordingly, the undominated
menu options\footnote{A menu option is dominated by another if the
  buyer prefers the latter to the former regardless of his value for
  the first item.} correspond to higher first-stage prices being coupled
with lower second-stage prices and vice versa.

In the remainder of this section, we use the notation $\mechl$ to
represent a two-stage posted-price mechanism that offers menu options
$(p, \lp)$ for every price $p$ in some range, where $\lp[\cdot]$ is a
non-increasing function mapping the first-stage price to the
corresponding second-stage price.\footnote{Observe that this notation
  captures menus with a finite number of options. In particular, if
  the function $\lp$ is constant over a range of prices $p$, then all
  options other than the smallest price in that range are dominated,
  and effectively not present in the menu.}

\paragraph{The buyer's optimization problem.} Fix a posted-price mechanism
$\mechl$, and consider a buyer with probability weighting function $\wt$ and
first-stage value $v_1$. Observe that if the buyer purchases the menu option
$(p,\lp)$, he gets utility of $v_1-p$ with probability 1, and expects to obtain
some (random) utility from the second-stage posted price of $\lp$. The
risk-averse expectation of the buyer's second-stage utility from posted price
$p_2$ can be written as
\begin{align*}
  \g(p_2) & := \rae{\max(0,v_2-p_2)} \\ &= \int_{0}^{\infty}
  \wt(1-F_2(z+p_2)) dz \\
  &= \int_{p_2}^{\infty} \wt(1-F_2(z)) dz.
\end{align*}
Accordingly, the buyer's risk-averse utility in the first stage from
purchasing option $(p,\lp)$ is
\begin{align*}
%  U(v_1, p) = 
v_1-p+\g(\lp). %= v_1-H(p)
\end{align*}
%where $H(p)$ is defined as above to be the ``effective price'' paid by
%the buyer in the first stage.
The menu option $(p,\lp)$ gives the buyer the same utility as offering an
``effective price" of $p-\g(\lp)$ in a single-shot mechanism.  Accordingly, the
buyer chooses the menu option corresponding to the minimal $p - \g(\lp)$ over
all prices that he can afford, that is, with $p\le v_1$. Then without loss of
generality, the mechanism contains menu options with effective prices that are
non-increasing in the first-stage price, as otherwise they would be dominated.
We assume without loss of generality that the buyer breaks ties across menu
options with equal effective prices in favor of the largest first-stage price.

% Independent of the buyer's first-stage value $v_1$, the buyer
% will choose a menu option with minimal $p - \g(\lp)$, where this 
% quantity gives the buyer the same utility as offering an ``effective price" of 
% $p-\g(\lp)$ in a single-shot mechanism.  Then without loss
% of generality, the mechanism contains menu options that all have
% equal effective prices, as otherwise they would be dominated.
% We assume without loss of generality that the buyer purchases 
% the remaining option corresponding to the largest price.

% commented out
\begin{comment}
The buyer chooses a menu option with the smallest effective
price. Observe that this choice is independent of his first-stage
value. Let $\peff$ denote the set of all first-stage prices
corresponding to the smallest effective price. If $|\peff|>1$, we assume
without loss of generality that the buyer purchases the option
corresponding to the largest price $p\in\peff$ with $p\le v_1$.
\end{comment}
%

\begin{example}
    Consider a buyer with $v_1, v_2 \sim U[0,1]$ and probability weighting
    function $\wt(x)=x^2$.  Consider the mechanism that offers menu options
    $(0,1)$, $(\frac{1}{6}, \frac{1}{2})$, and $(\frac{1}{3}, 0)$.  Note that
    $\g(\lp) = \int _{\lp} ^1 (1-v_2)^2 dv_2$.  Then $\g(1) = 0$, $\g(0) =
    \frac{1}{3}$, and $\g(\frac{1}{2}) \approx 0.04$.  This gives
    \begin{center}
    \renewcommand{\arraystretch}{1.3}
    \begin{tabular}{@{}M{2cm}M{5cm}@{}}
        \toprule 
        Option & Utility  \\ \midrule
        $\displaystyle\left(0,1\right)$ &
                $\displaystyle v-0+\g(1) = v$ \\
        $\displaystyle\left(\tfrac16, \tfrac12\right)$ &
                $\displaystyle v-\tfrac16+\g\left(\tfrac12\right)\approx v-0.13$ \\
        $\displaystyle\left(\tfrac13, 0\right)$ &
                $\displaystyle v - \tfrac13 + \g(0) = v$ \\ \bottomrule
    \end{tabular}
    \end{center}
    Then a buyer with $v \in [0, \frac{1}{3})$ will purchase the option
    $(0,1)$; a buyer with $v \in [\frac{1}{3}, 1]$ will purchase the option
    $(\frac{1}{3}, 1)$; and no buyer will purchase the option $(\frac{1}{6},
    \frac{1}{2})$, as it is dominated by the other options with cheaper
    effective prices of $0 < 0.13$.
\end{example}

\subsubsection*{The seller's revenue.} We now present an upper bound on the
revenue achievable via two-stage posted-price mechanisms. 

\begin{theorem}
\label{lem:pp-ub}
  The revenue of any two-stage posted-price mechanism for a buyer with
  value distribution $F_1\times F_2$ and probability weighting
  function $\wt$ is upper-bounded by 
  \[\myer(F_1) + \myer(F_2) +
  \expect[v_1\sim F_1]{\min(v_1, \rae{v_2})}. \]
 % where the first two terms
 %  denotes the single-shot revenue on each stage, and $\g(0)=\rae{v_2}$
 %  is the risk-averse expectation of the buyer's second-stage value.
\end{theorem}
 
\begin{proof}
  We will account for the seller's revenue in the two stages
  separately. Observe first that regardless of the buyer's first-stage
  value, the revenue obtained by the seller in the second stage is no
  more than $\myer(F_2)$.

  %commented out old version
  \begin{comment}
  Now let's consider the seller's first-stage revenue. Let $\pmin$
  denote the smallest price in the set $\peff$. Observe that since the
  effective price at each first-stage price in $\peff$ is equal, for any
  $p\in\peff$, we have
  \begin{align*}
    p = \pmin - \g(\lp[\pmin]) + \g(\lp) \le \pmin + \g(0),
  \end{align*}
  where $\g(0) = \rae{v_2}$ is the risk-averse expectation of the
  buyer's second-stage value. 
  \end{comment}
  
  %Kira's new version:
  Now let's consider the seller's first-stage revenue. Let $\pmin$
  denote the smallest price offered in stage one.
  Because the ``effective price'' is non-increasing as a function of the
  first-stage price, we have %all menu options have equal ``effective price," then
  $p -  \g(\lp) \le \pmin - \g(\lp[\pmin])$, hence
%  If the buyer prefers the option $(p,\lp)$ to $(\pmin,\lp[\pmin])$, 
%  it must be that
%  $$v_1 - p + \g(\lp) \geq v_1 - \pmin + \g(\lp[\pmin]),$$
%  hence,
  $$p \le \pmin - \g(\lp[\pmin]) + \g(\lp) \le \pmin + \g(0),$$
  where $\g(0) = \rae{v_2}$ is the risk-averse expectation of the
  buyer's second-stage value.

  On the other hand, the buyer never pays more than $v_1$ in the first
  stage. Therefore, the seller's first-stage revenue, when the buyer's
  first-stage value is $v_1\ge\pmin$, is bounded by $\min(v_1,
  \pmin+\g(0))$. We can now bound the seller's first-stage revenue by
  \begin{align*}
    & \expect[v_1\sim F_1]{\min(v_1, \pmin+\g(0))} \\ & \quad \le
    \pmin(1-F_1(\pmin))  + \expect[v_1\sim F_1]{\min(v_1, \g(0))} \\
    & \quad \le \myer(F_1) + \expect[v_1\sim F_1]{\min(v_1, \g(0))} 
  \end{align*}
\end{proof}

We will now show that there exists a simple posted-pricing mechanism
that achieves a 2-approximation to the upper bound in
Theorem~\ref{lem:pp-ub}.

\begin{theorem}
\label{lem:pp-lb}
  For the two-stage setting described above, there exists a posted-price
  mechanism $\mechl$ that obtains revenue at least
  \[\frac 12\left(\myer(F_1) + \myer(F_2) +
  \expect[v_1\sim F_1]{\min(v_1, \rae{v_2})} \right). \]
\end{theorem}

\begin{proof}
  Charging the optimal single-shot posted-price in each stage already
  obtains revenue $\myer(F_1) + \myer(F_2)$. We will now describe a
  posted-price mechanism $\mechl$ that obtains revenue at least
  $\expect[v_1\sim F_1]{\min(v_1, \g(0))}$. The intuition that a 
  mechanism can achieve this is as follows:
  if every menu option charges a price $p$ but guarantees utility
  equal to $p$ back in the next stage, then the buyer will be willing to pay 
  any price subject to ex-post IR.  The better of these two
  mechanisms achieves the bound stated in the lemma.
  % Myerson's prices give \rev1+\rev2. Remains to show that we can get
  % the third term. Offer lotteries $(p, G^{-1}(p))$. All bring the same
  % utility. So agent buys the most expensive one. This has price
  % min(v_1, G(0)).

  The mechanism $\mechl$ offers menu options $(p,\lp)$ with $\lp =
  \g^{-1}(p)$ for all $p\in [0,\g(0)]$. Observe that since $\g$ is
  continuous and ranges from $\g(\infty)=0$ to $\g(0)$, for every $p$
  in the range $[0,\g(0)]$, a second-stage price $\lp=\g^{-1}(p)$
  exists, and therefore the mechanism is properly defined.

  Furthermore, for every menu option, $(p,\lp)$, we have %$H(p) = 
  $p - \g(\lp) = p-p=0$. So all menu options bring the same effective
  utility to the buyer on the first stage, and by default the buyer
  purchases the most expensive one that he can afford. Consequently,
  the seller's first-stage revenue is given by $\min(v_1, \g(0))$, and
  the theorem follows.
\end{proof}

%\subsection{Beating the revenue of posted price mechanisms}

\subsection{Risk-robust approximation.}
\label{sec:risk-robust-lb}
We now turn to risk-robust approximation in the two-stage
setting. Observe that the results of Section~\ref{sec:single-shot}
already imply that we can obtain a risk-robust approximation to the
single-shot revenue achievable in each stage independently, when the
buyer's weighting function is bounded
(Definition~\ref{def:boundedness}). Can we obtain a risk-robust
approximation to the last term in the bound given by
Lemma~\ref{lem:pp-ub}, namely, $\expect{\min(v_1, \rae{v_2})}$?

In this section we argue that this last term cannot be extracted via a
posted-pricing mechanism in a risk-robust manner even if all of the
possible weighting functions for the buyer are
bounded.\footnote{Observe, of course, that if the risk averse
  expectation of the buyer's second-stage value does not differ much
  across the different weighting functions, then we can use ideas from
  Section~\ref{sec:single-shot} to extract the optimal revenue in a
  risk-robust manner.} This fact leads to Theorem~\ref{thm:2day-lb}.
\begin{theorem}
  \label{thm:2day-lb}
  No posted-price mechanism can obtain a constant-factor risk-robust
  approximation to revenue in the two-stage dynamic setting. This
  holds even if all of the relevant weighting functions
  are $\Theta(1)$-bounded.
\end{theorem}

At a high-level, the idea behind our construction is as follows. We choose the
family of weighting functions and the second-stage value distribution in such a
way that although all of the weighting functions satisfy the boundedness
property, they cover a large range of weighted expectations for the
second-stage value, placing different constraints on the first-stage menu.
Intuitively, in order to extract enough revenue, the seller must offer a menu
with many different prices, indeed a continuum of first-stage prices. Then, to
incentivize the buyer to pay as high a price as he can afford in the first
stage, the seller must provide a discount over the second stage's price. The
extent of discount provided depends on the most risk-averse weighting function
for which the effective menu contains the corresponding option.  As the buyer
goes from being very risk-averse to almost risk-neutral, the seller needs to
offer a bigger and bigger discount for higher and higher first-stage prices,
and eventually runs out of discounts to offer.

We now make this formal.  Consider a two-stage mechanism design setting
where the buyer's value for the first stage is distributed according to the
unbounded equal revenue distribution, that is, $F_1(v_1) = 1-1/v_1$ for $v_1\ge
1$, and his second-stage value is distributed according to the equal revenue
distribution bounded at $e^n$, that is, $F_2(v_2) = 1-1/v_2$ for $v_2\in
[1,e^n]$. We will consider a family $\wtfam$ of weighting functions
parameterized by $\eps\in[0, 1]$ as follows.
\begin{align*}
  \wte(x) & = \begin{cases}
      x^2 &\text{for } x\in [0, \eps] \\
%      \frac{1-x}{1-\eps}\eps^2+\frac{x-\eps}{1-\eps} &\text{for } x \in [\eps,1]
      (1+\eps)x - \eps &\text{for } x \in [\eps,1]
   \end{cases}
\end{align*}
In words, $\wte(x)$ is equal to $x^2$ up to $\eps$, and then rises linearly to
$\wte(1)=1$. Observe that each function in this family is convex and at least
$1/8$-bounded.

Suppose that $\menu=\{(p,\lp)\}_{p\in P}$ for some $P \subset \R_{\ge 0}$ is a
menu that achieves a risk-robust $c$-approximation, $c > 1$, with respect to
the family $\wtfam$. Let $P_\eps \subseteq P$ index the ``effective menu" when
the buyer's weighting function is $\wte$, i.e., the set of first-stage prices
corresponding to menu options that the buyer actually purchases at some value.
% Observe that $P_\eps$ is a subset of first-day prices, and not a subset of the
% menu options in $\menu$; for each $p \in P_\eps$, the second-stage price is
% given by $\lp$.
% In particular, the
% effective price for any $p\in\menu_\eps$ is no larger than the effective price
% at $p'<p$.
Any option indexed by $p \in P \setminus P_\eps$ is dominated.

Recall that $\g_{\eps}(\lp)$ is the risk-averse utility that a buyer with
weighting function $\wte$ obtains from the second-stage mechanism when he
chooses first-stage price $p$:
\begin{equation*}
    \g_{\eps}(\lp) = \begin{cases}
        \int_{\lp}^{e^n} \wte(1/v_2) dv_2 &\text{for } \lp \geq 1 \\
        1 - \lp + \int_1^{e^n}\wte(1/v_2) dv_2 &\text{otherwise.}
    \end{cases}
\end{equation*}

\noindent 
We make the following observations about effective menus. See
Appendix~\ref{sec:app-risk-robust} for proofs.
\begin{lemma} 
\label{lem:lb-observations}
For the setting described above, if $\left\{(p,\lp)\right\}_{p\in P}$ gives a
risk-robust $c$-approximation, the following properties hold without loss of
generality:
\begin{enumerate}
    % \item For $\eps<\eps'$, $P_{\eps'}\subseteq P_{\eps}$. 
    % \item \label{lem:contiguous} We may assume without loss of generality that
    %     effective menus are connected; adding missing points only improves
    %     revenue at all risk profiles in $\wtfam$.
    % \item \label{lem:starts-at-1} It is without loss of generality to assume
    %     that the effective menu is of the form $[1,p]$.
    \item \label{lem:lp-lb} For all $p \in P$, $\lp \geq 1$.
    \item \label{lem:peps} For any $\eps$, $P_{\eps} \supseteq [1,p_\eps]$ where
        $p_\eps$ is defined such that 
        \[ \label{eq:peps}
             1 + \expect[v\sim F_1]{\min(v,p_{\eps})}
            \;=\; \frac1c \expect[v\sim F_1]{\min(v,\g_{\eps}(0))}.
        \]
    \item \label{lem:pstar} For every $\eps$, $p_\eps = \alpha_c
        \g_{\eps}(0)^{1/c}$ for some constant $\alpha_c > 0$ depending only
        on $c$.
    \item \label{lem:derivative} For every $\eps$, the left derivative of
        $\g_{\eps}(\lp)$ with respect to $p$ at $p=p_\eps$ must be $\ge 1$. 
\end{enumerate}
\end{lemma}
Informally, the first property holds because second-stage prices below
1 lose as much second-stage revenue as they gain on the first stage;
the second property is a consequence of minimizing the required
second-stage discounts; the third follows by solving
Equation~\eqref{eq:peps}; the fourth follows from the assumption that
prices in $P_\eps$ are undominated.

We can now derive a differential equation for the function $\lp$. First,
\begin{align}
    \notag
    \frac{d\, \g_\eps(\lp)}{dp} &= \begin{cases}
        -\wte(1/\lp) \lpp &\text{when } \lp\ge 1 \\
        -\lpp & \text{ otherwise.}
    \end{cases}
\intertext{Then, Lemma~\ref{lem:lb-observations}~\eqref{lem:derivative} implies}
    \label{eq:DE}
    -\lpp &\ge \begin{cases}
        \frac{1}{\wt_{\eps_p}(1/\lp)} &\text{when } \lp\ge 1\\
        1 &\text{ otherwise,}
    \end{cases}
\end{align}
where, for any price $p$, $\eps_p$ is the value of $\eps$ for which
$p=p_\eps$ as given by Equation~\eqref{eq:peps}. We have the following two
boundary conditions:
\begin{align}
    \label{eq:boundary-low}  \lp[1] & \le e^n & & \text{and} \\
    \label{eq:boundary-high}  \lp & > 1 & & \text{for } p < p_0.
\end{align}
The first enforces that the second-stage price at $p=1$ be no more than the
maximum value that $v_2$ takes. The second follows from
Lemma~\ref{lem:lb-observations}~\eqref{lem:lp-lb}.

We use $\lpb$ to denote the solution to Equations~\eqref{eq:DE} and
\eqref{eq:boundary-low}, each with inequality replaced by equality. So $\lpb$
gives an upper bound on $\lp$. We will show that, for large enough $n$, a
solution to \eqref{eq:DE} which satisfies \eqref{eq:boundary-low} cannot also
satisfy \eqref{eq:boundary-high}.

The following claim will be useful; a proof appears in
Appendix~\ref{sec:app-risk-robust}.
\begin{lemma}
    \label{lem:g0-lb}
    For all $\eps$, $\g_\eps(0) \geq \min\{\ln1/\eps, n\}$.
\end{lemma}

\noindent
We are now ready to prove Theorem~\ref{thm:2day-lb}.

\begin{proof}{Theorem~\ref{thm:2day-lb}}
    By Lemma~\ref{lem:lb-observations}~\eqref{lem:pstar}, $p_\eps = \alpha_c
    \g_\eps(0)^{1/c}$.  Fix $\eps^* = e^{-(2/\alpha_c)^c}$, and let $N =
    (2/\alpha_c)^c$. Then, by Lemma~\ref{lem:g0-lb}, for all $n > N$,
    \begin{equation*}
        p_{\eps^*} \geq \alpha_c (\ln 1/\eps^*)^{1/c} = 2.
    \end{equation*}
    Note that $\eps^* \le \eps_p \le 1$ for all $p \in [1,2]$.

    We now show that $\lpb[2]$ is at most $1/\eps^*$ for all $n > N$.
    Suppose not. Then, for all $p \in [1,2]$, $\lpb \geq \lpb[2]$ and
    $\eps_p \geq \eps^*$ implies that $\lpb>1/\eps_p$, or
    $1/\lpb<\eps_p$. So by the definition of $\wte$ we have
    $\wte[\eps_p](1/\lpb) = \lpb^{-2}$. Thus the differential
    equation~\eqref{eq:DE} simplifies to $-\lpbp = \lpb^2$ for
    $p\in [1,2]$. The solution, incorporating the boundary condition
    $\lpb[1]= e^n$, is
    \[
        \lpb = \frac1{e^{-n} + p - 1},
    \]
    and so $\lpb[2] = 1/(e^{-n} + 1) < 1$, a contradiction.

    But if $\lpb[2]$ is bounded above by $1/\eps^*$, we can argue that
    $\lp\le\lpb = 1$ at some $p < p_0$, contradicting the boundary
    condition~\eqref{eq:boundary-high}.  Specifically, Equation
    \eqref{eq:DE} gives $\lpbp \leq -1$ for all $\lpb \geq 1$, so
    $\lpb[1/\eps^* + 1] \leq \lpb[2] - (1/\eps^* - 1) \le 1$. On the
    other hand, by Lemma~\ref{lem:lb-observations}~\eqref{lem:pstar},
    $p_0 \propto n^{1/c}$, so for large enough $n$ we have $p_0 >
    1/\eps^* + 1$.
\end{proof}

%% file: app-proofs.tex
\section{Deferred proofs}
\label{sec:app-proofs}

\subsection{Utility curves and a characterization of optimal
  single-shot mechanisms.}

  The proofs in this section lead up to the conclusion of 
  Theorem~\ref{thm:single-shot-opt}:
  both the optimal mechanism and the risk-averse utility 
  that the buyer gets from it have a simple structure.

\begin{numberedlemma}{\ref{lem:concaveUtil}}
%\begin{lemma}{\ref{lem:concaveUtil}}
  For any profile $\wt$ and lottery $\lotteryp$, $\lrau{v}$ is a
  concave function of $v$. The slope of this
  function lies between $1 - \wt(1-x)$ and $\wt(x)$, where $x =
  \prob{\alloc = 1}$.
\end{numberedlemma}
%\end{lemma}
\begin{proof}
    We assume that $\lotteryp$ satisfies $\prob{\paymt=0
    \given \alloc=0} = 1$. Let $\pdist(p) = \prob{P \leq p \given \alloc=1}$.
    We further assume that $\pdist$ is differentiable, with p.m.f. $\ppmf$. We
    make these assumptions for simplicity; both can be relaxed. By definition,
    \begin{align*}
       \lrau{v} = -\int_0^\infty \left(1-\wt(\prob{vX-P \geq -z}) \right) dz + 
                \int_0^\infty \wt(\prob{vX-P \geq z}) dz.
    \end{align*}
    For $z \geq 0$,
    \begin{align*}
        \prob{vX-P \geq -z} &= 1 - \prob{X=1}\prob{P \geq v+z \given X=1} \\
            &= 1 - x(1 - \pdist(v+z)),
    \intertext{and}
        \prob{vX-P\geq z} &= 1-\prob{X=1}\prob{P\geq v-z \given X=1} -
                \prob{X=0} \\
            &= 1 - x(1 - \pdist(v-z)) - (1-x) \\
            &= x\pdist(v-z).
    \end{align*}

    Therefore,
    \begin{align*}
        \drau{v} &= \int_0^\infty \wt'\left(1 -
                x(1-\pdist(v+z))\right)x\ppmf(v+z)dz \\ & \quad +
                \int_0^\infty \wt'\left(x\pdist(v-z)\right)x\ppmf(v-z)dz \\
            &= \left.\wt\big(1-x+x\pdist(v+z)\big)\right|_{z=0}^\infty \;-\;
                \left.\wt\big(x\pdist(v-z)\big)\right|_{z=0}^\infty \\
            &= 1 - \wt(1-x+x\pdist(v)) + \wt(x\pdist(v)).
    \end{align*}

    Note that when $\pdist(v) = 1$ (i.e., when $v$ is greater than the maximum
    price charged), $\drau{v} = \wt(x)$. Similarly, when $\pdist(v) = 0$,
    $\drau{v} = 1-\wt(1-x)$. In either case, $\drau{v} \geq 0$.

    Finally, to show concavity, we show $\ddrau{v} \leq 0$:
    \[
        \ddrau{v} = x\ppmf(v)\left[\wt'(x\pdist(v))-\wt'(1-x+x\pdist(v))\right].
    \]
    The first term, $x\ppmf(v)$, is always nonnegative. Note that $1-x +
    x\pdist(v) \geq x\pdist(v)$. Since $\wt$ is convex and increasing,
    $\wt'(1-x + x\pdist(v)) \geq \wt'(x\pdist(v))$, and so $\ddrau{v} \leq 0$.
\end{proof}

\begin{lemma}
    \label{lem:util-dominance}
    Fix any (allocation, payment) pair $\lotteryp$. Let $(x,p)$ be the lottery
    that sells with probability $x = \prob{X=1}$ and charges $p = \expect{P}/x$.
    Then, assuming payments are non-negative, $\lrau{v}{x,p} \geq \lrau{v}$ for all $v$.
\end{lemma}

\begin{proof}    
    First observe that $\expect{vX-P} = x(v-p)$, but we can also write it as
    \[
        \expect{vX-P} = \int _0 ^\infty \prob{Z > z} dz  -
                \int_{-\infty}^0(1-\prob{Z > z})dz
    \]
    where $Z = vX-P$ and it may range from $-\infty$ to $\infty$.
    Then
    \begin{align*}
        \lrau{v}{x,p} &= y(x)(v-p) \\
        &= \frac{y(x)}{x} x(v-p) \\
        &= \frac{y(x)}{x} \expect{vX-P}.
    \end{align*}
    Recall that $y(x)/x$ is non-decreasing in $x$ by convexity of $y$.  As in
    the proof of Lemma~\ref{lem:concaveUtil}, we assume $\lotteryp$ satisfies
    $\prob{\paymt=0 \given \alloc=0} = 1$ (this can be relaxed). Then if $Z >
    0$, it must be that $X = 1$, hence $\prob{Z > 0} \leq x$.  Then for the
    positive values that $vX-P$ takes on, we have
    \begin{align*}
        \int _0 ^{\infty} \frac{y(x)}{x} \prob{Z > z} dz &\geq
                \int _0^\infty \frac{y(\prob{Z > z})}{\prob{Z > z}}
                \prob{Z > z} dz \\
            &= \int _0 ^\infty y(\prob{Z > z}) dz.
    \end{align*}
    
    For the negative values that $Xv-P$ takes on, by monotonicity of $y(x)/x$
    in $x$ and because the utility is negative, we have
    \begin{align*}
        -\frac{y(x)}{x} \int_{-\infty}^0(1-\prob{Z > z})dz &\geq 
                - \int_{-\infty}^0 \frac{y(\prob{Z > z})}{\prob{Z > z}}
                (1-\prob{Z > z})dz \\
            &\geq -\int_{-\infty}^0(1-\wt(\prob{Z > z}))dz
    \end{align*}
    where the the second inequality follows from 
    $-\frac{y(\prob{Z > z})}{\prob{Z > z}} \geq -1$.
    All together, 
    \begin{align*}
        \lrau{v}{x,p} &\;=\; \frac{y(x)}{x} \expect{Xv-P} \\
            &\;\geq\; \int_0^\infty y(\prob{Z > z})dz -
                \int_{-\infty}^0(1-\wt(\prob{Z > z}))dz \\
            &\;=\; \lrau{v}.
    \end{align*}
\end{proof}

\begin{lemma}
    \label{lem:rev-dominance}
    Fix $\wt$, and let $\lotteryp$ be any (allocation, payment) pair. For any
    lottery $(x,p)$ such that there exists $v$ with $0 \leq \lrau{v}{x, p} \leq
    \lrau{v}$ and $x \geq \prob{\alloc = 1}$, the expected revenue of $(x,p)$
    is at least as large as the expected revenue of $\lotteryp$. If
    $\lrau{v}{x, p} < \lrau{v}$, then the revenue is strictly larger.
\end{lemma}
\begin{proof}
    Let $x' = \prob{\alloc=1}$ and $p' = \expect{\paymt}/x'$. Note that the
    expected revenue from $(x',p')$ is exactly $\expect{\paymt}$. By
    Lemma~\ref{lem:util-dominance}, $\lrau{v} \leq \lrau{v}{x',p'}$. Since $0
    \leq \lrau{v}{x,p} \leq \lrau{v}{x',p'}$, we have 
    \begin{align*}
        \wt(x')(v-p') &=    \lrau{v}{x',p'} \\
                      &\geq \lrau{v}{x,p} \\
                      &= \wt(x)(v-p).
    \end{align*}
    Since $x \geq x'$ by assumption, $\wt(x) \geq \wt(x')$. Therefore $p \geq
    p'$, and so $xp \geq x'p' = \expect{\paymt}$. Note that if $\lrau{v} >
    \lrau{v}{x,p}$, then $xp > \expect{\paymt}$.
\end{proof}

For any IC/IR mechanism $\lotteryp$ defined on the interval $[a, b]$, let
$\convU(v)$ be the lower convex envelope of $\lrau{v}{\lottery[v]}$.  That is,
$\convU$ is the maximal convex function upper bounded by the utility curve. 

\begin{definition}
    \label{def:differentials}
    For any convex function $f : I \to \R$, the {\em subdifferential} of $f$ at
    $x \in I$ is
    \[
        \subdif f(x) = \{m : f(x') - f(x) \geq m(x' - x) \; \forall x' \in I\}.
    \]
    Likewise, for any concave function $g : I \to \R$, the {\em
    superdifferential} of $g$ at $x \in I$ is
    \[
        \supdif g(x) = \{m : g(x') - g(x) \leq m(x' - x) \; \forall x' \in I\}.
    \]
\end{definition}
Let $\subdif^*f(x) = \max\{m \in \subdif f(x)\}$ be the maximal slope of a line
tangent to $f$ at $x$. Similarly, define $\supdif^*g(x) = \min\{m \in \supdif
g(x)\}$ to be the minimal slope of a line tangent to $g$ at $x$.

\begin{lemma}
    \label{lem:util-conv-env}
    $\subdif\convU(v) \subseteq [0, 1]$ for all $v \in [a, b]$.
\end{lemma}
\begin{proof}
    First, $\convU(v)$ is an nondecreasing function of $v$ because, by
    Lema~\ref{lem:concaveUtil}, $\lrau{v}$ is nondecreasing for all $\lotteryp$
    in $\menu$. So $\subdif\convU(v) \subseteq [0,\infty)$ for all $v$.

    Let $v^*$ be any value in $[a,b]$. Since $\convU(v)$ is the lower convex
    envelope of $\lrau{v}{\lottery[v]}$, there exists $v_0 \leq v^*$ such that
    for all $v' > v^*$,
    \begin{align*}
        \subdif^*\convU(v^*) &\leq \frac{\lrau{v'}{\lottery[v']} -
                \lrau{v_0}{\lottery[v_0]}}{v' - v_0} \\
            & \leq \frac{\lrau{v'}{\lottery[v']} -
                \lrau{v_0}{\lottery[v']}}{v' - v_0} \\
            & \leq \frac{v' - v_0}{v' - v_0} = 1
    \end{align*}
    The second inequality follows by the definition of $\lotteryp[v_0]$, and
    the third follows from Lemma~\ref{lem:concaveUtil} together with the fact
    that $\wt(\prob{\paymt[v']=1}) \leq 1$.
\end{proof}

\begin{numberedtheorem}{\ref{thm:single-shot-opt}}
%\begin{theorem}{\ref{thm:single-shot-opt}}
  For any revenue optimal IC mechanism $(\allocs, \paymts)$ in the
  single-shot setting, the buyer's utility function
  $\lrau{v}{\lottery[v]}$ is convex and nondecreasing. Furthermore,
  there exists an optimal ex-post IR mechanism that can be described
  as a menu of binary lotteries.
\end{numberedtheorem}
%\end{theorem}

%\begin{figure}[htbp]
%    \begin{center}
%    \scalebox{.6}{\input{figures/cvx_env_a.pgf}}
%    \scalebox{.6}{\input{figures/cvx_env_b.pgf}}
%    \caption{{\em Left:} The risk-averse utility for a menu of lotteries as a
%        function of the buyer's value, where the IC utility for the preferred
%        menu option is the dashed line.  {\em Right:} The lower convex
%        envelope of the risk-averse utility, achievable using binary lotteries
%        for each linear segment, and earning the seller more revenue.}
%    \label{fig:cvx_env}
%    \end{center}
%\end{figure}

\begin{proof}
    First, we show that we can find a menu of lotteries $(x(v), p(v))$
    which obtain a utility curve equal to $\convU$.  Fix $v_0 \in [a,b]$.  Let
    $m_0 = \subdif^*\convU(v_0)$.  Note that if $m_0 = 0$, then
    $\expect{\alloc[v_0]} = 0$ by Lemma~\ref{lem:concaveUtil}, and so
    $\expect{\paymt[v_0]} = 0$ by IR. So assume that $m_0 > 0$.  Let $p = v_0 -
    \convU(v_0) / m_0$ and let $x = \wtinv(m_0)$.  Since $\convU(a) \leq a$
    and $\convU(v)$ is convex and nondecreasing, $\convU(v_0) \leq m_0v_0$, so
    $p \geq 0$. By Lemma~\ref{lem:util-conv-env}, $m_0 \in [0, 1]$, so $x$ is a
    well-defined probability. So $(x,p)$ is a feasible lottery with
    utility curve tangent to $\convU$ at $v_0$: $\lrau{v_0}{x, p} = (v_0 -
    p)\wt(x) = \convU(v_0)$ and $\dlrau{v_0}{x, p} = m_0$.

    It remains to show that $xp \geq \expect{\paymt[v_0]}$. For ease of
    notation, let $x_0 = \prob{\alloc[v_0] = 1}$. We will show that the
    conditions of Lemma~\ref{lem:rev-dominance} are satisfied: namely,
    that $x \geq x_0$ and there exists $v$ such that $0 \leq \lrau{v}{x, p}
    \leq \lrau{v}{\lottery[v_0]}$. The latter condition is satisfied at $v_0$
    by construction, and the second inequality is strict if $\convU(v_0) <
    \lrau{v}{\lottery[v_0]}$.

    To show $x \geq x_0$, it suffices to show $m_0 \geq \wt(x_0)$.  Suppose
    $\convU(v_0) = \lrau{v_0}{\lottery[v_0]}$.  It follows that
    \begin{align*}
        m_0 &  =  \subdif^*\convU(v_0) \\
            &\geq \supdif^*\lrau{v_0}{\lottery[v_0]} \\
            &\geq \wt(x_0).
    \end{align*}
    The first inequality holds because $\convU(v') \geq
    \lrau{v'}{\lottery[v_0]}$ for all $v' \geq v_0$. The second follows by
    Lemma~\ref{lem:concaveUtil}. 

    Otherwise, if $\convU(v_0) < \lrau{v_0}{\lottery[v_0]}$, there exists a
    point\footnote{Possibly $v' = b$. Note that $\convU(b) =
    \lrau{b}{\lottery[b]}$.} $v' > v_0$  such that $\convU(v') =
    \lrau{v'}{\lottery[v_0]}$ and $\convU(v') = \convU(v_0) + m_0(v' - v_0)$.
    Thus,
    \begin{align*}
        \lrau{v'}{\lottery[v_0]} &= \convU(v_0) + m_0(v' - v_0) \\
            &\leq \lrau{v_0}{\lottery[v_0]} + m_0(v' - v_0).
    \end{align*}
    Rearranging and appealing to Lemma~\ref{lem:concaveUtil}, we have
    \begin{align*}
        m_0 &\geq  \frac{\lrau{v'}{X_{v_0}, P_{v_0}} -
                \lrau{v_0}{X_{v_0}, P_{v_0}}}{v' - v_0} \\
            &\geq \wt(x_0)
    \end{align*}
\end{proof}

%%========================
%%I changed the title of this section since it only contains the monotonicity requirement for the famility of Y's
%%=========================
%\subsection{Single-shot risk robust revenue }
\subsection{Monotonicity is required for Lemma~\ref{lem:rev-risk-monotonicity}.}
\label{sec:app-monotone}

%%%===============
%% Put the Proof to the main body, lemma statement is no longer needed
%%===============
%%\begin{numberedlemma}{\ref{lem:rev-risk-monotonicity}}
%\begin{lemma}{\ref{lem:rev-risk-monotonicity}}
%    Let $\wtfam$ be a monotone non-crossing family of weighting
%    functions, and let $\wt_1\ge \wt_2$ be any two weighting functions in
%    $\wtfam$. Then, for any IC mechanism $\mech$ composed of binary
%    lotteries, we have $\rev_{\wt_2,F}(\mech) \geq \rev_{\wt_1,F}(\mech)$.
%%\end{numberedlemma}
%\end{lemma}
%\begin{proof}
%    Fix a value $v$. We abuse notation and write $\wt_1(v)$ to mean $\wt_1(x(v))$, and similarly
%    with profile $\wt_2$ and bid $b$. By incentive-compatibility of $\mech$, $\wt_1(v)(v-p(v)) \geq \wt_1(b)(v-p(b))$
%    for all $b$.  By assumption and monotonicity of $x(v)$ in $v$, for any $b < v$, $\wt_2(v)/\wt_1(v) \geq \wt_2(b)/\wt_1(b)$. Multiplying
%    these inequalities gives
%    $$\frac{\wt_2(v)}{\wt_1(v)}\wt_1(v) (v-p(v)) \geq \frac{\wt_2(b)}{\wt_1(b)}\wt_1(b)(v-p(b)),$$
%    or equivalently,
%    $$\wt_2(v)(v-p(v)) \geq \wt_2(b)(v-p(b)).$$
%    Therefore, a buyer $v$ will not underreport his value, and so the revenue
%    of $\mech$ can only increase under profile $\wt_2$.
%\end{proof}

The following example shows that the condition of monotonicity is necessary for Lemma~\ref{lem:rev-risk-monotonicity}.
\begin{example}
    Suppose $\dist$ is a point mass at $v=1$, and $\mech$ is a menu consisting of two
    lotteries: $x_1 = 1-2\epsilon/3$, $p_1 = 3/4$; and $x_2 = 1/2$, $p_2 =
    \epsilon$. Let $\wt_1(x) = \max(3x/2 - 1/2, x^2)$ and $\wt_2(x) = x^2$. Observe that $\wt_1$ and $\wt_2$ are non-crossing but not monotone. 

    Under $y_1$, the buyer chooses the first lottery:
    \begin{align*}
        y_1(x_1)(1-p_1) &= \frac32(1-2\epsilon/3)\frac14 \\
            &= (1/4)(1-\epsilon) \\
            &= y_1(x_2)(1-p_2).
    \end{align*}
    The revenue under $y_1$ is therefore $\approx 3/4$.
    
    However, under $y_2$, the buyer chooses the second lottery:
    \begin{align*}
        y_2(x_1)(1-p_1) &= (1-3\epsilon/2)^2\frac14 \\
            &< \frac14(1-\epsilon) \\
            &= y_2(x_2)(1-p_2).
    \end{align*}
    The revenue under $y_2$ is therefore $\epsilon/2$.
\end{example}

\subsection{Lower bound for risk-robust approximation.}
\label{sec:app-risk-robust}

We will now prove Lemma~\ref{lem:lb-observations}.

\begin{numberedlemma}{\ref{lem:lb-observations}}
%\begin{lemma}{\ref{lem:lb-observations}}
    For the setting described in Section~\ref{sec:risk-robust-lb}, if
    $\left\{(p,\lp)\right\}_{p\in P}$ gives a risk-robust $c$-approximation,
    the following properties hold without loss of generality:
    \begin{enumerate}
        \item For all $p \in P$, $\lp \geq 1$.
        \item For any $\eps$, $P_{\eps} \supseteq [1,p_\eps]$ where
            $p_\eps$ is defined such that 
            \begin{align*}
                1 + \expect[v\sim F_1]{\min(v,p_{\eps})} =
                        \frac 1c \expect[v\sim F_1]{\min(v,\g_{\eps}(0))}.  
            \end{align*}
        \item For every $\eps$, $p_\eps = \alpha_c \g_{\eps}(0)^{1/c}$ for some
            constant $\alpha_c > 0$ depending only on $c$.
        \item For every $\eps$, the left derivative of
            $\g_{\eps}(\lp)$ with respect to $p$ at $p=p_\eps$ must be $\ge 1$. 
   \end{enumerate}
\end{numberedlemma}
%\end{lemma}

    % \item For $\eps<\eps'$, $P_{\eps'}\subseteq P_{\eps}$. 
    % \item \label{lem:contiguous} We may assume without loss of generality that
    %     effective menus are connected; adding missing points only improves
    %     revenue at all risk profiles in $\wtfam$.
    % \item \label{lem:starts-at-1} It is without loss of generality to assume
    %     that the effective menu is of the form $[1,p]$.

\begin{proof}
We prove the statements in sequence:
\begin{enumerate}
      \item Suppose there is a menu option $(p,\lp)$ for $p\in P$ with
       $\lp<1$. Consider replacing this menu option with the option
       $(p+1-\lp,1)$. Observe that the buyer's risk-averse utility
       under the two options is identical---relative to the original
       option, the buyer loses an additive amount of $1-\lp$ in his
       first-stage utility but gains the same additive amount of
       $1-\lp$ in his second-stage utility in the new menu option. On
       the other hand, the seller's revenue under the two options is
       also identical---the seller's first-stage revenue is higher by
       an additive $1-\lp$ amount under the new option, but his
       second-stage revenue is lower by the same additive $1-\lp$
       amount. Therefore, without loss of generality, we may replace
       $(p,\lp)$ with the new option $(p+1-\lp,1)$ without affecting
       the buyer's utility or the seller's revenue.
    \item We show below that the effective menu $M_{\eps}$ must be of the form 
    $[1, p]$. Then, the revenue from such an effective menu is $1 +
    \expect[v\sim F_1]{\min(v,p)}$ because the buyer purchases the
    option with price $\min(v,p)$ in the first stage, and the
    mechanism gets a fixed revenue of $1$ in the second stage. In
    order to obtain  a $c$-approximation, this quantity must be at
    least $\frac 1c \expect[v\sim F_1]{\min(v,\g_{\eps}(0))}$,
    implying, by the definition of $p_\eps$ that $p\ge p_\eps$.
%  Then if $p_\eps$ is defined such that 
% \begin{align}
% \notag
% 1 + \expect[v\sim F_1]{\min(v,p_{\eps})} =
% \frac 1c \expect[v\sim F_1]{\min(v,\g_{\eps}(0))},\end{align}
%          by definition of $p_\eps$, if $p < p_\eps$,
%         then this mechanism does not give a $c$-approximation.
    \begin{itemize}
    \item First, for $\eps<\eps'$, $\menu_{\eps'}\subseteq\menu_{\eps}$.  For
        any price $p$ in $\menu_{\eps'}$, and any $p' < p < 1/\eps$, it must be
        the case that the effective price for $p'$ is at least as large:
            \[
                p' - \g_{\eps'}(\lp[p']) \geq p - \g_{\eps'}(\lp).
            \]
        However, as $\g_{\eps}(\lp) = \int _{\lp} ^{1/\eps}
            \frac{1-F_2(v_2)-\eps}{1-\eps} dv_2$, then
        \begin{align*}
            \g_{\eps}(\lp)-\g_{\eps}(\lp[p']) &=
                \int_{\lp}^{\lp[p']} \frac{1-F_2(v_2)-\eps}{1-\eps} dv_2 \\
                &\geq \int _{\lp} ^{\lp[p']} \frac{1-F_2(v_2)-\eps'}{1-\eps'} dv_2 \\
                &= \g_{\eps'}(\lp)-\g_{\eps'}(\lp[p']) \\
                &\geq p - p'
        \end{align*}
        Hence the effective price of $p - \g_{\eps}(\lp)$ is preferable to any
            smaller first-stage price under $\eps$; a buyer with value $p$
            would prefer this menu option to all others.
    \item Second, we claim that without loss of generality effective menus are
        contiguous; adding missing points only improves revenue at all risk
            profiles in $\wtfam$.  Suppose not, and consider two first-stage
            prices in the effective menu, $p_1$ and $p_2 > p_1$, where no
            intermediate first-stage price is in the effective menu.  Then $p_1
            - \g(\lp[p_1]) \ge p_2 - \g(\lp[p_2])$.  For every $\alpha \in
            (0,1)$, we can add a menu option $(p', \lp[p'])$ with first-stage
            price $p' = \alpha p_1 + (1-\alpha) p_2$.  To have a non-increasing
            effective price, we note that it is possible to set $\lp[p'] =
            \g^{-1}(\alpha \g(\lp[p_1]) + (1-\alpha) \g(\lp[p_2])$ since
            $\lp[p_1] > \lp[p_2]$ and $\g(\lp[p_1]) < \g(\lp[p_2])$.  Then a
            buyer with value $v \in (p_1, p_2)$ will pay $v$ instead of $p_1$,
            earning only more revenue for the seller.
      \item Finally, it is without loss of generality to assume that the
          effective menu is of the form $[1,p]$.  From the first two
            observations we know that the menu is of the form $[l,h]$, earning
            revenue in the first stage equal to $\expect[v \sim F_1]{\min(v,h)
            \cdot \mathbbm{1}_{[v > l]}}$.  Suppose that $h-l = p$.
        % Because $l$ and $h$ are both in the effective menu, then $$l -
        % \g(\infty) \geq l - \g(\lp[l]) = h - \g(\lp[h]) \geq h - \g(0).$$
        % Since $\g(\infty) = 0$, then $h \leq l + \g(0)$.  
        Because the revenue is the area above the c.d.f from $l$ to $h$, a
            window of width $p$, this is strictly increased by shifting the
            window to the left, at $[1, p]$.
        %\kgnote{Add figure with area above the CDF?}
        \end{itemize} 
    \item This follows by recalling that for any $p$, $\expect[v\sim
      F_1]{\min(v,p)} = 1+\ln p$, and then solving for $p_\eps$.
    \item For any $p < p_\eps$, because $p_\eps$ is in the effective menu, then
        for all $p < p_\eps$,
        \begin{align*}
              & p_\eps - \g_\eps(\lp[p_\eps]) \leq p - \g_\eps(\lp). \\
            \intertext{Therefore}
              & \lim _{p \rightarrow p_\eps} \frac{\g_\eps(\lp[p_\eps]) -
              \g_\eps(\lp)}{p_\eps - p} \geq 1.\\
        \end{align*}%
%\kgfatal{TBD}        %For all $p \in P$, $\lp \geq 1$.
    %second-stage prices below 1 lose as much second-stage revenue as they gain on the first stage.
    % Prefer (p_eps, 1) to (p, \lp) for $\lp < 1$.
    % Rev lost on day 2: 1 - \lp
    % Rev gained on day 1: 
\end{enumerate}
\end{proof}

\begin{proofof}{Lemma~\ref{lem:g0-lb}}
%\begin{proof}{Lemma~\ref{lem:g0-lb}}
    For any $\eps \ge e^{-n}$,
    \begin{align*}
        \g_\eps(0) &= 1 + \int_1^{e^n}\wte(1/v)dv \\
            &= 1 + \int_1^{1/\eps}\left[\frac1v(1+\eps) - \eps\right]dv +
                \int_{1/\eps}^{e^n}\frac1{v^2}dv \\
            &= 1 + (1+\eps)\ln 1/\eps - \eps(1/\eps-1) + \eps - e^{-n} \\
            &= 2\eps - e^{-n} + (1+\eps)\ln 1/\eps.
    \end{align*}
    Since $\eps \ge e^{-n}$, this is at least $\ln 1/\eps$. If $\eps < e^{-n}$,
    a similar argument shows $\g_\eps(0) \geq n$.
\end{proofof}
%\end{proof}